%% file: ms.tex
\documentclass[twocolumn]{aastex701}
\usepackage{amsmath,amssymb}
\usepackage[frozencache,cachedir=.]{minted}
\usepackage{xcolor}
\hypersetup{colorlinks=true,linkcolor=linkcolor,citecolor=linkcolor,urlcolor=linkcolor}
\definecolor{linkcolor}{rgb}{0,0,0.25}
\input{macros}

\shorttitle{galpy with jax and PyTorch backends}
\shortauthors{Bovy}

\begin{document}

\title{\galpy\ with differentiable, accelerated backends: galactic dynamics in \jax\ and \torch}

\author[0000-0001-6855-442X]{Jo Bovy}
\affiliation{David A. Dunlap Department of Astronomy and Astrophysics, University of Toronto, 50 St.~George Street, Toronto, ON M5S 3H4, Canada}
\email{bovy@astro.utoronto.ca}

\begin{abstract}
\input{abstract}
\end{abstract}


\input{introduction}
\input{design}
\input{potentials}
\input{orbits}

\input{actionangle}
\input{dfs}
\input{streams}
\input{process}
\input{examples}

\input{conclusion}

\input{acknowledgments}

\appendix
\input{appendix_dxdv}

\bibliography{refs}
\bibliographystyle{aasjournalv7}

\end{document}

%% file: macros.tex
\allowdisplaybreaks
\newcommand{\eg}{e.g.}

\newcommand{\dd}{\mathrm{d}}
\newcommand{\sectionname}{$\mathsection$}

\newcommand{\eqnname}{equation}

\renewcommand{\vec}[1]{\ensuremath{\mathbf{#1}}}

\newcommand{\vecx}{\ensuremath{\vec{x}}}
\newcommand{\vecv}{\ensuremath{\vec{v}}}
\newcommand{\vecw}{\ensuremath{\vec{w}}}
\newcommand{\vecj}{\ensuremath{\vec{J}}}

\newcommand{\veca}{\ensuremath{\boldsymbol\theta}}

\newcommand{\flag}[1]{\texttt{#1}}
\newcommand{\galpy}{\flag{galpy}}
\newcommand{\numpy}{\flag{numpy}}
\newcommand{\scipy}{\flag{scipy}}
\newcommand{\matplotlib}{\flag{matplotlib}}
\newcommand{\astropy}{\flag{astropy}}
\newcommand{\jax}{\flag{jax}}
\newcommand{\torch}{\flag{PyTorch}}
\newcommand{\python}{\flag{Python}}
\newcommand{\cc}{\texttt{C}}
\newcommand{\diffrax}{\flag{diffrax}}
\newcommand{\torchode}{\flag{torchode}}

\newcommand{\github}{\texttt{GitHub}}
\newcommand{\pot}{\texttt{Potential}}
\newcommand{\orb}{\texttt{Orbit}}

\newminted{python}{fontsize=\footnotesize,autogobble,breaklines,breakanywhere}
\newenvironment{code}{\VerbatimEnvironment\begin{pythoncode}}{\end{pythoncode}}
\newenvironment{codefig}{\VerbatimEnvironment\begin{minted}[mathescape,linenos,numbersep=5pt,autogobble,frame=lines,framesep=2mm,fontsize=\small]{python}}{\end{minted}}

%% file: abstract.tex
We describe an extension of \galpy, the \python\ library for galactic
dynamics, that adds \jax\ and \torch\ as computational backends
alongside its existing \numpy\ and \cc\ implementation. With it,
\galpy's gravitational potentials, orbit integration, action--angle
transformations and their inverses, and distribution functions,
including those for tidal streams, run on GPUs, can be compiled as part
of a user's program, and can be differentiated with respect to their
inputs and parameters, from the force at a point to the samples drawn
from a distribution function. The user-facing design is kept simple: the
active backend follows from the type of the arrays that are passed in,
the \galpy\ interface is unchanged, and the \numpy\ path gives the same
results as before. We describe the design of the
backend framework and its implementation in each of \galpy's modules,
with examples of use, and we discuss its limitations. On a GPU, batches
of orbits, samples from distribution functions, and stream models run
one to five orders of magnitude faster than on a single CPU core. To
illustrate the new capabilities, we present simple examples of fitting
the Galactic potential with a hypervelocity star and of learning the
inverse action--angle transformation with a neural network trained
through \galpy. An AI coding agent wrote
essentially all of the code under the detailed direction and review of
the human author, and we describe how we organized this work and how we
established the correctness of the result. This work is the start of
\galpy's version~2, which is available at
\url{https://github.com/jobovy/galpy} and documented at
\url{https://docs.galpy.org/en/v2/}.

%% file: introduction.tex
\section{Introduction}\label{sec:intro}

Software for galactic dynamics has become a standard part of the toolkit
of Galactic astronomy. Codes such as the NEMO Stellar Dynamics Toolbox
\citep{Teuben1995}, \galpy\ \citep{Bovy2015}, \texttt{gala}
\citep{PriceWhelan2017}, and \texttt{AGAMA} \citep{Vasiliev2019}
provide gravitational potentials, orbit integration, action--angle
coordinates, and distribution functions (DFs) in a form that can be
applied directly to observed stars, both to characterize the orbits of
individual objects and collections of objects, and to build dynamical
models that are fit to data. Such codes are useful because the motion of
a star in a given gravitational potential is a solved problem that
nevertheless requires careful numerical work, which a shared,
well-tested implementation saves every user from repeating, and because
they let developers build sophisticated models within a well-defined
framework that users can apply without implementing them; \galpy's
action--angle methods are an example, because their implementation is
far from trivial and they have been used in many analyses that would
otherwise not have been attempted. The kinematic data from the Gaia
mission \citep{Gaia2016,Gaia2018,Gaia2023} have made such codes
indispensable, because for the first time orbits can be computed for
hundreds of millions of stars.

\citet{Bovy2015} introduced \galpy\ as a \python\ package, with \cc\
implementations of the computationally-expensive parts, for computing orbits,
action--angle coordinates, and DFs in general galactic potentials. It has
since been used in more than a thousand papers. %
Most of
these use \galpy\ as a tool: orbits are integrated, and orbital
parameters and actions computed, in a model of the
Milky Way to place objects in the Galaxy and to characterize their
motion, for stars in the disk with
measured abundances and ages \citep[\eg,][]{Frankel2020,Xiang2022},
the substructure of the stellar halo \citep[\eg,][]{Belokurov2020,
Lane2023}, globular clusters and dwarf galaxies \citep[\eg,][]{Fritz2018,
Pace2022}, hypervelocity and runaway stars and the natal kicks of
pulsars and compact objects \citep[\eg,][]{Boubert2018,Marchetti2022,
ODoherty2023}, open clusters and the star-forming regions near the
Sun \citep[\eg,][]{Zucker2022,Swiggum2024}, the hosts of exoplanets
and interstellar objects \citep[\eg,][]{Stefansson2023,Forbes2025},
and galaxies other than our own \citep[\eg,][]{Tonnesen2019}. A
smaller set of papers, perhaps one in six, uses \galpy\ as a modeling
framework in which a DF or a potential is fit to data: models of tidal
streams and of their perturbations by dark-matter subhalos
\citep[\eg,][]{Bovy2016,Malhan2019,Banik2021}, action-based models of
the disk \citep[\eg,][]{Trick2016,Mackereth2019}, DF and Jeans models
of the halo \citep[\eg,][]{Lane2023}, models of the Galactic bar
\citep[\eg,][]{Hunt2018,Lucey2023}, and measurements of the Milky Way's
mass and potential, from its rotation curve and halo tracers
\citep[\eg,][]{Cautun2020}, from the local escape velocity
\citep[\eg,][]{Necib2022}, and from the clustering of stream stars in
action space \citep[\eg,][]{Yang2020,Reino2021}. It is for this second
kind of use that derivatives matter, because fitting a model to data
requires the likelihood to be evaluated many times over the model's
parameters, and the efficiency of this search is set by whether its
derivatives are available.

\galpy\ has changed and grown a lot since 2015.
The number of potentials grew from 22 to 72, with the growth
shifting from specific density profiles to general solvers (basis-function,
multipole, and interpolated potentials) and to wrappers that make any
potential time-dependent, rotating, or tilted, and a library of
Milky-Way models from the literature and velocity-dependent forces,
such as dynamical friction and the fictitious forces in non-inertial
reference frames, were added. The
\orb\ class was rewritten as an arbitrarily-shaped array of orbits that
are integrated in parallel, such that millions of orbits can be analyzed
at once; orbits can be initialized by name through a query to SIMBAD,
and their eccentricities, peri- and apocenters, and maximum heights can
be estimated from the actions at a small fraction of the cost of an
integration \citep{MackerethBovy2018}. DFs for tidal streams were extended to stream gaps
\citep{Sanders2016} and to particle-spray models \citep[][implemented
in \galpy\ by \citealt{Qian2022}]{Fardal2015,Chen2025}, a general
framework for isotropic and anisotropic spherical DFs was added
\citep{Lane2022}, as were tools for Jeans modeling, inverse
action--angle transformations, chaos diagnostics, and surfaces of
section. Over the same period the code grew from 38,000 to 95,000 lines
of \python\ and from 8,000 to 26,000 lines of \cc, and the test suite
grew from 560 to 2,360 tests that cover $99.94\,\%$ of the code.

Over the same decade, numerical computing has been reshaped by
frameworks built for machine learning, chiefly \jax\
\citep{Bradbury2018} and \torch\ \citep{Paszke2019}. These provide
array programming that runs on CPUs and GPUs alike, just-in-time
compilation of whole programs, and automatic differentiation of
arbitrary code \citep[see][for a review]{Baydin2018}. A dynamics code
written in such a framework gains three things: it runs on a GPU, where
large batches of orbits or stars are processed at once; it can be
compiled into a single program rather than executed one array operation
at a time; and every quantity that it computes---a force, an orbit, an
action, the value of a DF---can be differentiated with respect to every
input, be it an initial condition or a parameter of the potential.
Derivatives are what gradient-based inference requires, whether this is
Hamiltonian Monte Carlo \citep{Duane1987,HoffmanGelman2014}, variational
inference, or the optimization of models with many parameters, such as
neural networks that stand in for a potential or a DF. Codes built on
these frameworks are now in use in cosmology
\citep[\eg,][]{Modi2021,Campagne2023,Li2024} and, over the past two
years, in galactic dynamics. \texttt{galax} \citep{Starkman2026}, a
re-implementation of \texttt{gala} in \jax, provides gravitational
potentials, orbit integration, and particle-spray stream models, all of
them differentiable and able to run on a GPU; it supplied the halo
potentials whose acceleration directions \citet{Wu2026b} and
\citet{Euclid2026} compare with the curvature of stellar streams around
other galaxies to constrain the shapes of their halos, the orbits with
which \citet{Starkman2026b} forecast how well such curvature tests
separate dark-matter models, and the GPU integration of streams
perturbed by subhalos that \citet{AnauMontel2026} differentiate to
obtain Fisher forecasts. With other \jax\ codes, \citet{Viterbo2025}
fit the progenitor and the host potential of a mock stream by gradient
descent through an $N$-body simulation, \citet{Chemaly2026} generate
the many particle-spray models needed to infer the distribution of halo
flattenings from a population of streams, and \citet{Nibauer2025}
compute the response of streams to subhalos by integrating linearized
equations of motion. With gradient-based samplers in \jax,
\citet{Nibauer2025b} fit a spline model of the track of GD-1 to measure
the accelerations along it, and \citet{PriceWhelan2025} fit a flexible
model of vertical orbits to the variation of stellar abundances.
Normalizing flows for the DF, differentiated automatically, recover the
potential from phase-space snapshots of simulated galaxies
\citep{Green2023,An2021,Buckley2023} and from 5.6 million \textit{Gaia}
stars \citep{Kalda2025}. Differentiating an entire simulation
\citep{Viterbo2025,AnauMontel2026}, or fitting a neural-network
potential to millions of stars \citep{Kalda2025}, is not practical with
codes that only return values.

In this paper, we describe an extension of \galpy\ that adds \jax\ and
\torch\ as \emph{backends} to its existing \numpy\ (and \scipy) and
\cc\ implementation. We support both frameworks deliberately: a
solution that works for both cannot depend on the particulars of
either, which keeps the implementation agnostic of the backend and
should make it easy to add a third framework, or to replace one, as the
ecosystem of array frameworks changes. The principle
that guided this work is that nothing should change for the existing
user: the \numpy\ path produces results that are byte-for-byte
identical to those of the previous version, with a handful of approved
exceptions that we list, the interface is unchanged, and \galpy's
existing test suite now runs under every backend, in eager and in
compiled mode, with the same tolerances, again with very few
exceptions. Backends are selected by the type of the arrays that are
passed in or through a context manager; \galpy's code is written
against a backend-agnostic array namespace; and \galpy\ does not
compile anything itself, but is written such that it can be compiled
and differentiated as part of a user's own program, of which a \galpy\
call is generally only a part. All of \galpy's core functionality is
covered: the potentials and their forces and derivatives, orbit
integration either inside the backend or through the \cc\ integrators
with their variational equations, the action--angle transformations
and their inverses, and the DFs, including those for tidal
streams. All of these are differentiable with respect to their inputs
and parameters, and they run on GPUs. We list the few things that are
not covered where they come up; the main gap is the use of \astropy\
units inside compiled programs, which a \jax-aware units package such
as \texttt{unxt} \citep{Starkman2025} may make possible in the future.

Claude, an AI coding agent developed by Anthropic, wrote essentially all
of the code described in this paper, working in the \texttt{Claude Code}
environment under the detailed direction and review of the human author (JB)
between June and October 2026. We discuss how this was done, how we
organized the work, and how we established the correctness of the code
in \sectionname~\ref{sec:process}; a statement on the use of AI, which
also explains the use of ``we'' in a paper with a single human author,
appears at the end of the paper.

The outline of this paper is as follows. \sectionname~\ref{sec:design}
describes the design of the backend framework: how a backend is chosen,
how we made \galpy's code backend-agnostic, and how we made the test
suite run under every backend.
\sectionname s~\ref{sec:potential}--\ref{sec:streams} go through
\galpy's main modules in turn---potentials, orbit integration and its
derivatives, action--angle coordinates, DFs, and models of tidal
streams---each starting from the general approach and then giving the
details that are specific to the module, with examples of use.
\sectionname~\ref{sec:process} describes the development process,
\sectionname~\ref{sec:examples} presents fuller examples that combine
several parts of the code in a single analysis, and
\sectionname~\ref{sec:conclusion} concludes. An Appendix gives the
variational equations behind the orbit derivatives, which are also
useful outside of the backend frameworks, for computing the sensitivity
of an orbit to its initial conditions and Lyapunov exponents.

%% file: design.tex
\section{Design of the backend framework}\label{sec:design}

\subsection{Principles}\label{sec:design-principles}

We set six rules before writing any code, and the rest of this paper
is largely a description of how we kept them. First, the \numpy\ path had to remain
byte-identical to the previous version of \galpy: every backend helper is
a strict pass-through when the active array library is \numpy, such that
migrated code written as \texttt{xp.sqrt(x)}, where \texttt{xp} is the
resolved namespace, runs the identical \numpy\ function on the identical
object. We enforce this rule with tests and, for each pull request (PR), by comparing
checksums of outputs between a checkout with and one without the change;
where the \numpy\ result did change, we flagged and approved the
change, and we point these places out where they arise. There are
three deliberate changes: the King model now solves its Poisson
equation to a relative tolerance of $10^{-10}$ rather than \scipy's
default of $10^{-3}$, which changes its potential and DF at the
$10^{-5}$ to $10^{-3}$ level, and the velocity sampler of the
quasi-isothermal disk DF, the samplers of the two-dimensional disk DFs,
and the sampler of the stream DF draw by inverse transformation rather
than by rejection, which changes the particular draws but not their
distribution, except for the two-dimensional disk DFs, whose previous
draws had a mean azimuthal velocity that was biased by a few per cent
and whose new draws do not (\sectionname~\ref{sec:df-disk}). In addition, the forces of the Einasto profile and of
the power law with an exponential cutoff were rewritten in forms that
are accurate and differentiable at small radii, such that their
\numpy\ values differ from the previous version by a few units in the
last place (at most $5\times10^{-15}$ relative), which propagates at
the same level to everything computed from these potentials' \python\
forces; the formulas of a dozen other potentials were restructured for
the same reasons, but evaluate in the same floating-point operation
order as before and remain byte-identical, which a test checks.
Accordingly, the tolerances of a handful of \numpy\ tests were
loosened slightly. Second, the \galpy\ interface had to stay unchanged: the same classes, methods,
and keywords, with the backend generally determined by the arrays that are passed
in (see below). Third, \galpy's code was to be written against a backend-agnostic array
namespace, with backend-specific code confined to a new \texttt{galpy.backend}
subpackage and never written as \texttt{if backend == `jax'} inside the
physics modules. Fourth, \galpy\ must use the basic array operations of \jax\ and
\torch\ and little else; we implemented the numerical methods that
\scipy\ supplies on the \numpy\ path---special functions, root finding,
quadrature, interpolation, sampling---in \texttt{galpy.backend} and
tested them against \scipy\ and against arbitrary-precision references,
rather than taking them from the frameworks' own, uneven, \scipy-like
modules. Fifth, we decided that \galpy\ should never compile anything internally itself: 
instead we required the code
to be traceable by \texttt{jax.jit} and \texttt{torch.compile}, and the
user decides what to compile. Sixth, the existing test suite, run
under every backend with the same tolerances, had to pass in full;
this was the main acceptance test, but not the only one, because the
new functionality---differentiation, compilation, running on a
GPU---has no existing tests, and we wrote new ones for it, as
described in \sectionname~\ref{sec:process-correctness}. Supporting
two frameworks from the start served the third rule
(\sectionname~\ref{sec:intro}).

\subsection{Choosing a backend}\label{sec:design-choosing}

The general design approach for using the backends is that the backend follows from the data: 
a \galpy\ function that receives a \jax\
array or a \torch\ tensor computes with that framework and returns an
array of the same kind, while a function that receives \numpy\ arrays or
\python\ scalars computes with \numpy\ as before. Thus, the radial force
of the standard Milky-Way model is evaluated on a \jax\ array as
\begin{code}
from galpy.potential import MWPotential2014
R= jnp.linspace(0.5,2.,1001)
FR= MWPotential2014.Rforce(R,0.) # jax array
\end{code}
without any further instruction, and with a \torch\ tensor the same
call returns a tensor. The data include an object's parameters: a
potential, distribution function, or action-angle object that is set up
with a backend array as one of its parameters computes with that backend
in every call, also at \python\ floats,
\begin{code}
pot= NFWPotential(a=jnp.array(2.))
pot(1.1,0.2) # jax array
\end{code}
and an \orb\ integrated in such a potential returns backend arrays as
well. In short, backends are \emph{greedy}: any backend array among
the inputs of a calculation, be it a coordinate, an initial condition,
or the parameter of an object, turns on the backend for that
calculation and its output. We chose this rule because we assume that
a user who passes a backend array wants the backend to perform the
computation and to receive backend arrays as results, whether to
differentiate, to compile, or to run on an accelerator. A parameter that
is not differentiated keeps the fast paths that a differentiated one
cannot take: an orbit in a potential with a concrete backend parameter
is still integrated by the \cc\ integrators, with the \cc-computed
sensitivities that make its result differentiable with respect to the
initial conditions. When there is no array to dispatch on, as when
a potential is evaluated at \python\ floats, the default
backend is used; it is \numpy\ unless a different one is set in the
configuration file or through a context manager,
\begin{code}
import galpy.backend
with galpy.backend.use('jax'):
    Phi= MWPotential2014(1.,0.) # jax
\end{code}
The configured or context default only applies in the absence of
backend data; the data always take precedence over it. Similarly, an
\orb\ that is set up from a tensor on a GPU is integrated there by an
in-backend solver (the default \cc\ integrators run on the CPU),
\begin{code}
o= Orbit(torch.tensor([1.,0.1,1.1,0.,0.05,0.],
                      device='cuda'))
o.integrate(ts,MWPotential2014,method='torchode')
o.R(ts) # a CUDA tensor
\end{code}
Internally, \galpy's migrated functions also accept an explicit
namespace as an \texttt{xp=} argument, which \galpy\ uses to keep
nested calls on one namespace; it is not meant for users. The full
resolution order is thus: an explicit \texttt{xp=} argument, the array
arguments and the objects' parameters, the context or configured
default, and finally \numpy. The context manager also accepts
\texttt{force=True}, which makes the default take precedence over the
data. Under a forced backend,
\galpy\ runs as much as possible on that backend, including the
internals that otherwise use \numpy\ and \scipy\ for \numpy\ inputs;
the test suite depends on this option to run \galpy's existing tests, which
are written with \numpy\ inputs, on the backend code, and it is not
meant for production code.

While a call with backend arrays returns backend arrays, a few
conventions are needed at the edges of the backend layer. \galpy\
computes in double precision, and its internal
constants are \python\ floats, such that under \torch, whose default
is single precision, \galpy\ enters a double-precision default for the
duration of a call; a user's single-precision tensor stays single
precision in and out. Outputs that are scalars in the \numpy\ sense,
such as a single potential value for scalar inputs under a forced
backend, are returned as \numpy\ values, because \astropy\ units cannot
wrap a backend array. Units are handled outside the backend layer: the
existing unit-handling decorators convert a \texttt{Quantity} input to
\galpy's internal units before it reaches the backend code and build a
\texttt{Quantity} output from the result afterwards. A potential
parameter that is given as a backend array carries no units and is
taken to be in internal units already, which is what makes it possible
to differentiate with respect to it. Because \texttt{jax.jit} cannot
trace \astropy\ \texttt{Quantity} objects, the use of units inside a
compiled program is not supported; a user-defined density that uses
units is evaluated outside the compiled program, through a callback,
and cannot be differentiated. A \jax-aware units package such as
\texttt{unxt} \citep{Starkman2025} could lift this restriction, and
we may add support for it in the future.

\subsection{Backend-agnostic code}\label{sec:design-agnostic}

The code in the physics modules is written in a backend-agnostic
manner wherever possible. To achieve this, every migrated compute
method starts by resolving the array namespace from its coordinate
arguments and then uses only that namespace; for example, the
potential of a Plummer sphere is computed as
\begin{code}
def _evaluate(self,R,z,phi=0.,t=0.):
    xp= get_namespace(R,z,phi,t)
    return -1./xp.sqrt(R**2+z**2+self._b2)
\end{code}
The namespace is the plain \numpy\ module for \numpy\ data,
\texttt{jax.numpy} for \jax\ arrays, and the array-API-compatible view of
\torch\ for tensors, obtained through the \texttt{array-api-compat}
package. This is the entire change for the majority of \galpy's
potentials, which are elementary functions of the coordinates; the work
is in the remaining potentials, which use \scipy\ or \numpy-specific functions and
which we describe in the module sections. The dispatch costs about
$3\,\mu$s per call when run eagerly and nothing under compilation,
because it is then resolved at trace time.

We placed the dispatch at \galpy's private compute layer
(\texttt{\_evaluate}, \texttt{\_Rforce}, and so on), such that the public
methods, their unit handling, and their documentation are untouched. The
public entry points carry one new decorator that declares which of the
arguments are coordinates, for example
\texttt{@backend\_input('R','z','phi','t')} for a method whose
coordinate arguments are $R$, $z$, $\phi$, and $t$. This decorator, which does nothing on \numpy\
data, coerces the declared arguments, and only those, onto the active
namespace, whether they are passed positionally or by keyword, and
leaves derivative orders, flags, masses, option strings, and grids
alone. It also resolves a single device for the call from the first
backend array it finds, such that a \python\ scalar passed alongside a
GPU array does not end up on the host CPU. The central coercion matters
more than it might appear: \torch, unlike \jax, rejects \python\ scalars
in most tensor operations, and about two thousand of the tests that
failed under \torch\ when we first ran the backends failed for this
reason alone. Coordinate transformations in \texttt{galpy.util.coords}
follow a similar pattern, where a marker decorator opts a migrated
function into framework-preserving scalar promotion, such that we could
migrate the functions one at a time.

\subsection{Numerical building blocks}\label{sec:design-numerics}

\galpy's \numpy\ code uses \scipy\ for special functions, root finding,
quadrature, and interpolation, and \numpy's global random-number
generator for sampling, none of which dispatch to \jax\ or \torch, while
the frameworks' own versions are incomplete and, in places, wrong. The
\texttt{galpy.backend} subpackage therefore provides \scipy-like
replacements that delegate to \scipy\ on \numpy\ data and run natively
otherwise.

For special functions, a router dispatches each function to
\texttt{scipy.special}, \texttt{jax.scipy.special}, or
\texttt{torch.special} unless the function is listed as missing or
unreliable for that framework, in which case we implemented a fallback directly in
\galpy. The list is short but important: \jax's hypergeometric
functions ${}_2F_1$ and ${}_1F_1$ currently return infinities with \texttt{nan}
gradients for arguments $z<-1$, which is exactly the regime in which
\galpy\ uses them; \jax's incomplete gamma function iterates per element,
which is about a hundred times slower than a vectorized series for large
arrays on the CPU and slow on a GPU at every size; and
\torch's incomplete gamma function has no derivative with respect to its
order and loses six digits near an order of 20, where its implementation
switches between a series and a continued fraction. We gave the
fallbacks fixed iteration counts, such that they trace and vectorize,
and checked them against \texttt{mpmath} at 30 to 50 digits; the worst
case
among them agrees with the reference to $7\times10^{-15}$, which is at
least as accurate as the \scipy\ path. A test asserts that the list of
unreliable functions still matches the installed frameworks, such that
entries are removed as the frameworks improve.

For root finding, \texttt{galpy.backend} provides a drop-in replacement
for \scipy's \texttt{brentq}, which on backend data runs a fixed number
of sign-preserving bisection steps (a rolled rather than an unrolled
loop under a \jax\ trace), set by the bracket width and the tolerance or
by the maximum number of iterations when the bracket is itself traced,
followed by one Newton step that polishes the value and, through the
implicit-function theorem, carries the derivative of the root with
respect to whatever the function depends on. This one root finder now
serves every action--angle and DF calculation, where \galpy\ previously
had several. Quadrature uses fixed-order Gauss--Legendre rules, with
semi-infinite, boundary-layer, finite-part, and nested variants, and is
therefore a differentiable approximation rather than an adaptive method;
we set the order for each use by comparing with the adaptive \scipy\
result, a comparison that in two cases found the \scipy\ result to be
the one in error (two of the fixes described in
\sectionname~\ref{sec:process-fixes}).
Interpolation comes in two forms: a fitted \scipy\ spline is converted to
a piecewise-polynomial table that is evaluated by namespace operations,
which is differentiable in the evaluation point and is used for tables
that do not depend on parameters; and cubic splines whose coefficients
are solved for in the backend, which are differentiable in the tabulated
values and are used where the table itself depends on a parameter, as
for the velocity dispersion in the dynamical-friction force.

Sampling follows the same logic: every sampler in \galpy\ that was a
rejection sampler with \numpy\ proposals has an inverse-cumulative
counterpart on the backends, in which uniform variates are transformed
through a tabulated cumulative distribution; this has static shapes,
traces, and makes the samples differentiable with respect to the
parameters of the distribution, because the transformation of a fixed
set of uniforms is a smooth function of those parameters. Random keys
are explicit: a \jax\ key or its \torch\ equivalent is passed as
\texttt{key=}, while \texttt{key=None} keeps the historical behavior of
\numpy's global generator, such that existing scripts produce the same
samples as before, except for the three samplers that were replaced on
the \numpy\ path as well (\sectionname~\ref{sec:design-principles}). For
a Hernquist sphere, this reads as follows:
\begin{code}
from galpy.backend import random as grandom
hp= HernquistPotential(amp=2.,a=1.3)
df= isotropicHernquistdf(pot=hp)
sam= df.sample(n=10**6)          # numpy, as before
key= grandom.key(42,backend='jax')
sam= df.sample(n=10**6,key=key)  # jax, differentiable
\end{code}

\subsection{GPUs}\label{sec:design-gpu}

\begin{figure*}[t!]
  \includegraphics[width=\textwidth,clip=]{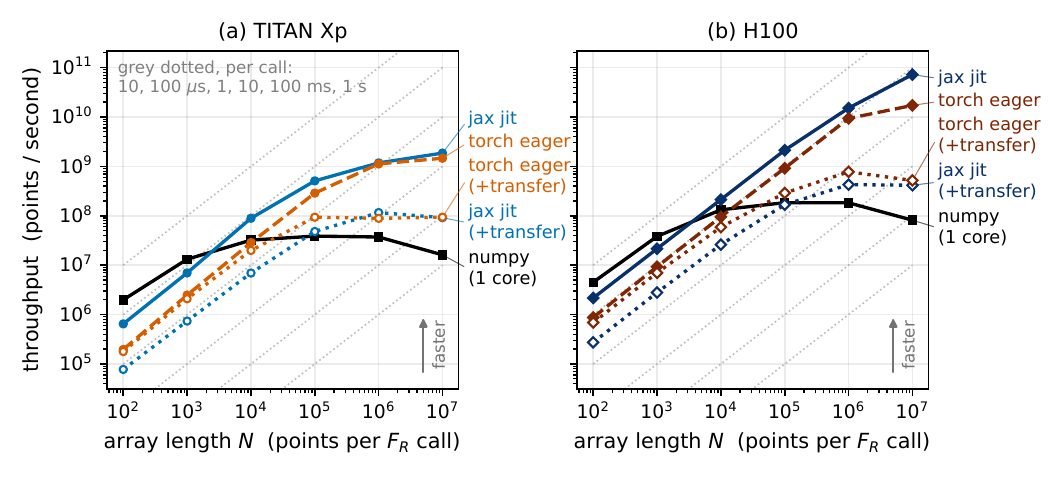}
  \caption{Throughput of the radial force of a Plummer potential as a
    function of the number of points per call. The left panel shows the
    development machine, with \numpy\ on one CPU core and \jax\ and \torch\
    on an NVIDIA TITAN~Xp; the right panel shows a node of the Trillium
    cluster, with an NVIDIA H100. Solid and dashed lines are compute-only (compiled \jax\ and eager
    \torch); dotted lines with open markers include the transfer of
    inputs and outputs between host and device. Light grey dotted lines mark constant wall time per call, which
    the GPU curves follow while launch-bound. The GPU gives its largest
    gains when the inputs and outputs stay on the
    device.}\label{fig:throughput}
\end{figure*}

Nothing in the above is specific to a device, and \galpy\ runs on a GPU
by being given arrays that live there. Throughout this paper we use
two GPUs, which stand in for the two kinds that a user is likely to
have: an NVIDIA TITAN~Xp in the development machine, a consumer card
from 2017 (about US\$1,200 at launch) with 12\,GB of memory that
executes double-precision arithmetic at one thirty-second of its
single-precision rate, and an NVIDIA H100 on a node of the Trillium
cluster, a data-center accelerator (of order US\$30,000 at the time of
writing) with 80\,GB of memory and a double-precision rate half of its
single-precision one. The one thing the library must
do is place the constants that it creates---rotation matrices, expansion
coefficients, quadrature nodes, spline tables---on the device of the
coordinates rather than on the host CPU, and we changed every site where a
stored table meets a coordinate array to do so. Because these sites are
invisible on a CPU, where host and device coincide, we found them by
running the test suite on a GPU, which also served to find code that
silently converted to \numpy. \figurename~\ref{fig:throughput} displays what the
GPU buys for a simple operation in \galpy, the evaluation of a force
at $N$ points: on either GPU the computation alone is launch-bound up to
$N\approx10^4$ and then scales until it saturates the device's memory
bandwidth, at about $10^9$ points per second on a consumer card and
$7\times10^{10}$ on an H100; including the transfer of the input and the
result over the PCIe bus, the gain over one CPU core is a more modest
factor of five to six at $N=10^7$. GPUs pay off when the data stay on the device across many
operations, which is what a compiled program does.

\subsection{Compilation and differentiation from the outside}\label{sec:design-jit}

\begin{figure*}[t!]
  \includegraphics[width=\textwidth,clip=]{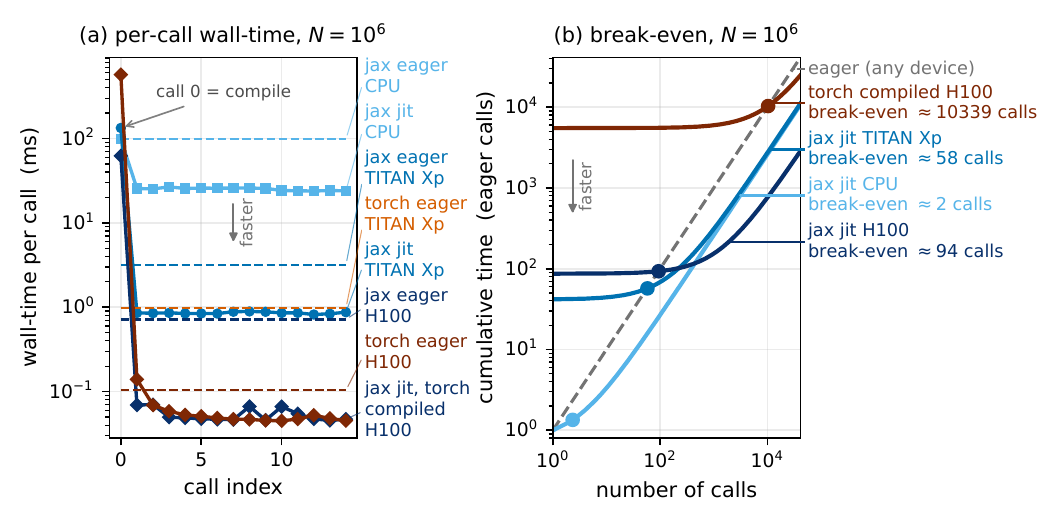}
  \caption{Cost and benefit of compilation for the radial force of a
    Plummer potential at $N=10^6$ points under \jax\ and \torch. Left:
    wall time of successive calls on one CPU core (\jax), the TITAN~Xp,
    and the H100; the first call compiles. Dashed lines are the eager
    cost per call; \texttt{torch.compile} does not support the TITAN~Xp,
    so \torch\ appears there eagerly only. Right: the cumulative time of
    compiled and eager calls, in units of each device's eager call, and
    the number of calls after which compilation pays for
    itself.}\label{fig:jitamort}
\end{figure*}

\begin{figure*}[t!]
  \includegraphics[width=\textwidth,clip=]{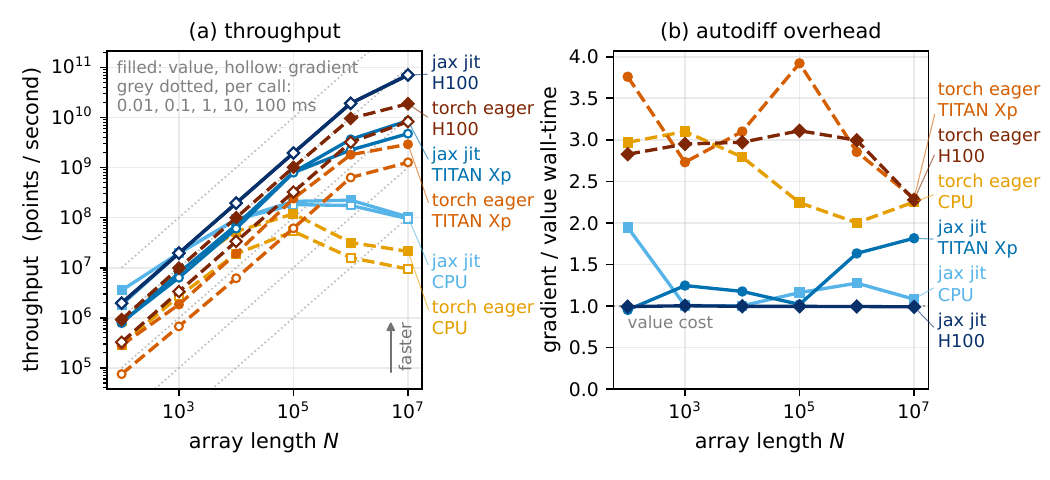}
  \caption{Cost of a derivative. Left: throughput of the Plummer
    potential (filled markers) and of its gradient with respect to the
    coordinates by reverse-mode automatic differentiation (open
    markers), under compiled \jax\ and eager \torch, on one CPU core and
    on the two GPUs. Right: the ratio of the wall time of the gradient to
    that of the value. The gradient costs a small multiple of the value
    itself, independent of the number of points and of the
    device.}\label{fig:adoverhead}
\end{figure*}

\galpy\ does not call \texttt{jax.jit} or \texttt{torch.compile} on its
own code, because compilation is most effective at the level of a
user's whole computation, in which a \galpy\ call is one step, and a
library that compiles its pieces separately defeats this. The user
therefore writes code as they would without \galpy, 
for example for a loss function that integrates an orbit in a potential and compares its position to an observed value,
\begin{code}
def loss(q):
    pot= LogarithmicHaloPotential(normalize=1.,q=q)
    o= Orbit(jnp.asarray(ic))
    o.integrate(ts,pot,method='diffrax')
    return jnp.sum((o.x(ts)-x_obs)**2)
grad_loss= jax.jit(jax.grad(loss))
\end{code}
and it is \galpy's job to be traceable. \figurename~\ref{fig:jitamort}
displays what compilation costs and what it buys for a simplest
operation, the radial force of a Plummer potential at $10^6$ points, on
one CPU core and on the two GPUs of \sectionname~\ref{sec:design-gpu}:
compiling it takes about a tenth of a second on each. Under \jax, a
compiled call is four times faster than an eager one on the CPU, such
that the compilation has paid for itself after two calls, and ten
times faster on the GPUs, where it pays for itself after 60 to 90
calls. \torch\ tells a different story: \texttt{torch.compile}
cannot be used on the TITAN~Xp at all, because its code generator
requires a GPU of compute capability 7.0 or higher, and on the H100 an
eager \torch\ call, which is seven times faster than an eager \jax\
call, is already within a factor of two of the compiled one, such that
the compilation pays for itself only after about ten thousand calls.
What \texttt{torch.compile} removes is the per-operation dispatch of a
longer computation, such as an orbit integration.
\figurename~\ref{fig:adoverhead} shows the cost of a derivative: under
compiled \jax\ the gradient of the potential with respect to the
coordinates at $N$ points costs between one and two times the value
itself, on the CPU as on the GPUs, and under eager \torch\ two to four
times, independent of $N$.

Making \galpy\ traceable required a set of changes that recur throughout
the code, which we list here once. We gave data-dependent loops fixed
iteration counts: the bisection discussed above, the arithmetic--geometric
mean in the elliptic integrals, the series and continued fractions of
the special functions, and the refinement of a stream track, which
is iterated a fixed number of times rather than to convergence. We replaced branches on data by \texttt{xp.where}
with both sides evaluated, which requires clamping the unused side to a
safe argument, because a \texttt{nan} produced on the discarded side
still poisons the reverse-mode derivative; checks that raise on invalid
input skip themselves when the input is traced. We made shapes static
and, where an array shrinks as a loop proceeds, pad it to the next power
of four, because eager \jax\ recompiles for every new shape; one sampler
spent 47 minutes compiling before this change and two after. We had to
distinguish between a value that is traced and one that is merely a
backend array, because \texttt{torch.compile} turns \texttt{float(x)}
into a symbolic scalar, and a \texttt{float} probe silently took the
\scipy\ branch of a surface-density integral inside a compiled region.
We wrap lazily built \numpy\ tables such that their first construction
inside a compiled region is not traced. A compiled function that
receives a \galpy\ object, such as a potential, as a static argument
decides from a hash of that argument whether an existing compilation
can be reused; \python's default hash of an object is its identity,
which would recompile for every newly created potential even when it
is identical to one seen before and, worse, would reuse a compilation
for a potential whose amplitude has been changed in place. We
therefore key \galpy\ objects by their type and attributes, with
\galpy\ sub-objects keyed recursively, such that a changed amplitude,
which as a \python\ float is a constant of the compiled program,
causes the re-trace that it must and a freshly created but identical
wrapper does not cause a needless one. The \torch\ compiler has its
own difficulties: because a read of a \python\ context variable severs
its graph, \galpy's backend selection, which is such a variable, reads
a thread-local mirror while \torch\ is tracing. None of these changes
affects what the \numpy\ path computes; they are the price of the
fifth rule, that the user, and not \galpy, decides what to compile.

\subsection{The test suite under every backend}\label{sec:design-tests}

We run \galpy's test suite unchanged under five modes: \numpy, \jax\ and
\torch\ in eager mode, and \jax\ and \torch\ in compiled mode. For
the compiled modes, \galpy\ has an optional global trace mode that
wraps every call at the coercion boundary of
\sectionname~\ref{sec:design-agnostic} in \texttt{jax.jit} or
\texttt{torch.compile}; it exists for the test suite and is not needed
in normal use, where the user compiles their own program. The
mechanism is a fixture that wraps every test in the forced-backend
context manager of \sectionname~\ref{sec:design-choosing} and, for the
compiled modes, in the global trace mode; it is selected by the
\texttt{--backend} and \texttt{--jit} options of the test runner, and a
\texttt{--device cuda} option additionally places every tensor on the
GPU; \sectionname~\ref{sec:process-islands} describes what makes a pass
in these modes meaningful and measures how much of the suite computes on
the backend, and \sectionname~\ref{sec:process-ledger} how we brought
the suite to pass. Because a compiled-mode run can
pass while every call falls back to eager execution, the test runner
counts the fraction of calls that were traced and fails the run if this
fraction drops below a floor, which is $99\,\%$ for a typical test
file. The dedicated tests of
the backend code, 18,807 tests in 100 files, are parametrized over the
installed frameworks and additionally exercise the mixed case in which
\numpy\ and backend arrays meet in one call, which the forced mode
cannot reach. In continuous integration, the two frameworks, the two
modes, and 22 shards of the test suite form a matrix that runs on every
PR; we run the GPU tests by hand, because the continuous integration framework 
available to us has no GPU.

%% file: potentials.tex
\section{Potentials}\label{sec:potential}

The \pot\ classes are the foundation of \galpy\ and were the first part
of the code that we migrated. The approach is the one of
\sectionname~\ref{sec:design-agnostic}: every private compute method
(\texttt{\_evaluate}, \texttt{\_Rforce}, \texttt{\_zforce},
\texttt{\_phitorque}, \texttt{\_dens}, \texttt{\_surfdens}, and the six
second derivatives) resolves the array namespace from its coordinates
and uses it throughout, and the public methods, through the coercion
decorator, bring scalar and \numpy\ inputs onto the namespace of any
backend input. For the majority of \galpy's potentials this was the
whole change, because they are elementary functions of the coordinates.
The remainder use special functions or tabulated quantities: the
basis-function and multipole expansions (\texttt{SCFPotential},
\texttt{MultipoleExpansionPotential}, and their disk variants) use the
associated Legendre and Gegenbauer polynomials; the power-law potential
with an exponential cutoff and the Einasto profile use incomplete gamma
functions; the double-exponential disk uses Bessel functions in a
Hankel-type integral; and the interpolated potentials and the King
profile evaluate splines. These go through the special-function router
and the interpolation layer of \sectionname~\ref{sec:design-numerics}.
All of the potentials, forces, and wrappers in \texttt{galpy.potential}
run on both backends, with the exception of the potentials that read
$N$-body snapshots through \texttt{pynbody} \citep{Pontzen2013} and
the interface \citep{Webb2020} to \texttt{AMUSE}
\citep{PortegiesZwart2009,Pelupessy2013,PortegiesZwart2018}, which
have no backend code. Even the wrapper that adiabatically contracts a
dark-matter halo in response to the baryons, following
\citet{Cautun2020}, whose construction is a root find for the
contracted profile, builds that profile on the backend when one of its parameters, or a parameter of
one of the potentials that it contracts, is being differentiated. The
convenience functions that act on potentials (the circular velocity
and epicycle frequencies, the radius of a circular orbit of given
angular momentum or energy, the enclosed mass, and the tidal radius
and tidal tensor of \citealt{Webb2019}) are backend-aware as well;
those that require a root are solved with the shared root finder and
those that require an integral use Gauss--Legendre quadrature.

\begin{figure*}[t!]
  \includegraphics[width=\textwidth,clip=]{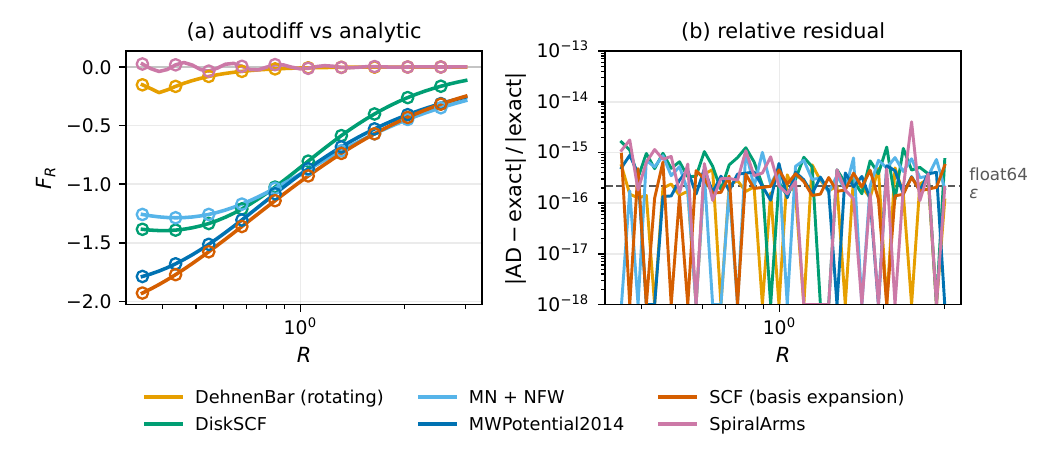}
  \caption{The radial force by automatic differentiation of the
    potential compared with the hand-coded force, for the Milky-Way
    model, a basis-function expansion and its disk variant, spiral arms,
    a rotating bar, and a disk-plus-halo composite. Left: the force along
    $R$ in the mid-plane, lines from automatic differentiation and
    circles from the analytic implementation. Right: the relative
    difference between the two, which is at the level of the
    double-precision rounding error.}\label{fig:adforces}
\end{figure*}

\begin{figure*}[t!]
  \includegraphics[width=\textwidth,clip=]{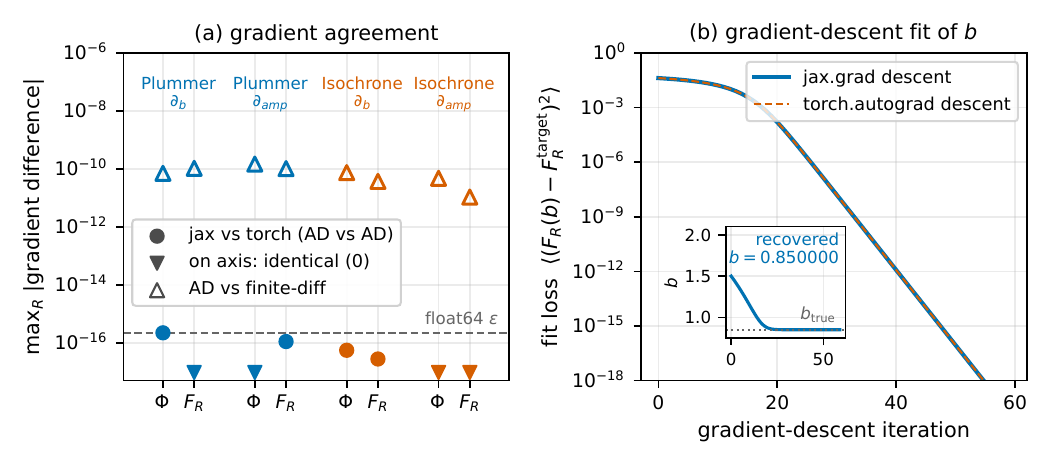}
  \caption{Derivatives of potentials with respect to their parameters.
    The left panel shows, for the Plummer and isochrone potentials, the
    largest difference between the derivatives of the potential and of
    the radial force with respect to the amplitude and scale parameters
    computed with \jax\ and with \torch\ (circles, at the level of the
    double-precision rounding error; where the two are bit-identical, the
    difference is zero and cannot be shown on the logarithmic axis, which
    the filled triangles on the axis mark) and between either of these and a
    central finite difference (open triangles, at the level of the
    finite-difference error). The right panel displays the recovery of
    the scale length of a Plummer potential from the radial force at a
    set of points by gradient descent on the derivative, starting from
    $b=1.5$ and converging to the true $b=0.85$, with both
    backends.}\label{fig:paramgrad}
\end{figure*}

The derivatives of a potential with respect to the coordinates are, of
course, its forces and second derivatives, which \galpy\ implements by
hand. Under a backend, these can also be obtained by differentiating
the potential, and the agreement between the two is a strong test of
both:
for the basis-function expansions, the spiral arms, a rotating bar, and
composite potentials, the force by automatic differentiation agrees
with the hand-coded force
to a few parts in $10^{15}$, within twenty units in the last place
(\figurename~\ref{fig:adforces}), that is, the derivative that
automatic differentiation
produces is the analytic force up to rounding. While the test suite
checks these identities for every potential, under both backends and in
both modes, the derivatives that matter to the user are those with
respect to the parameters: a potential constructed with a backend array
as a parameter is differentiable with respect to that parameter through
every method. For example, the derivative of the radial force of a Miyamoto--Nagai 
potential with respect to its scale length $a$ reads with \jax\ as
\begin{code}
from galpy.potential import MiyamotoNagaiPotential
def Rforce_of_a(a):
    mp= MiyamotoNagaiPotential(amp=1.,a=a,b=0.3)
    return mp.Rforce(1.2,0.1)
dFR_da= jax.grad(Rforce_of_a)(0.5)
\end{code}
and the same with \torch\ is
\begin{code}
a= torch.tensor(0.5,requires_grad=True)
mp= MiyamotoNagaiPotential(amp=1.,a=a,b=0.3)
mp.Rforce(torch.tensor(1.2),
          torch.tensor(0.1)).backward()
a.grad # dFR/da
\end{code}
All 47 potentials that can be constructed from parameters alone are
differentiable with respect to their amplitude (or, for the King
model, its mass) in this way, both eagerly and under
compilation. \figurename~\ref{fig:paramgrad}
displays the accuracy of these parameter derivatives and a small
example of their use, the recovery of the scale length of a Plummer
potential from its force by gradient descent. The parameters of a
potential constructed from a density, as the basis-function and
multipole expansions and their disk variants can be, carry the
derivative through the expansion coefficients.

The backend potentials differ from the \numpy\ ones in a few respects.
Where \galpy\ computes an integral over a potential adaptively on the
\numpy\ path, as for the enclosed mass of a general density or the
surface density, the backend computes a fixed-order Gauss--Legendre
approximation; we set the order such that the two agree, but the backend
result is an approximation of fixed cost rather than an adaptive one. Interpolated potentials evaluate
the same spline as \scipy\ does, to within one unit in the last place,
by converting the fitted spline to a piecewise-polynomial table; they are
differentiable in the evaluation point and, when they are built from a
potential or function with a parameter that is being differentiated,
also in the tabulated values, because the table and the spline
coefficients are then computed on the backend and the derivative with
respect to the parameter passes through them.
A potential that moves along a precomputed orbit, as the
\texttt{MovingObjectPotential}, interpolates that orbit on the backend
when the orbit was integrated there, such that the field of the moving
object is differentiable with respect to everything that the orbit
depends on, and lives on the GPU when the orbit does; an orbit
integrated with \numpy\ stays on \numpy\ and the field is computed at
backend coordinates. And a \python\ branch on the
value of a stored parameter, such as the special case of a vanishing
scale length, is harmless when the parameter is differentiated eagerly
but fails when the parameter is traced inside a compiled ODE solve; we
replaced such branches where we found them, but authors of new
potentials should keep this behavior in mind.

New potentials are written as before: a class that inherits from
\pot\ and implements \texttt{\_evaluate}, \texttt{\_Rforce}, and
\texttt{\_zforce}. If these are written with the resolved namespace
rather than with \numpy\ directly, as illustrated in \figurename~\ref{fig:newpot},
the new potential runs on every backend and can be differentiated with
respect to its parameters without any further work, and the test suite's
automatic checks of new potentials are run under every backend. A test
checks that no migrated potential calls \numpy\ directly in a compute
method.

\begin{figure}[t!]
\begin{codefig}
from galpy.potential import Potential
from galpy.backend import get_namespace
class MNLikePotential(Potential):
    # a Miyamoto-Nagai disk written for every backend
    def __init__(self,amp=1.,a=1.,b=0.1,
                 normalize=False):
        Potential.__init__(self,amp=amp)
        self._a= a
        self._b2= b**2
        if normalize: self.normalize(normalize)
    def _evaluate(self,R,z,phi=0.,t=0.):
        xp= get_namespace(R,z)
        s= self._a+xp.sqrt(z**2+self._b2)
        return -1./xp.sqrt(R**2+s**2)
    def _Rforce(self,R,z,phi=0.,t=0.):
        xp= get_namespace(R,z)
        s= self._a+xp.sqrt(z**2+self._b2)
        return -R*(R**2+s**2)**-1.5
    def _zforce(self,R,z,phi=0.,t=0.):
        xp= get_namespace(R,z)
        sqb= xp.sqrt(z**2+self._b2)
        s= self._a+sqb
        return -z*s/sqb*(R**2+s**2)**-1.5
# differentiate an orbit in it w.r.t. the scale length a
def final_z(a):
    o= Orbit(jnp.asarray([1.,0.1,1.1,0.1,0.05,0.]))
    mp= MNLikePotential(a=a,b=0.3,normalize=1.)
    o.integrate(ts,mp,method='diffrax')
    return o.z(ts[-1])
jax.grad(final_z)(0.5)
\end{codefig}
\caption{A new \pot\ class written for every backend. The only
  difference from a class written for \numpy\ is that the array
  functions are taken from the namespace resolved from the coordinates
  (\texttt{xp}) rather than from \numpy. The class can then be used, and
  differentiated, like any other potential; the last lines take the
  derivative of the final height of an orbit with respect to the scale
  length of the new potential.}\label{fig:newpot}
\end{figure}

%% file: orbits.tex
\section{Orbit integration}\label{sec:orbit}

\subsection{Three ways to integrate an orbit}\label{sec:orbit-paths}

\begin{figure*}[t!]
  \includegraphics[width=\textwidth,clip=]{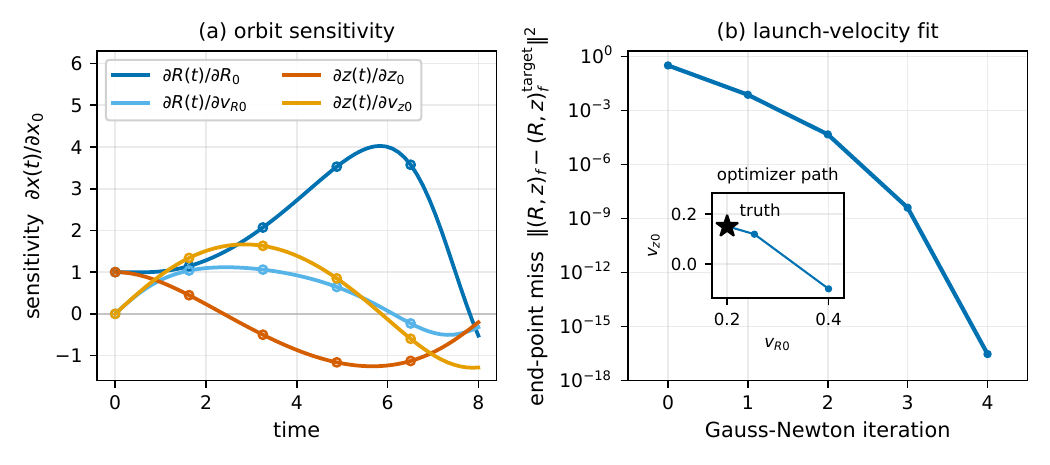}
  \caption{Derivatives of an in-backend orbit integration. Left: the
    sensitivities $\partial R(t)/\partial R_0$, $\partial R(t)/\partial v_{R,0}$,
    $\partial z(t)/\partial z_0$, and $\partial z(t)/\partial v_{z,0}$ of an
    orbit in a Plummer potential by reverse-mode differentiation through
    the \diffrax\ solve (lines) and by central finite differences
    (circles). Right: the squared distance between the end point of the
    orbit and a target as the launch velocity $(v_{R,0},v_{z,0})$ is
    fit by Gauss--Newton iterations on the Jacobian of the integration;
    the inset shows the path of the fit. The derivative through the
    solver agrees with the finite differences, and the fit that uses it
    converges in a handful of iterations.}\label{fig:orbitsens}
\end{figure*}

\orb\ instances are set up and integrated as before, but an \orb\ instance whose
initial condition is a backend array, or whose potential carries a
backend parameter, is integrated in one of two new ways, depending on
what is to be differentiated. The first integrates the equations of
motion inside the backend, with the backend's own ODE solver:
\diffrax\ \citep{Kidger2021} for \jax, with an eighth-order
Dormand--Prince method, and \texttt{torchdiffeq} \citep{Chen2018} or
\torchode\ \citep{LienenGunnemann2022} for \torch, with a fifth-order method.
The equations of motion are those of \galpy's \cc\ integrators,
evaluated in rectangular coordinates through the backend force layer,
and the default tolerances are $10^{-12}$. Because the whole integration
is then a backend program, everything that enters it can be
differentiated: the initial condition, the potential's parameters, the
integration times. $N$ orbits that share a time grid are integrated as
a single solve along a batch axis, under \jax\ with one step-size
controller for the whole batch; given per-orbit time grids, an
$(N,n_t)$ array, \galpy\ instead maps independent solves, each with its
own controller, over the batch with \texttt{jax.vmap}, which is also
what a user's own \texttt{jax.vmap} over single-orbit integrations
gives and how \torchode\ treats every batch; a user thus has both
options. Either way, the solve runs on
a GPU when the initial conditions do. This is the path to use for
derivatives with respect to the parameters of the potential, 
as in the following example:
\begin{code}
def final_R(b):
    o= Orbit(jnp.asarray(ic))
    pp= PlummerPotential(amp=1.,b=b)
    o.integrate(ts,pp,method='diffrax')
    return o.R(ts[-1])
dR_db= jax.grad(final_R)(0.6)
\end{code}
and under \torch:
\begin{code}
b= torch.tensor(0.6,requires_grad=True)
o= Orbit(torch.as_tensor(ic))
pp= PlummerPotential(amp=1.,b=b)
o.integrate(torch.as_tensor(ts),pp,method='torchode')
o.R(ts[-1]).backward() # b.grad = dR/db
\end{code}
Second derivatives are available under \jax\ by selecting the direct
adjoint through the \texttt{inbackend\_kwargs} keyword, and
\texttt{jax.vmap} over initial conditions works. \figurename~\ref{fig:orbitsens} displays the simplest
use of these derivatives: the sensitivity of an orbit to its initial
conditions as a function of time, which the solver's derivative gives
for free and which finite differences confirm, and a Gauss--Newton fit
of the launch velocity of an orbit to a target end point using the
Jacobian of the integration, which converges quadratically to machine
    precision in a handful of iterations.

\begin{figure*}[t!]
  \includegraphics[width=\textwidth,clip=]{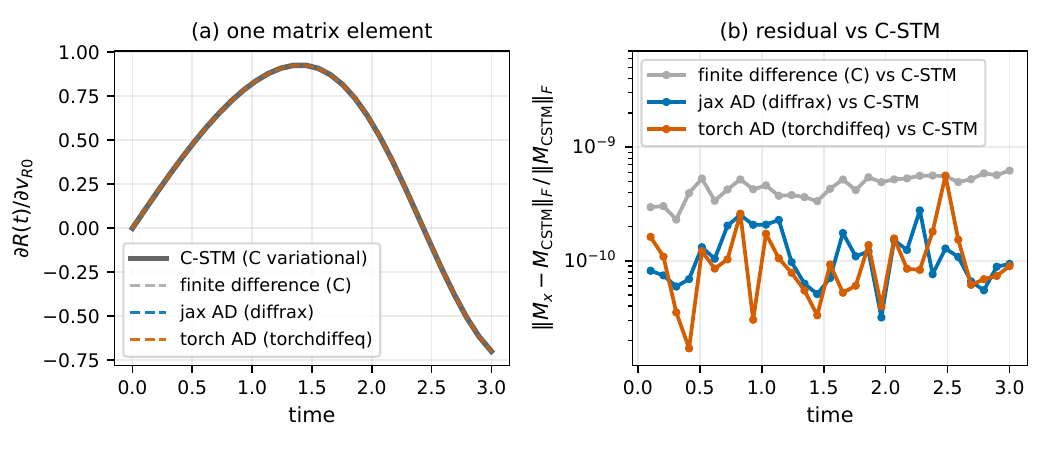}
  \caption{The derivative of an orbit with respect to its initial
    condition, in four ways. For a Sun-like orbit in \texttt{MWPotential2014}
    integrated for 3\,Gyr, the left panel shows one element of the
    state-transition matrix, $\partial R(t)/\partial v_{R,0}$, from the
    \cc\ variational equations, from a finite difference of the \cc\
    orbit, and from reverse-mode automatic differentiation through the
    \diffrax\ and \texttt{torchdiffeq} solvers. The right panel shows the
    relative Frobenius norm of the difference between each of the
    latter three and the variational result; all agree to a few parts in
    $10^{10}$.}\label{fig:icjacobian}
\end{figure*}

The second way uses \galpy's compiled \cc\ integrators, which are much
faster than a solver that runs inside \jax\ or \torch\ on a CPU (for a
single orbit in the Milky-Way model \texttt{MWPotential2014}, 
about ten times faster than the
\diffrax\ solve and a hundred times faster than \texttt{torchdiffeq}),
but which are opaque to automatic differentiation. What they can provide is the derivative of the
orbit with respect to its initial condition, through the variational
equations that \galpy\ integrates alongside the orbit in \cc\
(\appendixname~\ref{app:dxdv}). When a \galpy\ orbit with a backend
initial condition is integrated with a \cc\ method, the \cc\ integrator
returns both the orbit and the state-transition matrix
$\vec{M}(t) = \partial \vecw(t)/\partial \vecw(0)$, and a custom
derivative rule tells the backend that the derivative of the orbit with
respect to the initial condition is $\vec{M}$; the reverse-mode
derivative of any function of the orbit is then a single product of
$\vec{M}^T$ with the cotangent, without a second call to \cc. For a
six-dimensional orbit and a Runge--Kutta method, the six columns of
$\vec{M}$ are integrated together with the orbit as a 42-component
state; for the symplectic methods, the deviation vectors are propagated
through the same drift and kick steps as the orbit. The derivative is
exact to the precision of the solver: \figurename~\ref{fig:icjacobian}
compares the state-transition matrix of a Milky-Way orbit over 3\,Gyr
from the \cc\ variational equations with reverse-mode automatic
differentiation through
\diffrax\ and through \texttt{torchdiffeq} and with finite differences
of independently integrated orbits, and the four agree to a few parts in
$10^{10}$. In code, nothing marks the difference between the two paths except the
method:
\begin{code}
def end_point(ic):
    o= Orbit(ic)
    o.integrate(ts,MWPotential2014,
                method='dop853_c') # C + variational
    return jnp.stack([o.R(ts[-1]),o.z(ts[-1])])
M= jax.jacrev(end_point)(jnp.asarray(ic)) # STM rows
\end{code}
and a batch of orbits is differentiated with \texttt{jax.vmap},
\begin{code}
Ms= jax.vmap(jax.jacrev(end_point))(ics) # ics: (N,6)
\end{code}
This path gives derivatives with respect to the initial
conditions only, and to first order only; the variational equations do
not carry derivatives with respect to the parameters of the potential,
which must come from the in-backend solve.

\galpy\ chooses between the two automatically. An \orb\ whose
initial condition is being differentiated---a traced \jax\ array or a
\torch\ tensor that requires a gradient---and that is integrated with
a \cc\ method, in a potential that has \cc\ second derivatives, with
non-traced times and non-traced potential parameters, uses the \cc\
integrator and its variational equations. In every other case in
which something is differentiated, in particular whenever a potential
parameter is traced or when the integration is to run on a GPU, the
in-backend solver is used, under the name \texttt{method='diffrax'},
\texttt{'torchode'}, or \texttt{'torchdiffeq'}, which can also be
requested directly. When nothing is being differentiated, that is, for
a concrete backend initial condition, a \cc\ method integrates in
\cc\ as before, without the variational equations, and only its result
is a backend array; this is not a third way of integrating but the old
one, and it means that a backend initial condition costs nothing until
it is differentiated. The pure-\python\ integrators of \galpy\
(\texttt{method='odeint'}, the \python\ leapfrog, and so on) are not
routed at all: given a backend initial condition they integrate in
\numpy\ and return a backend array that is not differentiable, with a
warning.

After integration, the orbit's methods work as before on backend
arrays: positions and velocities at the integration times or, by
interpolation with an in-backend spline, between them; energies and
angular momenta; peri- and apocenter radii and maximum height, both
from the integrated orbit and analytically through the action--angle
code of \sectionname~\ref{sec:aa} \citep{MackerethBovy2018}; and the transformations to observed
coordinates. Initializing an orbit from observed coordinates with
backend arrays works as well, such that a derivative with respect to a
star's distance or proper motion is available.

\subsection{Backend symplectic integrators for many orbits}\label{sec:orbit-leapfrog}

The in-backend solvers of \sectionname~\ref{sec:orbit-paths} are
adaptive Runge--Kutta methods: accurate, but expensive, because a step
of \diffrax's eighth-order Dormand--Prince method evaluates the force
thirteen times. For large batches of
orbits in a smooth potential, where modest precision suffices and the
cost is dominated by the force evaluations, a symplectic integrator
with a fixed step is the better tool: it evaluates the force once per
step, its energy error oscillates without a secular drift, it keeps
little state per orbit, and its fixed steps make the derivative of a
batch straightforward. \galpy\ therefore also integrates inside the
backends with a kick-drift-kick leapfrog with a constant step,
\texttt{method='leapfrog\_jax'} and \texttt{method='leapfrog\_torch'},
as for example in the following:
\begin{code}
o= Orbit(jnp.asarray(ics)) # ics: (N,6)
o.integrate(ts,MWPotential2014,method='leapfrog_jax',
            inbackend_kwargs={'nsteps':8000})
\end{code}
The same route provides backend versions of the fourth- and
sixth-order symplectic integrators \citep{Yoshida1990,Kinoshita1991}
of \galpy's \cc\ library, \texttt{method='symplec4\_jax'} and
\texttt{'symplec6\_jax'} (and their \torch\ twins), which evaluate the
force three and seven times per step and agree with their \cc\
counterparts at the same step to rounding; they are the better choice
when more than a few digits are wanted: for the batch of disc orbits of
\sectionname~\ref{sec:orbit-speed}, 450 sixth-order steps over 2\,Gyr
(3150 force evaluations) reach the worst-case accuracy of the adaptive
Dormand--Prince solve at tolerances of $10^{-8}$, which takes about 250
steps (3250 force evaluations).
The leapfrog is second order: its error falls by a factor of four per
halving of the step, and in a static potential its energy error
oscillates without the secular drift that a non-symplectic
second-order method with the same steps shows; we verified both for a
batch of 2000 orbits in \texttt{MWPotential2014} with a moving
satellite and the acceleration of a non-inertial frame, against the
\cc\ integrator. For reverse-mode
derivatives, the steps are grouped into blocks under the backend's
checkpointing (\texttt{jax.checkpoint} or \texttt{torch.utils.checkpoint}),
such that the derivative stores the state at the start of each block and
the states within one block, about $2\sqrt{N_\mathrm{steps}}$ states
rather than $N_\mathrm{steps}$: for $10^5$ orbits and 8000 steps in
\texttt{MWPotential2014}, the compiled \jax\ derivative needs 1.4\,GB of
temporary memory instead of 39\,GB. The derivatives agree with
Richardson-extrapolated central differences of the same integration to
$10^{-10}$, and the \jax\ and \torch\ versions agree with each other to
$2\times10^{-14}$. An initial condition in single precision is
integrated in single precision throughout, including the times and the
forces, which matters on GPUs with a low double-precision rate
(\sectionname~\ref{sec:orbit-speed}). Forces that do not derive from a
potential, such as dynamical friction or the Coriolis force of a
rotating frame, which \galpy\ marks with the \texttt{isConservative}
attribute, raise an error, because the closing kick would evaluate
them at the half-step velocity and the method would then converge only
at first order. As a second-order method, the leapfrog needs small steps
to be precise: for $10^4$ disc orbits integrated for 0.36\,Gyr, 1000
leapfrog steps (1000 force evaluations) give a worst-case position
error of $3\times10^{-4}$ (in units of $R_0$), while 250 Dormand--Prince
steps (3250 force evaluations) give $4\times10^{-7}$. The leapfrog is
therefore the method of choice when the force is expensive to evaluate
and modest precision suffices, the sixth-order integrator when the
orbits share a timescale and precision matters, and the adaptive
Runge--Kutta solvers remain the choice for batches that mix
timescales.

\subsection{Variational equations and chaos}\label{sec:orbit-dxdv}

\begin{figure*}[t!]
  \includegraphics[width=\textwidth,clip=]{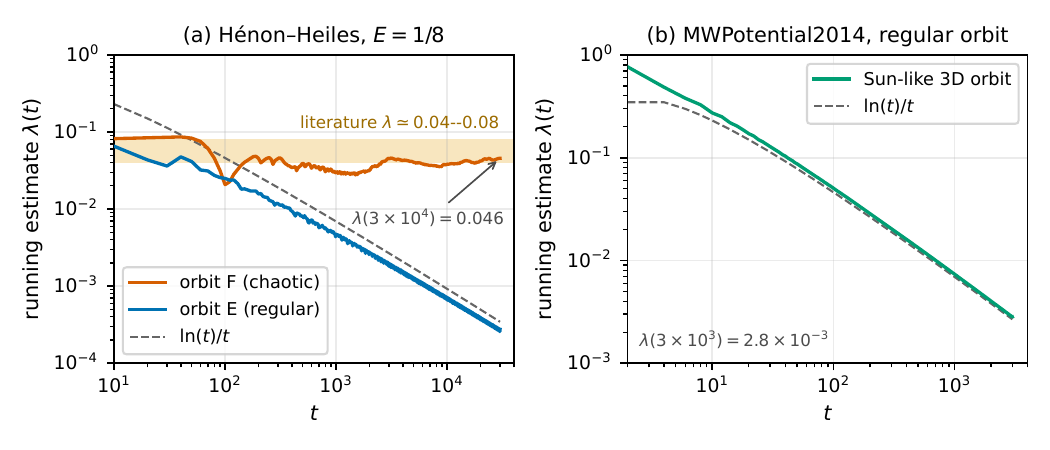}
  \caption{Largest Lyapunov exponent from \texttt{Orbit.lyapunov}, as a
    running estimate against time. Left: a chaotic and a regular orbit
    in the H\'enon--Heiles potential at $E=1/8$; the shaded band is the
    range of values in the literature for the chaotic orbit, and the
    dashed line is the $\ln t/t$ decay expected for a regular orbit.
    Right: a Sun-like orbit in \texttt{MWPotential2014}, which is
    regular. The estimate for the chaotic  H\'enon--Heiles orbit converges into the range
    found in the literature, while that for the regular orbit decays as
    expected.}\label{fig:lyapunov}
\end{figure*}

The variational equations that the \cc\ state-transition matrix relies
on were part of \galpy\ for planar orbits since 2011; as part of this
work, we extended them to three-dimensional orbits, to all of the \cc\
integrators including the symplectic ones, to one-dimensional orbits,
and to dissipative, velocity-dependent forces, and we released them in
\galpy\ \flag{v1.12}. \appendixname~\ref{app:dxdv} gives the equations
and the checks that we applied to them. Their user-facing form is
\texttt{Orbit.integrate\_dxdv}, which integrates an orbit together with a
deviation vector or with the full state-transition matrix, and
\texttt{Orbit.lyapunov}, which computes the largest Lyapunov exponent, or
the spectrum, by the renormalization method of \citet{Benettin1980a,Benettin1980b},
for example as
\begin{code}
o= Orbit([0.8,0.3,0.75,0.,0.2,0.])
o.integrate_dxdv([1.,0.,0.,0.,0.,0.],ts,
                 MWPotential2014,method='dop853_c')
o.getOrbit_dxdv() # the deviation along the orbit
o.lyapunov(ts,MWPotential2014,method='dop853_c',
           renorm_every=10)
\end{code}
For
example, for the H\'enon--Heiles potential at $E=1/8$, the exponent of a chaotic orbit
converges into the range quoted in the literature while that of a
regular orbit at the same energy decays as $\ln t/t$
(\figurename~\ref{fig:lyapunov}). Like the variational equations
themselves, these are \numpy\ and \cc\ features that do not require
the backends: they accept backend initial conditions, but run on the
CPU, return \numpy\ arrays, and are not differentiable.

\subsection{Limitations}\label{sec:orbit-limitations}

In-backend orbit integration has a few limitations. 
\texttt{torchdiffeq}
has no batching rule, such that a batch of orbits under \torch\ is a
\python\ loop unless \torchode\ is used. The eighth-order method of
\texttt{torchdiffeq} carries a gradient error of about $10^{-5}$ in its
backward pass, which is why \galpy\ uses its fifth-order method. Additionally, the
pure-\python\ integrators implemented in \galpy\ are not routed to the
backends (\sectionname~\ref{sec:orbit-paths}); they can run on a
backend under the forced mode of \sectionname~\ref{sec:design-choosing},
but are slow there, because each step is a separate set of array
operations, and we do not recommend it.
Finally, the \cc\ integrators cannot be used on a GPU at
all; their derivative rule runs the integration on the CPU.

\subsection{Speed}\label{sec:orbit-speed}

\begin{figure*}[t!]
  \includegraphics[width=\textwidth,clip=]{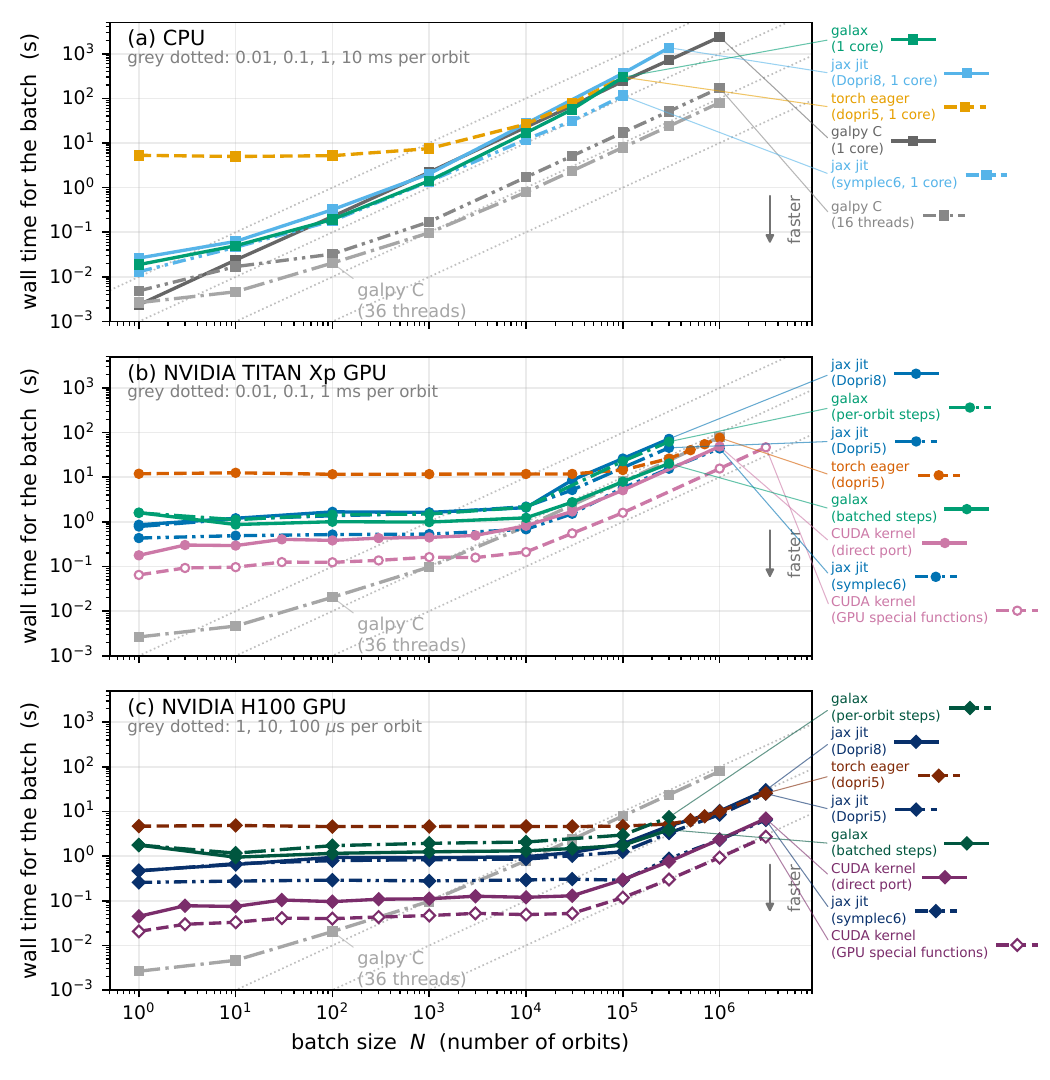}
  \caption{Wall time for integrating $N$ orbits for 2\,Gyr in
    \texttt{MWPotential2014} with relative and absolute tolerances of
    $10^{-8}$, on (a) the CPU, (b) the TITAN~Xp, and (c) the H100; each
    label repeats the style of its line. Grey: \galpy's \cc\ integrator
    \texttt{dop853\_c}, which parallelizes over orbits with OpenMP, on
    one core, 16 threads, and the 36 physical cores of the node; the
    latter is repeated in the GPU panels for reference. Blue: the
    in-backend \diffrax\ solve under \jax, compiled, with per-orbit
    step control, with the
    eighth-order method that \galpy\ uses by default and with a
    fifth-order method that performs the same number of force
    evaluations as the \torch\ solve, and the in-backend sixth-order
    symplectic integrator (\texttt{symplec6\_jax}) with 450 common
    steps over the 2\,Gyr, the step at which its worst-case position
    error over the batch equals that of the adaptive solve. Orange and
    red: the eager \texttt{torchdiffeq}
    solve. Green: \texttt{galax} with its default batched step-size
    control and with per-orbit control as in \galpy. Purple: an
    experimental CUDA port of \galpy's \cc\ forces and \texttt{dop853}
    that integrates one orbit per GPU thread (not part of \galpy\ and
    not differentiable), with a direct port of \galpy's \cc\ code and
    with special functions rewritten for the GPU, including the copies
    between host and device, which keep these lines flat up to
    $N\approx10^4$. Light grey dotted lines mark a constant cost per
    orbit, and the arrows point towards lower wall time.}\label{fig:orbitspeed}
\end{figure*}

\begin{figure*}[t!]
  \includegraphics[width=\textwidth,clip=]{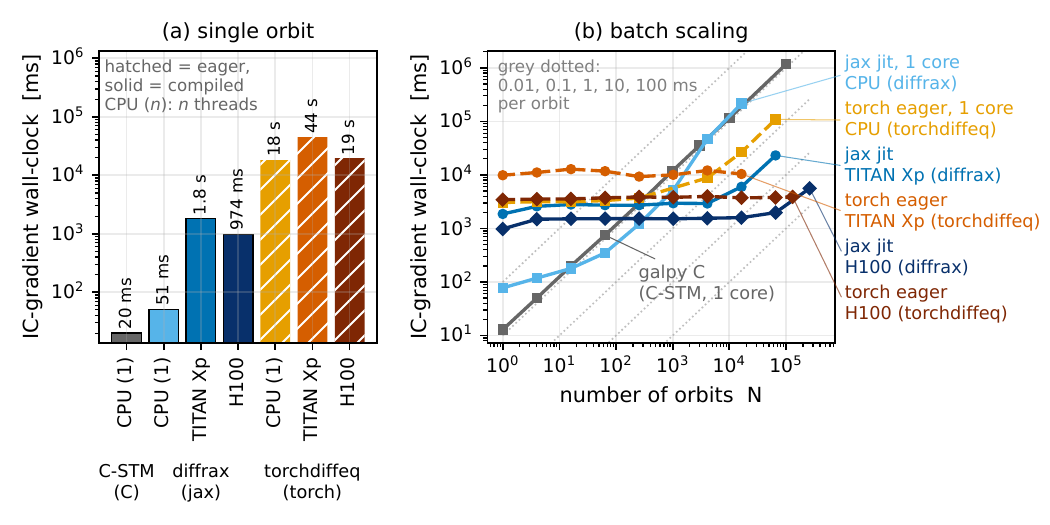}
  \caption{Cost of the derivative of an orbit with respect to its
    initial conditions, for a 2\,Gyr orbit in \texttt{MWPotential2014}.
    Left: wall time for a single orbit with the \cc\ variational
    integrator and by reverse-mode differentiation through the
    \diffrax\ and \texttt{torchdiffeq} solves on the CPU and the two
    GPUs; hatched bars are eager and solid bars compiled. Right: the same as a function of the
    number of orbits differentiated at once. The \cc\ variational
    integrator is the fastest way to obtain the derivative of a single
    orbit, while a GPU overtakes it for large batches. All CPU timings use
    one core.}\label{fig:icderivcost}
\end{figure*}

\figurename~\ref{fig:orbitspeed} compares the wall time of integrating
$N$ orbits for 2\,Gyr in \texttt{MWPotential2014} with the \cc\ integrator,
which parallelizes over orbits with OpenMP, on one core and on the 36
physical cores of the CPU node used in this work, and with the in-backend solvers on one CPU
core and on two GPUs. The absolute times depend on the specifics of the
problem, so we focus on the relative trends. On a single CPU core, the compiled \jax\
solve and the eager \torch\ solve cost about the same per orbit as the
\cc\ path for large batches, but they start from a floor of one to a
few seconds, set by compilation and kernel launches, that the \cc\ path
does not have, such that the \cc\ path is faster for up to a few
thousand orbits; the compiled solve additionally takes seconds to
compile, a cost that is paid once per shape of the problem. The \cc\
path is also the only one that makes use of more cores: its OpenMP loop
speeds up almost linearly to the 36 cores of our node for $10^4$ orbits
or more, eager \torch\ gains at most a factor of 3.5 (on 16 threads),
and compiled \jax, whose runtime keeps no more than a few cores busy,
gains nothing. The GPUs amortize a floor of half a second to a second
over the batch: the compiled \jax\ lines overtake one-core \cc\ at a
few hundred orbits and the \torch\ lines at a few thousand. The
consumer GPU, which executes double-precision arithmetic at one
thirty-second of its single-precision rate, never beats the 36-thread
\cc\ node with the adaptive solvers, while the H100 passes it at about
$10^4$ orbits and is 5 to 9 times cheaper per orbit for
$N\geq3\times10^5$. The
sixth-order symplectic integrator of
\sectionname~\ref{sec:orbit-leapfrog}, with the common step of that
section, is the cheapest in-backend route for this batch of disc
orbits, because its state per orbit is a fraction of the adaptive
solvers' and its steps are the same for every orbit: on one CPU core
it costs half of the adaptive \cc\ path, on the consumer GPU it beats
the 36-thread \cc\ node for $N\geq10^4$, which the adaptive solvers
cannot there, and on the H100 it is 3 to 7 times cheaper than the
adaptive solve on the same card and about 30 times cheaper than the
36-thread \cc\ node for $N\geq10^5$; for fewer
orbits, its compiled step loop is bound by its kernel launches, at
0.3--0.4\,s for the 450 steps on either GPU. The derivative of an orbit with respect to its
initial condition follows the same pattern
(\figurename~\ref{fig:icderivcost}): the \cc\ variational integrator
computes it fastest for a single orbit, by one to three orders of
magnitude over the backend solvers, and the GPU overtakes it at
batches of a few hundred orbits.

To gauge what GPU-native code could add, we also compiled \galpy's
\cc\ implementations of \texttt{MWPotential2014} and of its integrators
as a CUDA kernel that integrates one orbit per GPU thread; this is an
experiment and not part of \galpy. After restructuring the kernel, such
that the force is evaluated by a single out-of-line function and
threads pick up a new orbit when theirs is done, \texttt{dop853} on the
H100 is 4 to 11 times faster per orbit than \galpy's compiled \jax\
integrator on the same GPU and 30 to 90 times faster than the \cc\
integrator on the 36 cores of our CPU node (2.4\,$\mu$s per orbit
with a direct port of \galpy's \cc\ code and 0.9\,$\mu$s with special
functions rewritten for the GPU), while agreeing with the latter to
within its truncation error (\figurename~\ref{fig:orbitspeed}). The
kernel is limited by per-thread state: an adaptive eighth-order step
keeps $\approx$880\,B per orbit live, so few threads fit on each
multiprocessor. \galpy's sixth-order symplectic integrator with one
common step for the whole batch reaches the same accuracy with 15\%
fewer force evaluations and a fraction of the state, and runs another
3 to 4 times faster. These numbers hold for a smooth, cheap potential
and near-solar disc orbits with similar dynamical times, for which one
step suits every orbit; a batch that mixes very different timescales,
or has close encounters or a rapidly varying potential, has to step at
the rate of its fastest orbit, and there adaptive integration wins.
Such kernels are not differentiable, so they would complement the
backend integrators rather than replace them
(\sectionname~\ref{sec:conclusion}).

\galpy\ is not the only galactic-dynamics code that integrates orbits
with \diffrax: \texttt{galax} \citep{Starkman2026}, a re-implementation
of \texttt{gala} in \jax, does so by design, and comparing the two
checks that the cost of the in-backend solve is in the solver and the
potential rather than in the library around them. The green lines in
\figurename~\ref{fig:orbitspeed} show this comparison for the same
orbits, potential, and tolerances. When we make \texttt{galax} control
its step size per orbit, as the \galpy\ lines do, the two codes take the same
time to within 13 per cent on both GPUs for batches of $10^5$ orbits or
more; for smaller batches, whose cost is dominated by fixed overheads,
they differ by up to a factor of 1.6, in opposite directions on the two
GPUs. In its default mode, \texttt{galax} integrates the whole batch as
one system with a single step-size controller, whose error estimate is
the root-mean-square over all orbits, such that one sequence of steps
serves every orbit; this is cheaper per orbit for large batches, by a
factor of two to three, at the price of an error control that holds
only in the batch-averaged sense and that depends on which orbits
happen to be in the batch. \galpy's \jax\ solve does the same for a
batch on a shared time grid (\sectionname~\ref{sec:orbit-paths}); the
\galpy\ lines in \figurename~\ref{fig:orbitspeed} were measured with
per-orbit control, as \texttt{jax.vmap} over single-orbit solves, which
is the arrangement that the per-orbit \texttt{galax} line matches.

%% file: actionangle.tex
\section{Action--angle coordinates}\label{sec:aa}

\subsection{General approach}\label{sec:aa-general}

\begin{figure*}[t!]
  \centering\includegraphics[width=0.96\textwidth,clip=]{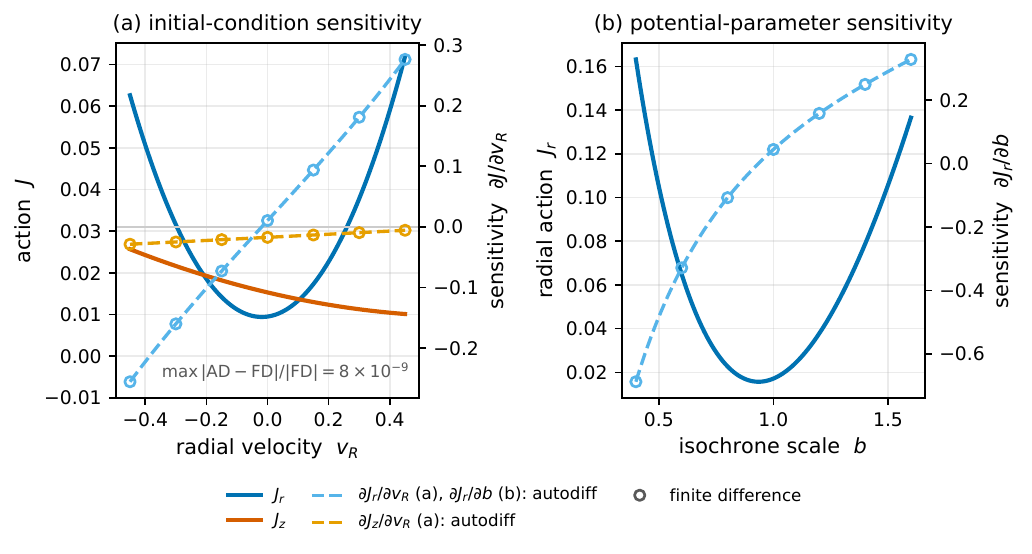}
  \caption{Derivatives of actions in the isochrone potential. Left: the
    radial and vertical actions of an orbit as a function of its radial
    velocity (solid, left axis) and their derivatives with respect to
    the radial velocity by automatic differentiation (dashed, right
    axis) with finite differences (circles). Right: the radial action as
    a function of the scale radius $b$ of the potential and its
    derivative with respect to $b$.}\label{fig:aasens}
\end{figure*}

The action--angle module is where \galpy\ makes heavy use of numerical
methods, all of which were discussed in
\sectionname~\ref{sec:design-numerics}: root finding for the turning
points of an orbit, quadrature of the action, frequency, and angle
integrals between them, and, for the isochrone approximation of
\citet{Bovy2014}, auxiliary orbit integrations with the in-backend
solvers of \sectionname~\ref{sec:orbit-paths}. All of
\texttt{galpy.actionAngle} is migrated onto the backends except for
\texttt{actionAngleTorus}, a convenience wrapper around the external
torus-mapping code of \citet{BinneyMcMillan2016} that is not a core
part of \galpy\ and that we excluded from this work. The public
interface is unchanged: an instance of one of the \texttt{actionAngle}
classes, such as \texttt{actionAngleStaeckel}, is set up for a
potential and called with phase-space positions, through
\texttt{\_\_call\_\_} for the actions, \texttt{actionsFreqs} for actions
and frequencies, and \texttt{actionsFreqsAngles} for all three; a call
with backend arrays, or with a potential that carries a backend
parameter, computes with that backend. Because these methods take a
variable number of positional arguments, they resolve the namespace and
promote scalars internally rather than through the coercion decorator
of the potentials. Derivatives are available with respect to the
phase-space position and, through the Python implementations, with
respect to the parameters of the potential; these parameter derivatives
agree with finite differences in which the potential and the
\texttt{actionAngle} instance are rebuilt from the perturbed parameter
to a few parts in $10^{10}$ for, e.g., the spherical, adiabatic, and
St\"{a}ckel methods. \figurename~\ref{fig:aasens} shows the two kinds of
derivative for the isochrone potential, where the actions are known in
closed form: the derivatives of the radial and vertical actions with
respect to the radial velocity, and the derivative of the radial action
with respect to the scale radius of the potential, which is the
derivative that an inference loop uses when its model is a function of
the actions, as a DF of the actions is
(\sectionname~\ref{sec:df-disk}).

\subsection{Spherical potentials and the isochrone approximation}\label{sec:aa-other}

The isochrone and harmonic-oscillator actions are closed forms, which we
migrated by swapping the namespace. \texttt{actionAngleSpherical}
replaces its \scipy\ root finds and adaptive quadratures with the
shared root finder and a fixed-order Gauss--Legendre rule, with special
handling of circular and near-circular orbits and of orbits that are
unbound. The
\texttt{actionAngleIsochroneApprox} method of \citet{Bovy2014}
integrates an auxiliary orbit, and does so with the in-backend solvers
of \sectionname~\ref{sec:orbit-paths}; it is therefore differentiable
with respect to the potential's parameters, twice so under \jax\ with
the direct adjoint, and runs on a GPU.

\subsection{Inverse transformations}\label{sec:aa-inverse}

\begin{figure*}[t!]
  \centering
  \includegraphics[width=0.9\textwidth,clip=]{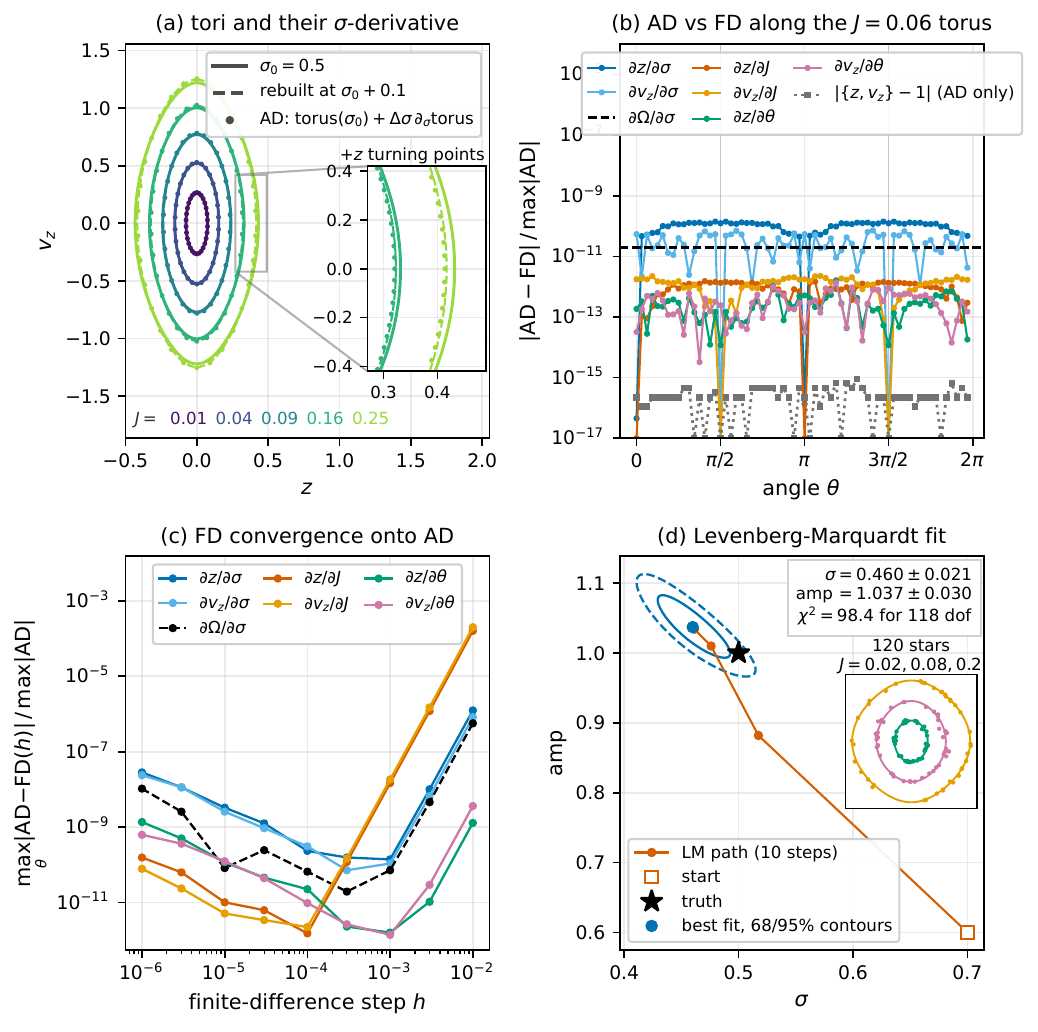}
  \caption{Differentiable inverse action--angle transformation in a
    self-gravitating isothermal disk with \texttt{actionAngleVerticalInverse}.
    (a) Tori at fixed actions for a disk with velocity dispersion
    $\sigma_0$ (solid), the same tori for a disk rebuilt with
    $\sigma_0+0.1$ (dashed), and the first-order prediction from the
    derivative of the tori with respect to $\sigma$ by automatic
    differentiation (points). (b) The difference between the derivatives
    of $z$, $v_z$, and the frequency with respect to $\sigma$, $J$, and
    $\theta$ along the $J=0.06$ torus and five-point finite differences
    with steps between $10^{-4}$ and $10^{-3}$ (the best step of the scan
    in panel c for each quantity), relative to the largest
    derivative. (c) The finite differences
    converge onto the automatic derivatives as $h^4$ until their
    round-off floor. (d) A Levenberg--Marquardt fit of $(\sigma,\,\mathrm{amp})$
    and 120 angles to 120 stars with noisy $(z,v_z)$ on three tori, using
    the automatic Jacobian; the fit converges in ten steps and the truth
    lies inside the 68\,\% contour.}\label{fig:vinverse}
\end{figure*}

The inverse transformation, from actions and angles to positions and
velocities, exists in \galpy\ for the isochrone and harmonic potentials
in closed form and for general one-dimensional potentials through
\texttt{actionAngleVerticalInverse}, which constructs a torus map by a
momentum-matched canonical transformation and tabulates it as a
function of the action. The construction---a root find for the maximum
excursion, a spectral sweep over the angle, a batched Gauss--Newton
solve for the anomaly, and Hermite-spline tables---and the evaluation
run natively on the backends, trace under compilation, and are
differentiable with respect to the action, the angle, and the
parameters of the potential. \figurename~\ref{fig:vinverse} demonstrates this for a
self-gravitating isothermal disk: the derivative of the tori with
respect to the velocity dispersion of the disk predicts the tori of a
nearby disk, agrees with finite differences to $10^{-10}$, and serves
as the Jacobian of a Levenberg--Marquardt fit of the disk's parameters to
stars scattered around three tori. In code, we differentiate the tori of
a family of disks parametrized by their velocity dispersion as
\begin{code}
def torus(sigma):
    disk= IsothermalDiskPotential(amp=1.,sigma=sigma)
    aAVI= actionAngleVerticalInverse(
        pot=disk,Es=Es,nta=128,setup_interp=True)
    return aAVI.xvFreqs(J,angles)[:2] # (z,v_z)
z,vz= torus(0.5)
dz_dsigma,dvz_dsigma= jax.jacfwd(torus)(0.5)
\end{code}
The older construction modes of this class, which use point
transformations and are not recommended, are built on \numpy\ and
cannot take a potential with backend parameters.

Other \galpy-native inverse action--angle transformations are under
active development and will be added with full backend support soon.

\subsection{The St\"{a}ckel and adiabatic approximations}\label{sec:aa-staeckel}

\begin{figure*}[t!]
  \centering\includegraphics[width=0.96\textwidth,clip=]{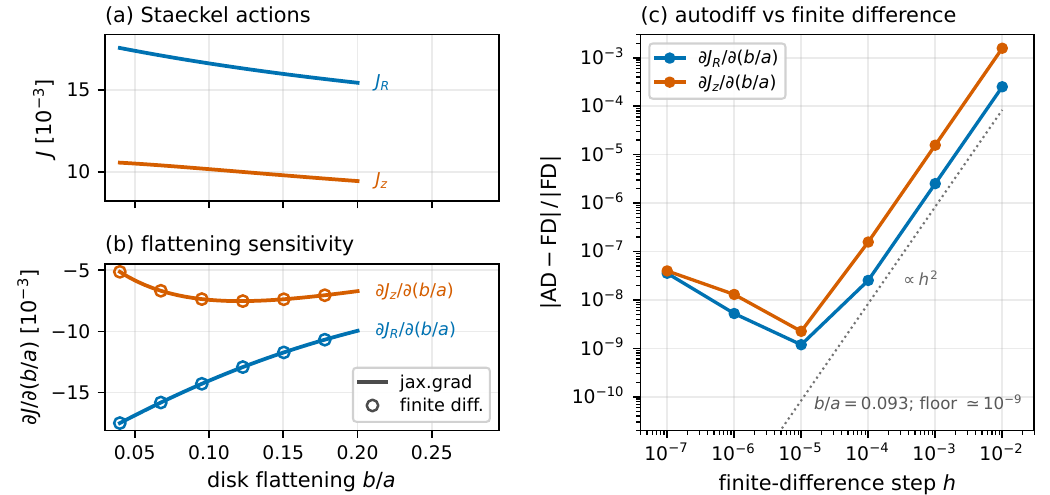}
  \caption{Derivatives of St\"{a}ckel actions with respect to a parameter
    of the potential. (a) The radial and vertical actions of a Sun-like
    orbit in \texttt{MWPotential2014} as a function of the flattening
    $b/a$ of its Miyamoto--Nagai disk (\texttt{MWPotential2014} has
    $b/a=0.093$), computed with \texttt{actionAngleStaeckel(c=False)} at
    fixed focal length. (b) Their derivatives with respect to $b/a$ by
    \texttt{jax.grad} (lines) and by central finite differences in which
    the potential and the action--angle object are rebuilt (circles). (c)
    The relative difference between the two at $b/a=0.093$ as a function
    of the finite-difference step: the finite difference converges as
    $h^2$ onto the automatic derivative until rounding error takes over
    below $10^{-9}$.}\label{fig:staeckelq}
\end{figure*}

\texttt{actionAngleStaeckel}, the workhorse for axisymmetric potentials
\citep{Binney2012,BovyRix2013},
has two implementations that behave differently under a backend. As
part of this work, we rewrote the pure-\python\ path (\texttt{c=False})
to be vectorized over phase-space points, where it previously looped
over them, and the same code now runs under \numpy, \jax, and \torch: it
finds the turning points with the fixed-count bisection and Newton
polish of \sectionname~\ref{sec:design-numerics} and evaluates the
action, frequency, and angle integrals with the same Gauss--Legendre
rule as the \cc\ implementation, in an anomaly variable that removes the
square-root singularities at the turning points, such that the two agree
to machine precision. Everything on this path is
differentiable with respect to the phase-space coordinates and the
potential's parameters, as for example in
\begin{code}
def jr_of_a(a):
    mp= MiyamotoNagaiPotential(a=a,b=0.3)
    aAS= actionAngleStaeckel(pot=mp,delta=0.4,c=False)
    return aAS(R,vR,vT,z,vz)[0].sum()
djr_da= jax.grad(jr_of_a)(0.5)
\end{code}
with the restriction that the derivative of the action integral is of
first order only. \figurename~\ref{fig:staeckelq} shows such a
derivative, of the radial and vertical actions of a Sun-like orbit in
\texttt{MWPotential2014} with respect to the flattening $b/a$ of its
Miyamoto--Nagai disk (its scale height over its scale length, at fixed
scale length), with the potential, its normalization, and the
St\"{a}ckel transformation rebuilt from $b/a$ inside the differentiated
function: the automatic derivative agrees with finite differences to
$10^{-9}$, at the bottom of the finite-difference error curve, and once
compiled it costs about a millisecond.

 The \cc\ path (\texttt{c=True}) is much faster for
small numbers of points and, on a backend, does more than before: the
\cc\ code computes, along with the actions, frequencies, and angles,
their Jacobian with respect to the five phase-space coordinates, which
\galpy\ attaches to the result as a custom derivative rule, such that
\texttt{jax.grad} or \torch's \texttt{backward} through a
\texttt{c=True} call costs one \cc\ call and a matrix product:
\begin{code}
aAS= actionAngleStaeckel(pot=MWPotential2014,
                         delta=0.45,c=True)
def total_jr(R):
    return aAS(R,vR,vT,z,vz)[0].sum()
dJ= jax.jit(jax.grad(total_jr))
djr_dR= dJ(jnp.asarray([1.1,0.9]))
\end{code}
As for the orbit integrators, the \cc\ path gives derivatives with
respect to the phase-space position only, to first order, and the focal
length \texttt{delta} is a fixed number; derivatives with respect to the
potential's parameters require \texttt{c=False}. The focal-length
estimate \texttt{estimateDeltaStaeckel} runs on the backends and
traces, such that a focal length estimated along an integrated orbit
can be part of a compiled program; for example, the following compiles
\begin{code}
R,z= o.R(ts),o.z(ts)
aAS= actionAngleStaeckel(pot=MWPotential2014,c=False,
     delta=estimateDeltaStaeckel(MWPotential2014,R,z))
jr,lz,jz= aAS(R,o.vR(ts),o.vT(ts),z,o.vz(ts))
\end{code}
The grid-based interpolated version,
\texttt{actionAngleStaeckelGrid}, builds its grid on the backend, such
that the grid lives on the GPU when the potential does and the
interpolated actions are differentiable with respect to the query
point; when the potential carries a parameter that is being
differentiated, the grid class delegates to the exact method, with a
warning, because the grid cannot be differentiated through its
construction. The adiabatic approximation \citep{Binney2010} follows the same pattern,
with a fused Jacobian in \cc\ for \texttt{c=True} and, for
\texttt{c=False}, the migrated spherical and vertical machinery for
its radial and vertical parts.

\subsection{Speed}\label{sec:aa-speed}

\begin{figure*}[t!]
  \includegraphics[width=\textwidth,clip=]{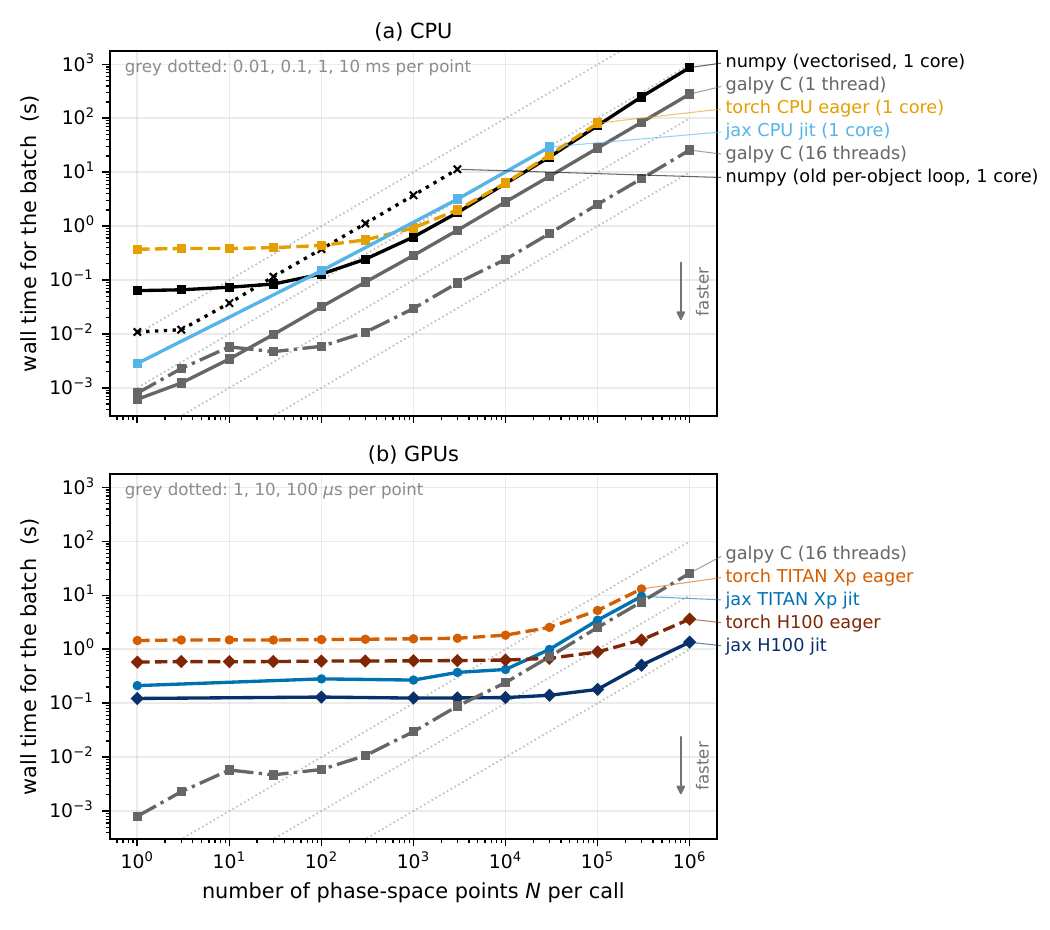}
  \caption{Speed of the St\"{a}ckel action--angle transformation
    (\texttt{actionAngleStaeckel.actionsFreqsAngles}) for
    \texttt{MWPotential2014} as a function of the number $N$ of
    phase-space points per call: wall time on (a) the CPU and (b) the
    GPUs. CPU lines use one core, except for the dash-dotted \cc\ line,
    which uses 16 OpenMP threads and is repeated in (b) for reference; GPU
    lines are for eager \torch\ (dashed; \texttt{torch.compile} is no
    faster here, see the text) and for compiled \jax\ (solid)
    on the TITAN~Xp and the H100; host--device transfers add at most a
    few per cent to the eager \torch\ times and are not shown. The black dotted line with
    crosses is the per-object \numpy\ loop that the vectorized path
    replaces; eager \jax, which sits at a floor of several seconds on
    every device, is not shown. Points where the TITAN~Xp runs out of
    memory are omitted. Light grey dotted lines mark a constant cost per
    point, and the arrows point towards lower wall time. After a launch
    floor, eager \torch\ on the H100 overtakes one \cc\ thread at
    $N\approx2000$ and 16 threads at $N\approx3\times10^4$; compiled
    \jax, with a floor of 0.12\,s, overtakes them at $N\approx400$ and
    $N\approx5000$.}\label{fig:staeckelspeed}
\end{figure*}

\figurename~\ref{fig:staeckelspeed} shows the speed of the full
St\"{a}ckel transformation (actions, frequencies, and angles). On a single
CPU core, the compiled \cc\ path remains the fastest. The new vectorized
\python\ path costs 2--3 times more per point in plain \numpy, but it
is about 6 times faster than the per-object loop that it replaces for
$N\gtrsim10^3$. Eager \torch\ on the CPU runs at the speed of \numpy,
while eager \jax\ is bound by dispatching hundreds of small operations
per call and stays at a floor of several seconds. Compilation removes
that floor, but for large $N$ compiled \jax\ is no faster than \numpy\
on the CPU, because this quadrature-heavy code is bound by arithmetic
that \numpy\ already executes in compiled loops. (\torch\ is shown
eagerly: \texttt{torch.compile} breaks this program into hundreds of
graphs at the root finder's data-dependent loops, which \jax\ stages
as compiled loops, and the result is no faster than eager execution on
the CPU and fails on the H100.) On a GPU, the same
code overtakes the \cc\ path once $N$ is large enough to amortize a
launch floor. For eager \torch\ that floor is about 0.6\,s on the
H100, where it overtakes one \cc\ thread at $N\approx2000$ and \cc\
on 16 OpenMP threads at $N\approx3\times10^4$, and evaluates $10^6$
points 78 times faster than one \cc\ thread. Compiled \jax\ on the
H100 starts from a floor of only 0.12\,s and is five times faster than
eager \torch\ up to $N=10^5$ and 2.7 times at $10^6$ (1.4\,s, 200
times faster than one \cc\ thread and 19 times faster than 16
threads); it overtakes one \cc\ thread at $N\approx400$ and 16 threads
at $N\approx5000$. The floor is set by the
incomplete gamma function of the bulge of \texttt{MWPotential2014}:
\jax's native version iterates per element, which is slow on a GPU at
every array size, so compiled programs on a GPU use \galpy's vectorized
series (\sectionname~\ref{sec:design-numerics}) at every size; with the
native function the floor was 0.7\,s. On the TITAN~Xp, with its slow
double-precision arithmetic, eager \torch\ beats single-threaded \cc\
by a factor of 5--6 at large $N$, but not \cc\ on 16 threads; compiled
\jax\ starts from 0.2\,s there and overtakes one \cc\ thread at
$N\approx10^3$, but not 16 threads. Thus the differentiable backends
are also the fastest way to evaluate the transformation for large
samples on a modern GPU.

%% file: dfs.tex
\section{Distribution functions}\label{sec:df}

\subsection{General approach}\label{sec:df-general}

\galpy\ contains DF models to represent spherical and disky galactic components.
These DF models are constructed from a potential and a set of parameters,
they can be evaluated at phase-space points, integrated to obtain their moments 
(densities, mean velocities, and velocity dispersions), and they can be sampled.
On the backends, evaluation and the moments are differentiable for
every DF; sampling is differentiable wherever it is done by inverse
transformation, which is everywhere except for the many-position
velocity sampler of the quasi-isothermal disk DF
(\sectionname~\ref{sec:df-disk}); and construction is differentiable
for every DF, including those whose construction is itself a numerical
computation---the spherical DFs obtained by inverting a density, the
King model, the quasi-isothermal disk DF, whose table of guiding-center
radii is then built on the backend, and the stream models of
\sectionname~\ref{sec:streams}---with the one exception of the
optional corrections of the two-dimensional disk DFs for the difference
between their input and output profiles, which we did not migrate
(\sectionname~\ref{sec:df-disk}). The parameters
of a DF, like those of a potential, can be given as backend arrays and
are then differentiated through everything that follows.

\begin{figure*}[t!]
  \includegraphics[width=\textwidth,clip=]{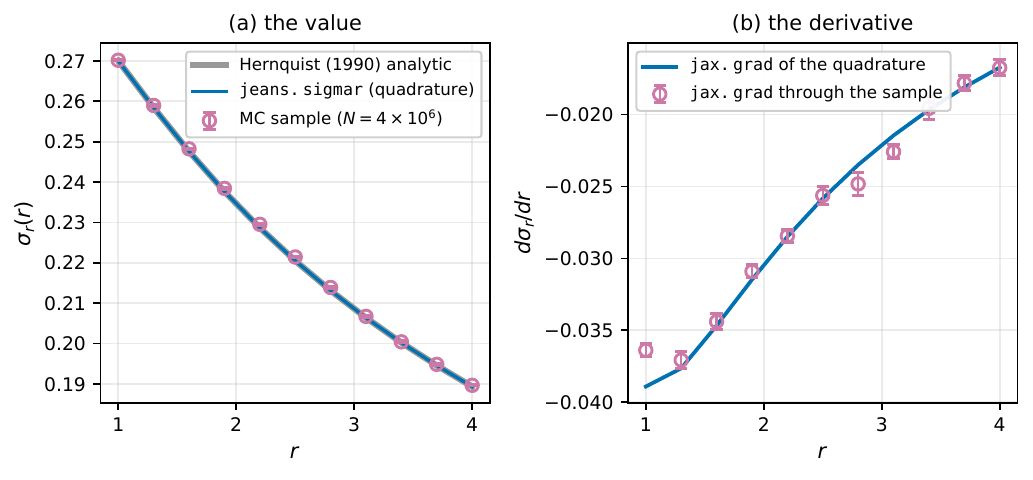}
  \caption{A spatial derivative of a kernel estimate over a fixed DF
    sample. Left: the radial velocity
    dispersion of the isotropic Hernquist DF from the analytic
    expression \citep{Hernquist1990}, from the Jeans equation by quadrature, and from a sample
    of $4\times10^6$ particles. Right: its derivative with respect to
    radius, from \texttt{jax.grad} of the Jeans quadrature and from
    \texttt{jax.grad} of a kernel estimator over the fixed sample, with
    error bars from eight sub-samples; the two agree to a median of
    $0.5\sigma$ of the Monte-Carlo noise, with the innermost point at
    $5\sigma$ from the bias of the kernel.}\label{fig:dfsample}
\end{figure*}

\begin{figure*}[t!]
  \includegraphics[width=\textwidth,clip=]{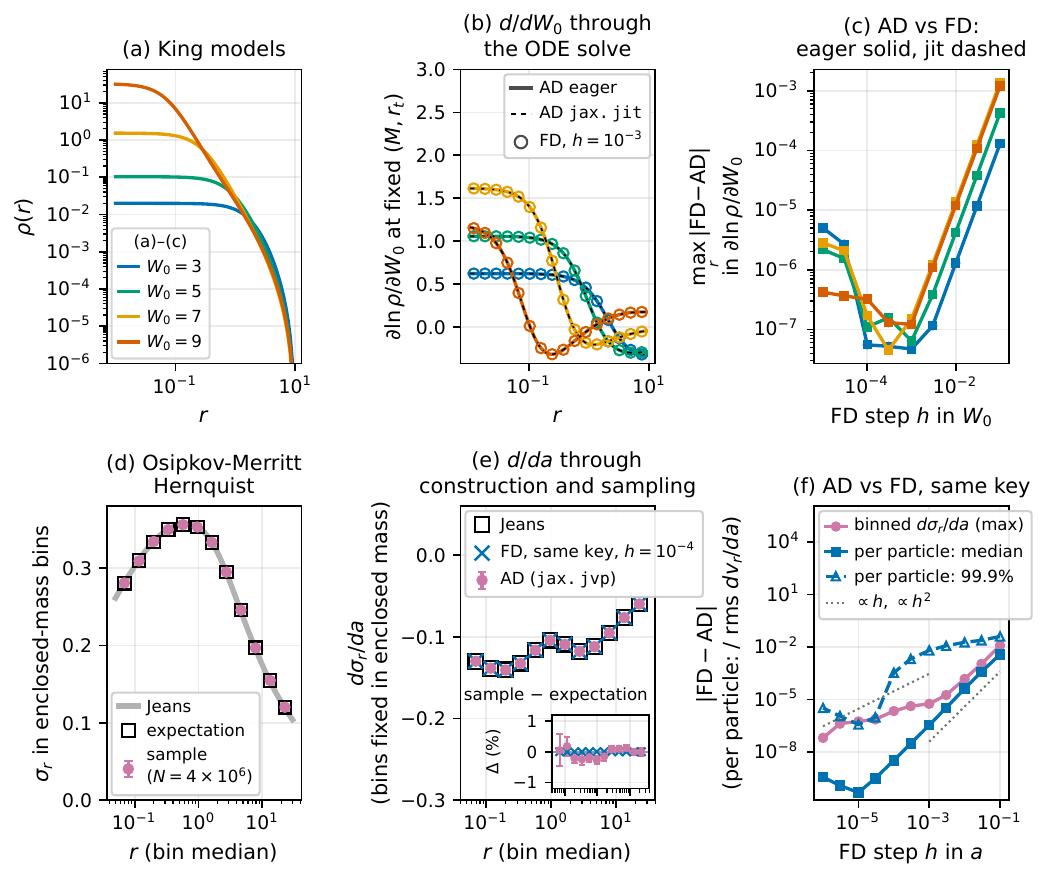}
  \caption{Derivatives through the construction and sampling of
    spherical DFs. Top row: (a) King models of unit mass and tidal radius
    $r_t=10$ for four values of $W_0$; (b) the derivative $\partial\ln\rho/\partial W_0$
    at fixed $(M,r_t)$ through the solution of the Poisson equation,
    eagerly with \scipy\ and under \texttt{jax.jit} with \diffrax, compared
    with a finite difference in which the model is rebuilt, and (c) the
    convergence of that finite difference onto the automatic derivative. 
    Bottom row: an Osipkov--Merritt DF of a Hernquist sphere with
    anisotropy radius $r_a=1.5$, constructed and sampled ($4\times10^6$
    particles) in one compiled function of the scale radius $a$; panel (d) shows that the
    sampled velocity dispersion matches the Jeans equation, (e) demonstrates that its
    derivative with respect to $a$ through construction and sampling
    matches both a finite difference of the same sample and the Jeans
    expectation within the Monte-Carlo noise, and panel (f) shows that the per-particle
    derivatives converge onto the automatic ones as $h$ and $h^2$.}\label{fig:dfparam}
\end{figure*}

Every sampler in \galpy\ takes a \texttt{key=} argument:
\texttt{key=None} draws from \numpy's global generator with the
previous algorithm, and therefore reproduces the previous draws, for
every sampler except the three of
\sectionname~\ref{sec:design-principles} that were replaced on the
\numpy\ path too, while a backend key selects the inverse-cumulative samplers of
\sectionname~\ref{sec:design-numerics} and returns backend arrays. For a
fixed key, the sample is then a smooth function of the DF's and the
potential's parameters, and its derivative is the pathwise, or
reparameterization, derivative: differentiating a statistic of the sample with respect to a
parameter gives an unbiased estimate of the derivative of the
expectation of that statistic. \figurename~\ref{fig:dfsample} shows a
first check that involves no derivative of the sampler itself: a kernel
estimate of the velocity dispersion over a large, fixed sample is
differentiated with respect to the radius at which it is evaluated,
which enters only through the kernel weights, and compared with the
derivative of the Jeans equation; the two share no code and agree
within the Monte-Carlo noise. \figurename~\ref{fig:dfparam} then shows
the pathwise derivative itself, that of a sample through the construction of the DF, for an
Osipkov--Merritt DF of a Hernquist sphere constructed and sampled
inside a single compiled function of the scale radius: the derivative of
the sampled velocity dispersion with respect to the scale radius agrees
with the derivative of the Jeans-equation prediction to within the
Monte-Carlo noise, and with a finite difference of the same sample to
$10^{-7}$. The same figure shows the other kind of derivative of a DF's
construction, that of the King model's density profile with respect to
the dimensionless central potential $W_0$ at fixed mass and tidal
radius, which passes through the solution of the Poisson equation. In
eager mode, \galpy\ solves this ODE with \scipy\ and attaches the
derivative from the forward sensitivity equations, while under
compilation it solves the same ODE in the backend.

\subsection{Spherical DFs}\label{sec:df-spherical}

The spherical DFs of \galpy---the Eddington inversion for isotropic
systems, the Osipkov--Merritt and constant-anisotropy inversions, the
King model, and the closed-form Hernquist and NFW DFs---share one
sampler, which draws radii by inverting the cumulative mass profile and
velocities by inverting a tabulated $p(v|r)$, and which on the backends
does both by interpolation in tables that are differentiable in their
knots. The inversions themselves, which are integrals of a derivative
of the density over the potential, use fixed-order Gauss--Legendre
quadrature in substituted variables and are differentiable through the
integrand, such that the DF can be differentiated with respect to the
parameters of the density and potential that it is built from. The
King model is the one DF whose \numpy\ result changed on purpose: its
Poisson equation is now solved to a relative tolerance of $10^{-10}$
rather than \scipy's default of $10^{-3}$, which left errors of order
$10^{-4}$ in its tables (\sectionname~\ref{sec:design-principles}). The
constant-anisotropy DF is the one place where \jax\ was already present
in \galpy, as an optional dependency for the fractional derivative in its
inversion; its construction is now much faster than it was, by about a
factor of four on \numpy\ and by orders of magnitude under
compilation, because the $f(E)$ inversion is a fixed-shape quadrature
that compiles well. As an example,
under \torch\ we obtain the derivative of a set of sampled radii with
respect to the scale radius of a Hernquist potential as
\begin{code}
a= torch.tensor(1.,dtype=torch.float64,
                requires_grad=True)
hp= HernquistPotential(amp=2.,a=a)
df= constantbetadf(pot=hp,beta=-0.2)
key= galpy.backend.random.key(7,backend='torch')
sam= df.sample(n=1000,key=key)
dr_da= torch.autograd.grad(sam.r().sum(),a)
\end{code}
and for a Hernquist sphere this derivative is $r_i/a$, which the sampler
reproduces to $10^{-13}$. Under \jax, construction, evaluation, and
sampling of these DFs run under \texttt{jax.jit}, and for the King
model so does the derivative with respect to $W_0$, as in for example
\begin{code}
def sigmar(W0):
    kdf= kingdf(W0=W0,M=2.3,rt=1.4)
    return kdf.sigmar(0.5)
dsig= jax.jit(jax.grad(sigmar)) # ODE by diffrax
dsig(3.)
\end{code}
A DF built from a density that is not the one generating the
potential is differentiable with respect to the parameters of both the
density and the potential; for example, with respect to those of the
density:
\begin{code}
def dispersion(a_tracer):
    hp= HernquistPotential(amp=0.1,a=a_tracer)
    df= eddingtondf(pot=NFWPotential(amp=1.,a=2.),
                    denspot=hp)
    return df.sigmar(1.)
jax.grad(dispersion)(0.7)
\end{code}
The closed-form DFs, including the analytic DFs of \citet{Evans1997}
for power-law spheres, and their differential energy distributions
$\mathrm{d}M/\mathrm{d}E$ are likewise differentiable.

\subsection{Disk DFs}\label{sec:df-disk}

\galpy\ contains models for two-dimensional and three-dimensional disk DFs.
The two-dimensional axisymmetric disk DFs of \citet{Dehnen1999},
\citet{Shu1969}, and Schwarzschild
and their moments are evaluated with nested Gauss--Legendre quadrature
on the backends and are differentiable with respect to the parameters of
their surface-density and velocity-dispersion profiles. Their samplers
draw by inverse transformation on every backend: the guiding-center
radius of a sample from the surface-density profile, its orbital phase
from the relation between radius and radial angle that the action-angle
quadrature of \sectionname~\ref{sec:aa-other} tabulates for every
sampled orbit at once, its velocities at a given position from the DF on
a velocity mesh, and its distance along a line of sight from the
projected density; each orbit is then replicated according to the ratio
of the epicycle frequency at its guiding radius to its radial frequency,
which is the DF's measure in the sampled variables. With a key, the
positions and velocities of the samples are differentiable with respect
to the profile parameters, the number of samples, set by the
replication, is not. This replaced adaptive rejection sampling and an
orbit integration per sample on the \numpy\ path too, and made sampling
several times faster. It also corrected the samples: the old sampler
weighted each orbit by the epicycle frequency at the sampled radius over
the radial frequency of a different orbit, the one with zero radial
velocity at the guiding radius, which biased the mean azimuthal velocity
of its samples by a few per cent; the new samples reproduce the DF's
surface density and velocity moments.
We did not migrate
the corrections of these DFs for the difference between their input and
output profiles. \galpy\ also contains
a two-dimensional non-axisymmetric disk DF that is a perturbation of a Dehnen DF by a
non-axisymmetric and/or time-dependent potential, which is evaluated by integrating orbits
backwards in time to an initial axisymmetric DF (in \texttt{galpy.df.evolveddiskdf}). 
The moments of
this evolved disk DF differentiate through that integration,
which on the backends is one vectorized in-backend solve over the
velocity grid rather than a loop. 

As a three-dimensional disk DF, \galpy\ implements the quasi-isothermal
DF of \citet{Binney2010,BinneyMcMillan2011}, in the form used by
\citet{BovyRix2013},
which is a function of the actions and is evaluated by first computing the actions with 
\texttt{actionAngleStaeckel} and then evaluating the DF.  The quasi-isothermal DF is now 
evaluated and its moments computed on the backends,
differentiably with respect to the parameters of the potential and the
DF's scale lengths, and its velocity sampler at a given position draws
by inverse transformation on a three-dimensional velocity mesh, such
that a sample drawn with a fixed key is a differentiable function of the
DF's parameters;
this sampler replaced a rejection sampler on the \numpy\ path as well,
which is one of the three places where the \numpy\ draws changed (the
distribution is the same, the particular draws are not). In code this looks as follows:
\begin{code}
qdf= quasiisothermaldf(1./3.,0.2,0.1,1.,1.,
                       pot=MWPotential2014,aA=aAS)
vs= qdf.sampleV(0.9,0.05,n=10**4,
                key=grandom.key(0,backend='jax'))
sigma_R= jnp.std(vs[:,0])
\end{code}
Its construction tabulates the guiding-center radius as a function of
angular momentum, which is built on the backend when the potential or a
scale length carries a backend parameter, such that the DF and its
moments are differentiable through the construction. Its interpolated
velocity sampler, which samples many positions at once, is a rejection
sampler with \numpy\ proposals that runs on the backends, with its
shrinking arrays padded to fixed sizes, but is not differentiable.

\subsection{Speed}\label{sec:df-speed}

\begin{figure*}[t!]
  \includegraphics[width=\textwidth,clip=]{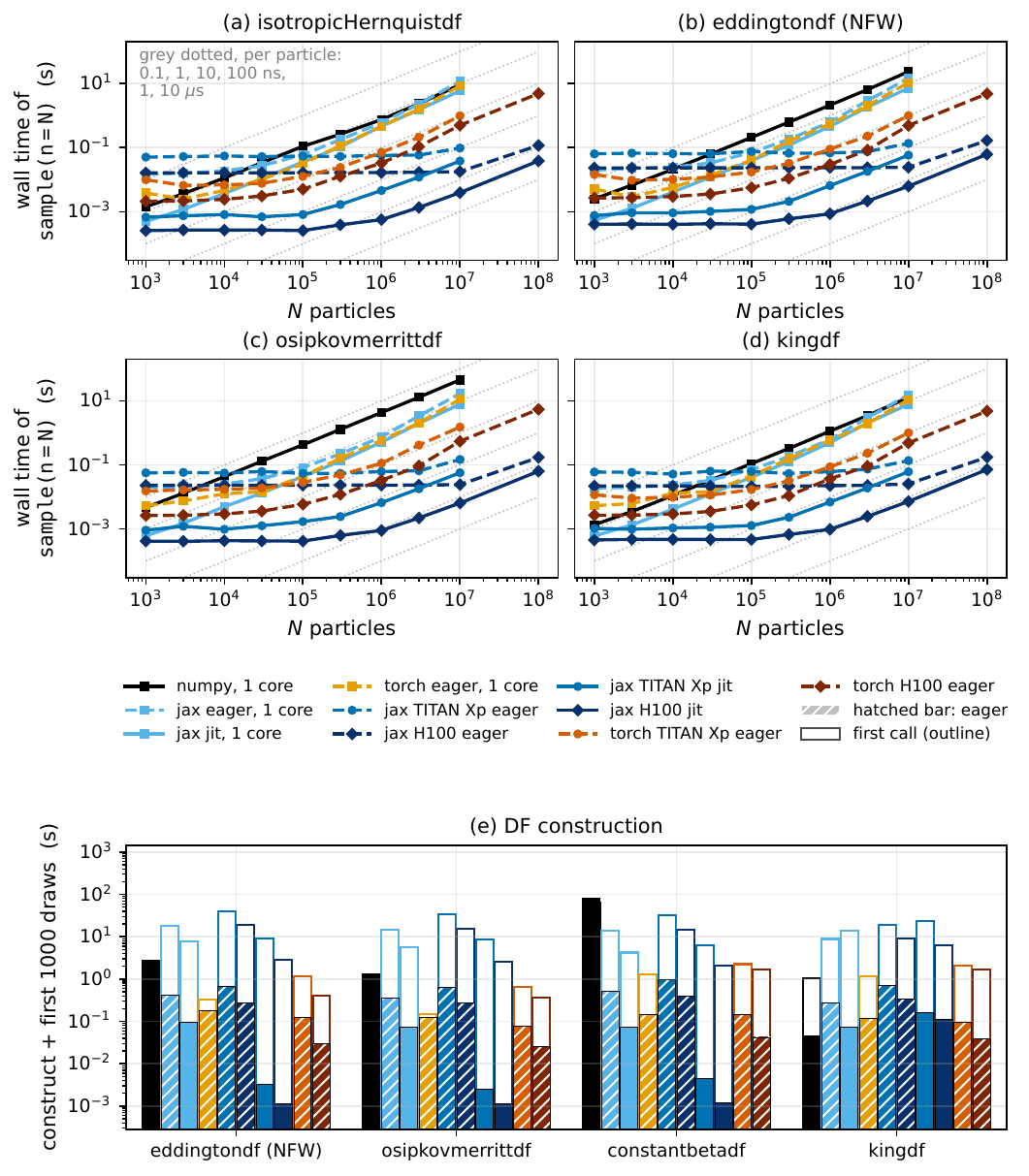}
  \caption{Speed of sampling and constructing spherical DFs. Top: wall
    time of \texttt{sample(n=N)} for the isotropic Hernquist DF, the
    Eddington inversion of an NFW profile, an Osipkov--Merritt DF, and a
    King model, on \numpy\ and on the backends on the CPU, the TITAN~Xp,
    and the H100; solid lines are compiled, dashed eager. Bottom: time to
    construct each DF and draw a first thousand particles, warm (filled)
    and including compilation on the first call in a fresh process
    (outline); hatched bars are eager. All CPU series use one core. Compiling with \texttt{jax.jit} or
    running on a GPU makes sampling much faster than with \numpy, by two
    to nearly four orders of magnitude on a GPU for large samples.}\label{fig:dfspeed}
\end{figure*}

Sampling is where the backends change what is practical, as is clear
from \figurename~\ref{fig:dfspeed}, which displays the time to draw $N$
particles from four spherical DFs. On the CPU, eager \jax\ and \torch\ are slower
than \numpy\ for small samples, because every one of the hundreds of
array operations in a draw is dispatched separately and each costs a few
to tens of microseconds, and they catch up only at $N\approx10^4$;
compiled \jax\ is faster than \numpy\ at every size, by a factor of
1.5 to 6 at $N=10^7$ on one core, because the compiler fuses the
hundreds of array operations of a draw into a few loops, where \numpy\
executes each separately and writes its intermediate results to
memory. On a GPU the sample of $10^7$ particles takes
tens of milliseconds on the consumer card and a few milliseconds on the
H100, two to nearly four orders of magnitude faster than \numpy\ on
one core. The lower panel
shows the cost of constructing a DF, which for the inversions is
dominated by the $f(E)$ integral: more than a minute on \numpy\ for the
constant-anisotropy DF, a fraction of a second once compiled, with the
compilation itself costing seconds and paid once.

%% file: streams.tex
\section{Tidal streams}\label{sec:streams}

Tidal streams are the application in which derivatives have been most
sorely missed, because a stream model is an expensive forward
computation with a handful of parameters, the potential's among them,
and fitting it to data has meant exploring that parameter space with a
sampler that knows nothing of the model's smoothness, a cost that has
limited such fits to a few parameters at a time. \galpy\ contains three stream models, all of
which now run on the backends: the action--angle model of
\citet{Bovy2014}, \texttt{streamdf}, with its extension to stream gaps
\citep{Sanders2016}; and the particle-spray models of \citet{Fardal2015}
and \citet{Chen2025}, \texttt{streamspraydf} \citep[see also][for the \galpy\ implementation]{Qian2022}.

\subsection{The action--angle stream model}\label{sec:streams-streamdf}

\begin{figure*}[t!]
  \includegraphics[width=\textwidth,clip=]{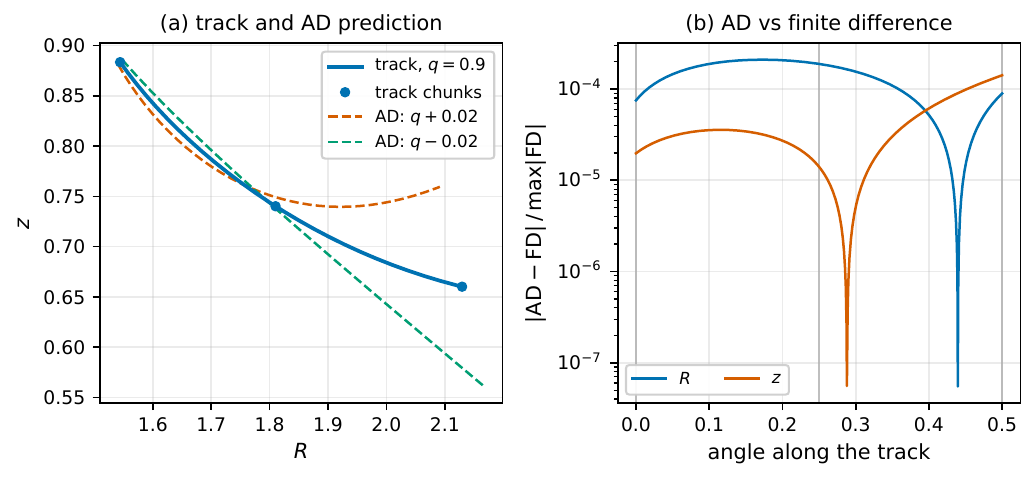}
  \caption{Derivative of a \texttt{streamdf} track with respect to the
    flattening $q$ of a logarithmic halo. Left: the track in the
    meridional plane at $q=0.9$ (the interpolated track through the
    three computed segments, which are marked) and its first-order
    extrapolations to $q\pm0.02$ from the derivative by automatic
    differentiation through the whole track computation, including the
    interpolation. Right: the difference between this derivative and a
    central finite difference ($h=10^{-3}$) of the same computation
    along the track, relative to the largest derivative of each
    coordinate; the vertical lines mark the computed segments. The two
    agree to a few parts in $10^4$, the level set by the fixed iteration
    count of the track refinement.}\label{fig:streamtrack}
\end{figure*}

\begin{figure*}[t!]
  \includegraphics[width=\textwidth,clip=]{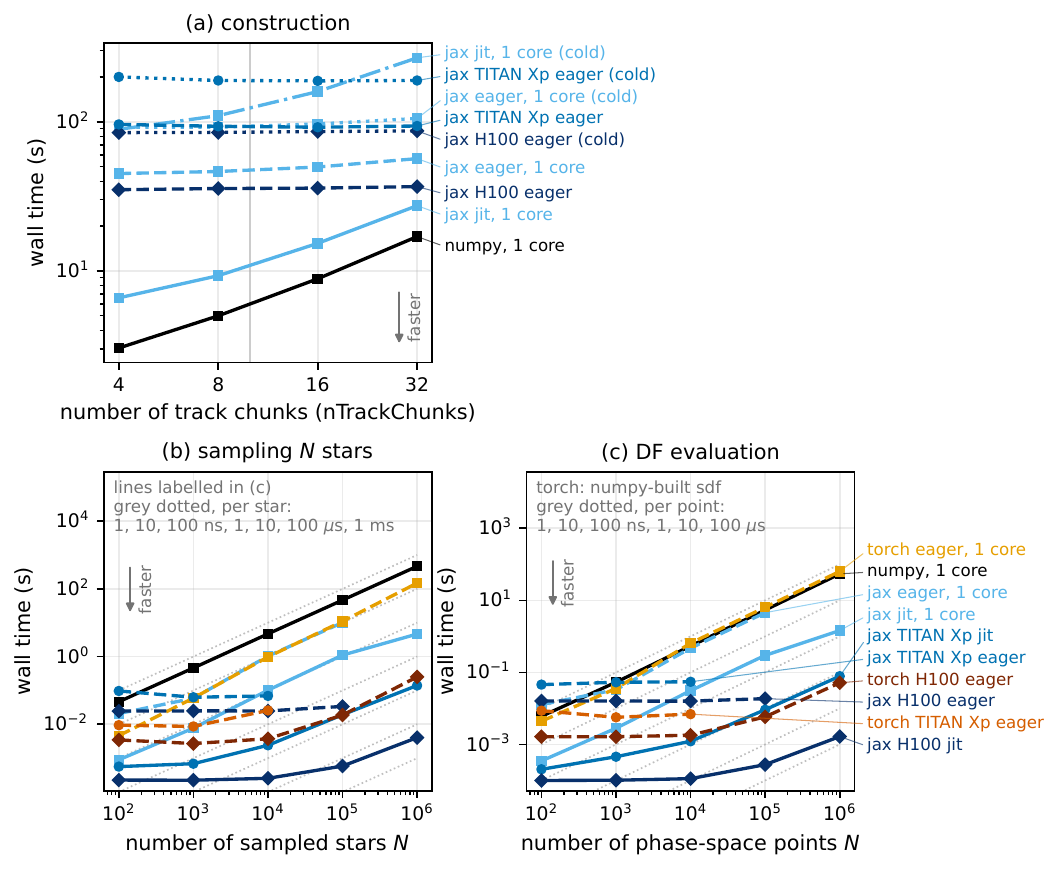}
  \caption{Speed of the action--angle stream model. (a) Wall time to
    construct a \texttt{streamdf}, including its track, as a function of
    the number of track segments, on \numpy\ with the \cc\ integrators
    and on the backends, warm and including compilation (cold). (b)
    Sampling $N$ stars and (c) evaluating the DF at $N$ phase-space
    points from the constructed model. All CPU series use one core. The
    compiled \jax\ series run to $N=10^6$ on every device; the eager
    series end at the last $N$ whose temporaries (arrays of $N$ by the
    number of track points) fit in memory: $10^4$ on the 12\,GB
    TITAN~Xp, and $10^5$ for eager \jax\ on the H100 (with \jax's
    default allocator settings) and on the CPU, where $10^6$ needs more
    than 50\,GB; eager \torch\ on the H100 reaches $10^6$ using
    50\,GB.}\label{fig:streamdfspeed}
\end{figure*}

\texttt{streamdf} models a stream as a distribution in frequency and
angle space, computes the mean track of the stream in configuration
space by iterating the transformation between the two, and samples and
evaluates the DF around that track. On the backends the track
computation is a pure function of the progenitor's orbit and the
potential that is vectorized over the segments of the track and is
differentiable from end to end: the auxiliary orbits are integrated in
the backend, and the Jacobian of the action--angle transformation along
the track, which was a finite-difference Jacobian in the \numpy\ code,
is the automatic Jacobian of the methods of \sectionname~\ref{sec:aa}.
The derivative of the track with respect to a parameter of the
potential agrees with a finite difference of the same computation to a
few parts in $10^4$ (\figurename~\ref{fig:streamtrack}), which is the
level set by the fixed iteration count of the track refinement. Construction of a \texttt{streamdf} is
differentiable with respect to the potential, and can be compiled,
although the compilation of a construction, which contains tens of
orbit integrations and the iteration of the track, takes a few minutes
on one CPU core and is only worthwhile when many streams are to be
built; the construction itself, once compiled, takes 7 to 30\,s on one
core, about twice as long as the \numpy\ and \cc\ version.
A \texttt{streamdf} is set up as before, with a progenitor orbit, an
\texttt{actionAngle} instance, and a velocity dispersion, as in for example:
\begin{code}
aAI= actionAngleIsochroneApprox(pot=lp,b=0.8,
        integrate_method='diffrax')
sdf= streamdf(sigv/220.,pot=lp,aA=aAI,
              progenitor=Orbit(jnp.asarray(prog)),
              tdisrupt=4.5/time_in_Gyr(220.,8.),
              leading=True,nTrackChunks=11)
sdf.sample(n=10**4,key=grandom.key(3,backend='jax'))
\end{code}
When \texttt{lp} in this code carries a backend parameter the track and the
sample are differentiable with respect to it. Sampling and evaluating
the DF, on the other hand, are 40--100 and 18--37 times faster under
compilation than on \numpy\ on one CPU core, and hundreds to thousands
of times faster on a GPU;
the sampler draws by inverse transformation on \numpy\ too, which is the
other place where the \numpy\ draws changed, in the draws but not in
their distribution. \figurename~\ref{fig:streamdfspeed} summarizes these
costs.

\subsection{Particle-spray models}\label{sec:streams-spray}

\begin{figure*}[t!]
  \centering\includegraphics[width=0.94\textwidth,clip=]{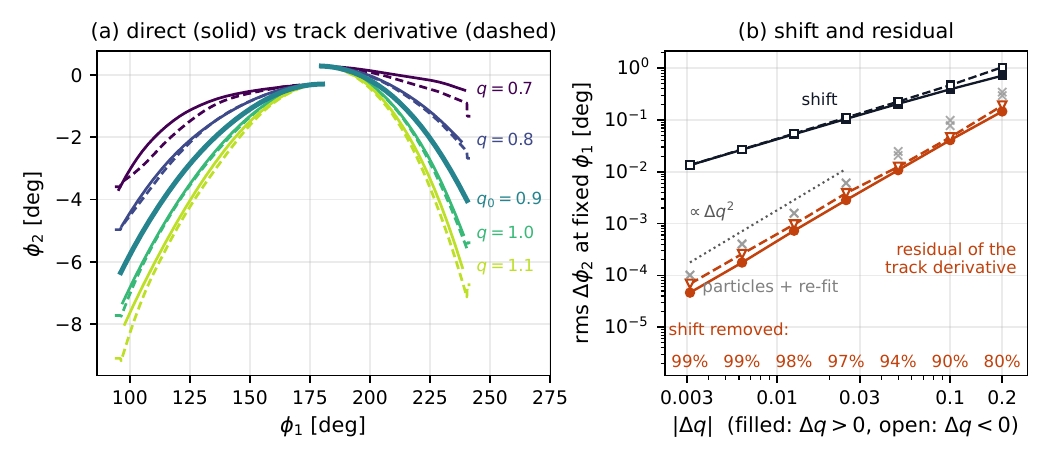}
  \caption{Predicting a stream's track in a different potential from one
    derivative. Left: the leading and trailing tracks
    (\texttt{streamTrack}) of a 6000-particle \texttt{fardal15spraydf}
    stream after 4.5\,Gyr in a logarithmic halo with flattening $q_0=0.9$
    (thick), the tracks in halos with $q=0.7$ to $1.1$ computed directly
    (solid), and their first-order prediction
    $\mathrm{track}(q_0)+\Delta q\,\mathrm{d\,track}/\mathrm{d}q$ from the
    forward-mode derivative of the sampled and fitted track at $q_0$
    (dashed), in a coordinate frame aligned with the progenitor's orbit.
    Right: the root-mean-square shift of the track at fixed $\phi_1$ over
    the central 80\,\% of each tail and the residual of the first-order
    prediction, for $\Delta q>0$ (filled) and $\Delta q<0$ (open), with
    the same random numbers, frame, and track fit throughout; the numbers
    give the fraction of the shift that the prediction removes, and the
    residual falls nearly as $\Delta q^2$ (dotted). Crosses show the
    residual when instead the particles are moved to first order and
    the track is re-fitted.}\label{fig:extrapolation}
\end{figure*}

The particle-spray models release particles from the progenitor at
random times with random offsets in position and velocity and integrate
them forward. On the backends, \texttt{sample()} and the track-fitting
method \texttt{streamTrack()}, which fits a smooth track through the
sampled particles, are pure functions of the progenitor's
orbit, the potential, the progenitor mass (or a mass history), and a
random key: the stripping times are drawn by inverse transformation of
uniform variates, the offsets by reparameterized normal variates, the
progenitor and the particles are integrated in the backend, and the
track is fit with a penalized spline. Every step of that fit is a smooth
function of the particles: each particle's closest point on the
progenitor's orbit is found by Newton iterations on a spline of the
orbit, the amount of smoothing by maximizing the restricted likelihood,
and the extent of the track from a kernel-smoothed quantile, so that the
fit, which is also the default for \numpy\ inputs, gives the same track
eagerly and compiled. The whole computation compiles
and is differentiable with respect to every parameter; forward-mode
derivatives, which are the efficient choice for a few parameters and
many particles, require \diffrax's direct adjoint, like in the following example
\begin{code}
key= galpy.backend.random.key(1,backend='jax')
def leading_arm(q):
    lp= LogarithmicHaloPotential(normalize=1.,q=q)
    spdf= fardal15spraydf(mass,progenitor=Orbit(ic),
        pot=lp,tdisrupt=td,
        integrate_kwargs={'adjoint':'direct'})
    return spdf.sample(n=1500,return_orbit=False,
                       tail='leading',key=key)
xv,dxv= jax.jvp(leading_arm,(q0,),(1.,))
\end{code}
where the forward-mode derivative gives the change of every particle's
position for a unit change in the halo flattening, with the random
numbers in common between the two. The fitted track can be
differentiated in the same way,
\begin{code}
def track(q):
    lp= LogarithmicHaloPotential(normalize=1.,q=q)
    spdf= fardal15spraydf(mass,progenitor=Orbit(ic),
        pot=lp,tdisrupt=td,
        integrate_kwargs={'adjoint':'direct'})
    tr= spdf.streamTrack(n=1500,tail='leading',key=key)
    return tr.x(tps)
track(0.9); jax.jacfwd(track)(0.9)
\end{code}
where the derivative includes the response of the fit itself: of each
particle's closest point on the progenitor's orbit, through the implicit
derivative of the Newton iterations, and of the amount of smoothing.
This is the derivative that matters for a generative model like this
one, in which the stream's stars are a sample from the model's DF: the
derivative of the track with respect to a parameter can be taken
either with the sample of stars held fixed, which is what one obtains
by differentiating the track code alone, or by letting the sample move
with the parameter, which is what a finite difference of the full
model computes. The two differ, and the first is not a physical
quantity; where \galpy\ provides derivatives of stream models, they
are of the second kind, and we test them against finite differences
whose step is shown to be converged: central finite differences of the
fitted track converge to the derivative above to $2\times10^{-6}$, and
the derivative of the sampled stream with respect to the potential
agrees with finite differences of the same sample to $10^{-6}$.
\figurename~\ref{fig:extrapolation} shows what such a derivative can
do: a single forward-mode pass through the sampling, integration, and
track fit of a 6000-particle stream after 4.5\,Gyr in a logarithmic
halo predicts the stream's track in a halo of different flattening as
$\mathrm{track}(q_0)+\Delta q\,\mathrm{d\,track}/\mathrm{d}q$. This
first-order prediction removes 80\,\% of the shift of the track for
$|\Delta q|=0.2$ and 99\,\% or more for $|\Delta q|\leq0.006$, a range
that covers the steps a fitter takes, with a residual that falls as
$|\Delta q|^{1.9}$; moving the particles to first order and re-fitting
the track instead leaves a residual that is 1.5 to 2.4 times larger at
every step. Each spray
particle's orbit is one more orbit in a batch, such that a sprayed
stream costs about 0.6\,ms per particle on one CPU core under compiled
\jax\ for batches of up to $3\times10^4$ particles (1.2--1.7\,ms for
$10^5$--$10^6$), a fifth of the cost
of the non-differentiable \cc\ version, and
about 0.15\,ms on the TITAN~Xp and 10\,$\mu$s on the H100 per particle
(\figurename~\ref{fig:spraysp}), consistent with the cost per orbit of
a large batch of in-backend orbit integrations
(\figurename~\ref{fig:orbitspeed}), of which a spray particle's orbit
is one.

\subsection{Stream gaps}\label{sec:streams-gaps}

The stream-gap model of \citet{Sanders2016}, \texttt{streamgapdf},
computes the velocity kick that a passing subhalo imparts along a stream
and propagates it through the action--angle model of
\texttt{streamdf}. The kick kernels for
the Plummer and Hernquist subhalos, the interpolation of the kicks along
the stream, and the transformation to the impact frame run on the
backends and are differentiable with respect to the parameters of the
impact, and the orbit integrations in the model go through the batched
in-backend solver when given backend inputs. The construction of the
model also runs on the backends and can be compiled: with a backend
progenitor and an \texttt{actionAngleIsochroneApprox} that integrates
with \diffrax, every stage of the set-up is a backend computation, and
the construction compiled with \texttt{jax.jit} reproduces the eager
one to $2\times10^{-11}$. The density along the stream, the mean
frequency, and the frequency--angle kicks are differentiable through
the construction with respect to the subhalo's mass, scale, impact
parameter, and velocity, the time of the impact, and the parameters of
the potential, and these derivatives agree with finite differences of
rebuilt models to better than $10^{-6}$. The same construction runs
under \torch, with \torchode\ for the orbit integrations, and gives the
same density and mean frequency along the stream, but eagerly: the
set-up of the model that takes a minute and a half under compiled \jax\
takes half an hour under \torch, because its many small array
operations are each dispatched separately, so that \jax\ is the
backend of choice for this model.

%% file: process.tex
\section{How the backends were built}\label{sec:process}

Because an AI agent wrote the code, the question of how one comes to
trust such code is as much a part of this paper as the code itself. In
this section, we describe the organization of the work, how we reviewed
it, the stretch during which the agent worked without supervision, how
we made the existing test suite pass under the new backends, and the
checks through which we established that the code is correct. Because
the division of labor between the human author and the agent is the
subject here, this section names them separately, as JB and the agent.

\subsection{Organization}\label{sec:process-organization}

Work started on 2026 June 1 with a design discussion between JB and
Claude, Anthropic's AI coding agent, used through the \texttt{Claude
  Code} environment, which has access to a shell, to the \galpy\
repository, and to \github. The discussion settled the scope (all of
\galpy, on both \jax\ and \torch) and the rules of
\sectionname~\ref{sec:design-principles}: byte-identity of the \numpy\
path, an unchanged interface, no compilation inside \galpy, and the
existing test suite as one of the main acceptance tests. The first
code, which took a day to write, was a pilot that made two potentials
run under all three array libraries and checked value parity, gradients
against finite differences, and the identities between the potential,
its forces, and its second derivatives under automatic differentiation.

\begin{figure*}[t!]
  \centering\includegraphics[width=0.94\textwidth,clip=]{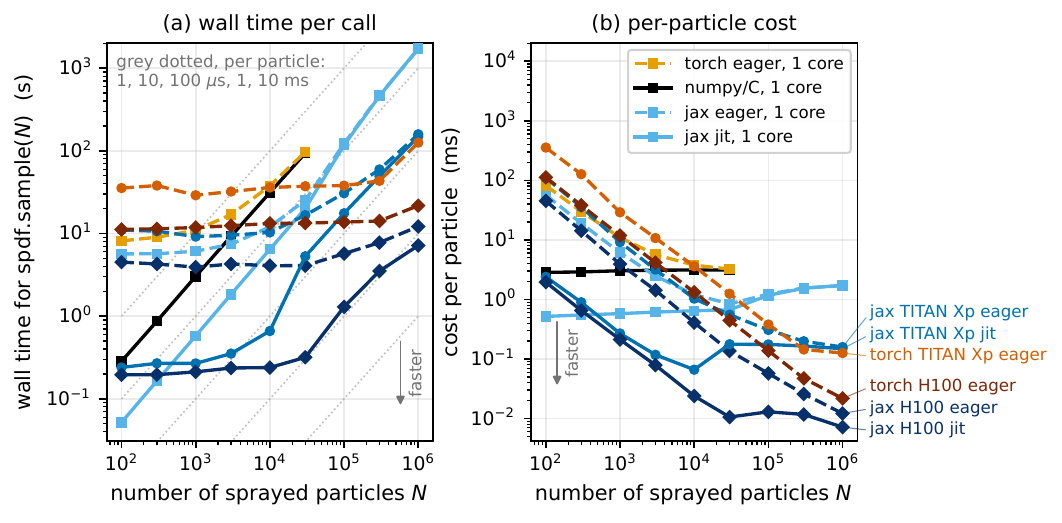}
  \caption{Speed of the particle-spray stream model: wall time (left)
    and cost per particle (right) of \texttt{fardal15spraydf.sample(N)}
    with 4.5\,Gyr of stripping, for the \numpy\ version with the \cc\
    integrator and for the differentiable backend version on the CPU,
    the TITAN~Xp, and the H100, compiled (solid) and eager
    (dashed). All CPU series use one core; the \numpy\ and eager \torch\
    CPU series end where one call exceeds a minute, while the \jax\ CPU
    series were continued to $N=10^6$, where one compiled call takes
    29\,min. Once compiled, the differentiable version is faster than the
    \numpy\ version with the \cc\ integrator at every $N$, and much
    faster on a GPU for large $N$.}\label{fig:spraysp}
\end{figure*}

The work was organized in branches of the \galpy\ repository. \galpy's
main branch, from which version~1.x is released, continued to receive
its usual maintenance and the fixes to the existing \numpy\ and \cc\
code that the backend work turned up
(\sectionname~\ref{sec:process-fixes}). The backend code was developed
on a long-lived \emph{backend branch}, to which every backend change
was proposed as a pull request (PR) and which was periodically rebased
onto the main branch, without merge commits, so that it always contained
the main branch's fixes; it has become \galpy's version~2, released as
pre-releases while the main branch continues for now as version~1.x.
Between 2026 June 1 and October 30, 723 PRs were opened for this work,
of which 696 were merged: 602 on the backend branch, 29 for the
three-dimensional variational equations, which had a branch of their
own, and 65 against \galpy's main branch, of which 40 fixed defects in
the existing \numpy\ and \cc\ code that the backend work had found
(\sectionname~\ref{sec:process-fixes}), 16 added capabilities or
speed-ups to that code, and 9 concerned the tests and the continuous
integration. The merged PRs have a median size of 182 changed lines
(quartiles 74 and 372 lines). We kept them small on purpose, because a
PR is reviewable when it does one thing, and we split larger pieces of
work into stacks of dependent PRs that are reviewed in order (101 PRs
are members of such stacks); the migration of the spherical DFs, for
example, is a base PR and four PRs that each port one family. Each PR
carries a description of what was changed, the measurements that
support it (timings, parity counts, gradient errors), and what was
explicitly left out, such that the description can be reviewed without
re-deriving it from the diff. The rate of PRs followed the kind of
work: 88 backend PRs in June, when the agent migrated the potentials
family by family from a fixed template; 176 in July, when it ported the
orbit, action--angle, and DF modules and the autonomous stretch
started; 114 and 112 in August and September, fewer and larger, when
the work turned to the compiled modes, the test ledger, and the harder
DFs, whose remaining problems were coupled; and 135 in October.

In round numbers, and in US dollars, the work cost the following. The
agent ran on a fixed-price subscription to Anthropic's Claude: until
2026 June 25 the ``Max'' plan at its 20$\times$ tier (\$200 per month),
which Anthropic provided free of charge, afterwards the 5$\times$ tier
at \$100 per month, and for the final two weeks the 20$\times$ tier
again, this time paid. Successive versions of the Claude Opus models
did almost all of the coding (Claude Fable did it on six days, about
4\,\% of the agent's output), and Claude Fable was otherwise used only
to draft this paper and its major revisions. Had the agent's usage
(\sectionname~\ref{sec:process-review}) been paid for through the
Claude API instead, it would have cost about \$31,000 at the list
prices of the time, 61\,\% of it for re-reading the cached context at
every step, 29\,\% for writing it to the cache, and 9\,\% for output.
The larger cost was the continuous integration, because every
backend PR runs \galpy's test suite in five modes over 22 shards, about
150 jobs of 20 to 90 minutes each; the GitHub Actions bill for the
repository at list price was \$13,100 in June, \$16,500 in July,
\$15,800 in August, \$10,200 in September, and \$8,600 in October 2026,
of which about \$59,000 is attributable to the backend project (in
August and September, the repository also received 58 PRs for another
\galpy\ project, and a PR to the main branch runs only the \numpy\
suite). GitHub's Education program covered this cost, which would
otherwise have been prohibitive. The continuous integration was
crucial to establishing that the PRs were correct: it ran every test
in every mode on every PR, and it often turned up errors that the local
runs before pushing had not, in modes, shards, or versions of the
dependencies that the agent had not tried. It was also the bottleneck
of the project, because a PR could not be merged before its matrix had
run, the shards queued behind each other when several PRs were open,
and a flaky job meant a two-hour wait; several of our engineering
decisions---cancelling the runs of superseded commits, running the
compiled modes only on demand until they were stable, running the
touched tests under forced backends locally before pushing---served to
spend less time in that queue.

JB fixed a few rules early and enforced them throughout, beyond those of
\sectionname~\ref{sec:design-principles}. A tolerance could be loosened
only for one test, only slightly, and with a justification in the PR.
Test coverage was not allowed to decrease, not by a single line; in the
end, the backend branch leaves 14 of its 56,950 executable statements
uncovered, against 24 of 41,704 on the main branch. Tests had to make
meaningful assertions at tight tolerances: values had to match their
reference to a relative $10^{-13}$ or better, often $10^{-14}$, and
derivatives a finite difference that is shown to converge; a test could
never merely check that a result is finite, and only tests of sampling
had looser tolerances, set by their Monte Carlo noise. A test that
passes under a backend only because the code silently fell back to
\numpy\ does not count (\sectionname~\ref{sec:process-islands}). Before
pushing any PR, the agent ran its tests locally under forced \jax\ and
\torch\ backends, which the \numpy-only checks cannot do, and measured
the \numpy\ byte-identity by comparing checksums of outputs between two
checkouts that differ only in the files of the PR.

\subsection{Review}\label{sec:process-review}

Every PR was reviewed before it was merged into the backend branch, and
JB merged all of them except during the stretch described in
\sectionname~\ref{sec:process-autonomous}, whose work JB reviewed before
it was merged into the backend branch. Much of the review was
conversational, in the session in which the agent worked, and
concerned design as much as code: that a hand-rolled root finder
belongs in a utility module or should be replaced by \scipy's, that
speed-ups are to be demonstrated by timing both versions rather than
asserted, that tests should go through public methods rather than
internal functions, and so on. Reviews on \github\ itself caught the
kinds of problems that reviews of human code catch: a variable omitted
from an OpenMP \texttt{private} clause, a coverage drop of one line that
the automated check had passed.

Claude Code records every message of a session, and the token counts of
every response, in a transcript, from which we can estimate how much
input the work took, if only roughly, because separating JB's own text
from the material that he passed on (reports from other agents,
reviews, logs) relies on simple heuristics. Between 2026 June 4 and
October 30, JB sent about 2,500 messages, on 141 of the 149 days,
almost a quarter of them while the agent was working. Most were short,
with a median length of 14 words, and in total they contain about
57,000 words of JB's own text, or roughly 110 single-spaced pages, plus
about 33,000 words that he passed on. Over the same period, the agent
produced about 114 million output tokens (including its reasoning and
its tool calls, a third of them in subagents that it started) and read
about 44 billion input tokens, almost all of them its own context
re-read from a cache at every step.

JB also used two other forms of review. First, at several points he ran
an independent AI code reviewer (OpenAI's Codex) over the agent's work and
passed its findings to the agent to address (18 reviews in all). These
reviews found real defects: a gradient of the incomplete gamma function
that was \texttt{nan} at the endpoint $x=0$ although its value was
correct, a surface-density routine that used the backend quadrature
only when the height above the plane was traced, such that a derivative
with respect to the radius at fixed height fell back to \scipy\ and
failed, and a branch-selection rule in the compiled path of the
hypergeometric function that returned infinity where the eager path was
correct. The agent reproduced each finding before fixing it and dropped
the findings that it could not reproduce with a note saying why.
Second, before merging complex PRs, JB instructed the agent to run
adversarial reviews of its own work by independent agents whose only
task was to break the PR against the real code; these found, for
example, that a stream sampler handed the same random key to the
position- and velocity-angle samplers, and that the parity tests of an
early PR compared at a tolerance four orders of magnitude too loose to
catch the $10^{-16}$ differences from re-associated arithmetic. The
latter finding set the pattern for all subsequent work: the original
\numpy\ expression is kept byte-for-byte and only the branching is
rewritten, never an algebraically equivalent form.

\subsection{The autonomous stretch}\label{sec:process-autonomous}

From 2026 July 26 to August 9, JB was away and the agent worked on its
own, with the goal of making the whole existing test suite pass under
compilation and of emptying the ledger of expected failures
(\sectionname~\ref{sec:process-ledger}). For this, the agent opened PRs
against a separate development branch, which it merged itself once the
continuous integration was green, under the understanding that JB would
review the accumulated branch on his return. It could not merge into
the backend branch or into \galpy's main branch, and it could not
change the continuous-integration configuration. These were hard
guardrails rather than instructions: the agent worked through its own
\github\ account, whose access token lacks the permission to modify the
workflow files that define the continuous integration, and branch
rulesets on \github\ blocked all updates of the main and backend
branches for the duration of the stretch. A scheduler woke the agent at
regular intervals to keep it going, and the agent kept a dated log of
what it did and why; this log, which runs to 60,000 lines for the
summer, is part of the repository that accompanies this paper.

In the two unsupervised weeks, the agent merged 74 PRs. JB was pleased
enough with the progress on his return that he let the unsupervised
procedure continue for a few more weeks, alongside his supervised work,
and 145 PRs in total went through the development branch before it was
merged into the backend branch on 2026 September 2, after JB had
reviewed it as a whole, with the help of a Codex review whose verdict,
that ``the work is coherent, the ledger is honest,'' matched JB's own.
The stretch made clear both what such an agent can and cannot do. It
burned the compiled-mode ledger down from 100 to 15 failing tests in
three days, found and fixed a performance problem in the coercion of
inputs that had cost 178\,$\mu$s per call, resolved the interaction
between \astropy\ units and compilation, and found two pre-existing bugs
in \galpy's \numpy\ code that it validated against a 30-digit reference
before filing them as issues. It also introduced regressions, all of
which it found itself within days, using the tools it had built for the
purpose, and fixed before JB returned; it misread a stale rule; it once
left conflict markers in a working tree, the type of error that a human
review would have spotted easily; and its summary table of PRs fell out
of date and had to be rebuilt from \github. The scorecard that it wrote
for itself judged the stretch against its goal honestly: the ledger
went from 392 to 140 entries against a target of zero. The autonomous
stretch succeeded in keeping the agent working productively while JB
was away, but it did so because the agent had a clear, limited, and
verifiable goal; work towards the broader aims of the project, which
requires judgment about what to build and how, would have been much
harder to leave unsupervised.

\subsection{Making the test suite pass under every backend}\label{sec:process-ledger}

\begin{figure*}[t!]
  \includegraphics[width=\textwidth,clip=]{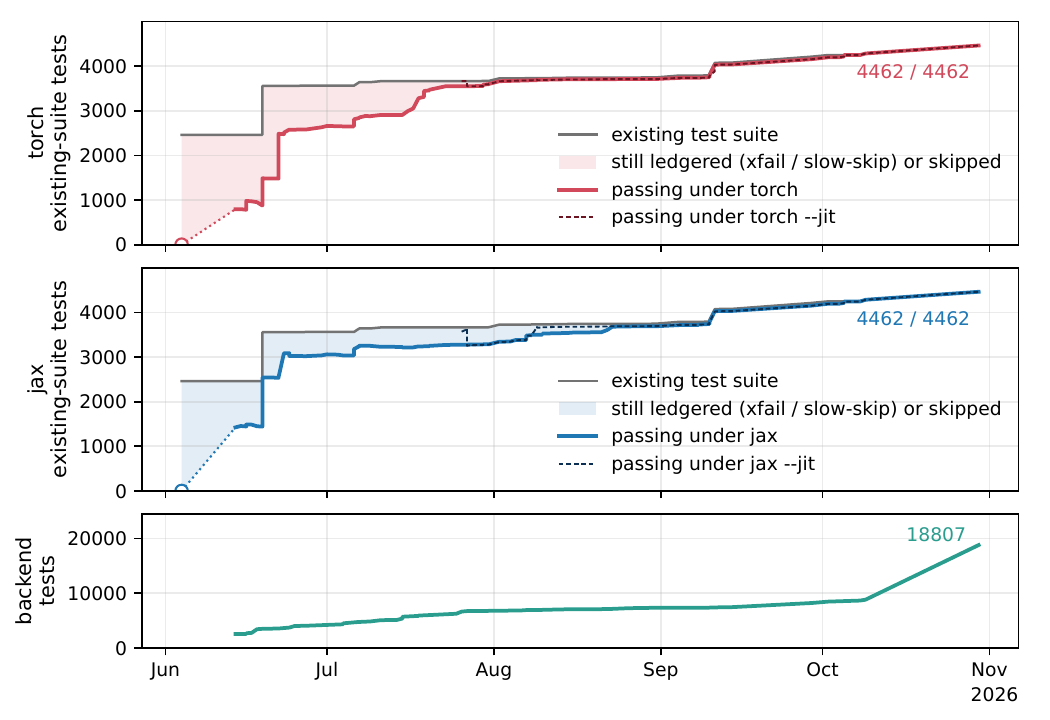}
  \caption{Tests of \galpy's existing test suite that pass under each
    backend, as a function of time, leaving aside the 27 tests of the
    out-of-scope torus-mapping code. The grey line is the size of the
    suite, which grows from 2,462 to 4,462 tests (most of the jump on
    June 19 comes from splitting the potential tests into one test per
    potential); the colored curves show the number of these tests that
    pass under \torch\ (top) and \jax\ (middle) in eager mode (solid)
    and under compilation (dashed), that is, the size of the suite minus
    the tests that are ledgered as expected failures or as too slow, or
    skipped, under that mode (the shaded band; a ledger entry that
    covers a whole file counts as the tests that it matched at the
    time). In August, \jax\ passes more tests under compilation than in
    eager mode because slow orbit tests that are skipped in eager mode
    run in reasonable time when compiled. The bottom panel shows the
    number of tests in the dedicated backend test modules; its jump in
    October is the two sweeps of
    \sectionname~\ref{sec:process-correctness}. At the end, all 4,462
    tests pass in every mode.}\label{fig:ledger}
\end{figure*}

One of the main acceptance tests for the backends was \galpy's own test
suite in the five modes of \sectionname~\ref{sec:design-tests}. Leaving
aside the 27 tests of the out-of-scope torus-mapping code (the 23 tests
of its own module and four tests of the stream DF that build the
stream's track with it), the suite grew from 2,462 to 4,462 tests over
the course of this work, mainly because we split tests that looped over
all of \galpy's potentials into one test per potential, such that each
failure could be traced to its potential. We did not delete or modify
tests that failed under a backend, but listed them in a \emph{ledger},
one line per test and mode, that marked them as expected failures, such
that the size of the ledger was the measure of what remained to be
done; a second ledger listed tests that passed but exceeded the time
limit of 300\,s per test in the continuous integration. We regenerated
the ledgers from the actual test results on demand, such that they
could not drift from reality, and the continuous integration posted a
report of each PR's results per file and mode, ledgered failures
included, to the PR, such that a reviewer saw at a glance what a PR
fixed or broke. \figurename~\ref{fig:ledger} displays how we brought
the suite to pass. On 2026 June 14, when the all-backend continuous
integration was first run, 1,042 tests failed under \jax\ and, a week
later, 2,080 under \torch; the difference between the two is almost
entirely due to \torch's rejection of \python\ scalars where \jax\
accepts them, and it disappeared when we centralized the coercion of
inputs (\sectionname~\ref{sec:design-agnostic}). The eager ledgers were
essentially empty by 2026 September 12, and we emptied the compiled-mode
ledgers, which started at 100 and 146 tests when we introduced those
modes, over the following month. At the time of writing, the ledgers
are empty in every mode.

\subsection{How much of the suite runs on the backend}\label{sec:process-islands}

A test that passes under a forced backend shows that the code returned
the right answer, not that the backend computed it: a code path that
converts its input to \numpy, computes with \numpy\ or \scipy, and
converts the result back passes every check of a value. In the first
months, such \emph{numpy islands} were common, because the natural way
to add a backend path to a function, a branch on whether its input is a
backend array, took the \numpy\ branch for the \numpy\ inputs of the
existing tests even under a forced backend; a first backend version of
the action--angle code passed its tests under \jax\ in this way and was
rejected in review. The forced mode of
\sectionname~\ref{sec:design-choosing} is meaningful because functions
branch on the resolved namespace rather than on the type of their
input, because objects that are constructed under a forced backend hold
their parameters as backend arrays, such that their methods compute on
the backend also when they are called from code that still uses
\numpy, and because on a GPU an implicit conversion of a device array
to \numpy\ raises an error rather than copying silently
(\sectionname~\ref{sec:design-gpu}).

To measure what remains, we instrumented a run of the existing suite
under forced \torch\ to record every place where \galpy\ hands backend
data to \numpy: implicit conversions by \numpy\ and \scipy,
explicit copies to the host, and regions in which \galpy\ deliberately
computes with \numpy.\footnote{We use \torch\ for this measurement,
  because \numpy\ reads a \jax\ array on the CPU through the buffer
  protocol, which bypasses the conversion method that the instrument
  intercepts.} We classified every such place by the reason for the
hand-off and every test by the least benign of its reasons
(\tablename~\ref{tab:islands}). Four reasons concern work that has to
happen on the host: \galpy's \cc\ extensions, for orbit integration and
the St\"{a}ckel actions, whose derivatives come from the state-transition
matrix and from custom derivative rules
(\sectionname~\ref{sec:orbit-paths}); decisions that need a value on the
host, such as a tolerance, a root bracket, the number of integration
steps, or a cache key; output to \astropy\ quantities or plots; and
\numpy\ functions that the test itself supplies, such as the density of
an expansion. Three are \numpy\ computations that could run on the
backend but have no reason to: constant tables built at construction
from parameters that are not differentiated, such as the spline of the
guiding radius or the coefficients of an expansion (a differentiated
parameter takes a backend path, as the sweep of
\sectionname~\ref{sec:process-correctness} verifies); sampling with
\numpy's global random-number generator, which the tests seed (a
backend random key runs the backend sampler); and the \python\
integrators of the \numpy\ code, which warn that their result is not
differentiable. The last is one documented exception that keeps
\numpy\ for \numpy\ inputs under a forced backend, while a backend input
or a differentiated parameter takes the backend path: a stream that the
tests build with \numpy\ before the backend is forced, whose lazily
computed track stays in \numpy\ like the stream itself; four tests
reach it.

\begin{table}
  \caption{Tests of the existing suite by how they hand backend data to
    \numpy\ under a forced \torch\ backend (4,353 tests that ran;
    cumulative).}\label{tab:islands}
  \centering
  \small
  \begin{tabular}{lrr}
    \hline
    hand-offs to \numpy\ at most & tests & share \\
    \hline
    none & 2,011 & $46.2\,\%$ \\
    + \galpy's \cc\ extensions & 2,321 & $53.3\,\%$ \\
    + decisions on the host & 3,153 & $72.4\,\%$ \\
    + output (units, plots) & 3,322 & $76.3\,\%$ \\
    + \numpy\ functions of the test & 3,389 & $77.9\,\%$ \\
    + constant tables & 3,949 & $90.7\,\%$ \\
    + \numpy's random numbers & 4,098 & $94.1\,\%$ \\
    + the \python\ integrators & 4,349 & $99.9\,\%$ \\
    + the documented exception (four tests) & 4,353 & $100.0\,\%$ \\
    \hline
  \end{tabular}
\end{table}

When we first ran the instrument, 220 tests ($5.3\,\%$) passed through a
genuine island, \numpy\ computation that could and should run on the
backend: the transformation of an integrated orbit in the isochrone
approximation (55 tests), the \numpy\ track tables of
\texttt{streamdf} (53), the profiles of the disk DFs (39), \scipy\ root
finding for orbit quantities such as the guiding radius and for the
Lindblad resonances (49), and smaller sites in about 25 other functions.
We moved all of these to the backend, in eight PRs, mostly with backend
code that already existed; the root finders, which now run on the
backend also for plain scalars under a forced backend, needed a
bracketing that evaluates the function at many points of the bracket at
once, because the 45 sequential evaluations of a bisection each cost a
full eager dispatch. The same instrumented run now finds no test that
passes through a genuine island (\tablename~\ref{tab:islands}):
$46\,\%$ of the tests never hand backend data to \numpy, $78\,\%$ do so
only for work that has to happen on the host, and the remaining
$22\,\%$ use \numpy\ only for constant tables, for sampling with
\numpy's random-number generator, in the \python\ integrators, or in
the documented exception.

\subsection{Fixes to \lowercase{\galpy}}\label{sec:process-fixes}

Running the same calculations through a second implementation, and
differentiating them, is an effective way to find bugs in the first, and
the backend work led to the 40 PRs against the main branch that fixed
\galpy's \numpy\ and \cc\ code (\sectionname~\ref{sec:process-organization}). The three-dimensional
variational equations, which require correct second derivatives of the
potential, found some of these: four potentials had incorrect
planar second derivatives in \cc, and the \cc\ second derivatives of
the basis-function potentials had the wrong sign in one component,
which the usual check that the determinant of the state-transition
matrix is one cannot catch. The comparison of the backend quadratures
with the existing ones found others: the plain Gauss--Legendre
quadrature of the action integrands in the adiabatic and St\"{a}ckel
approximations loses about four digits at the turning points, because
the integrands have square-root endpoints, and we regularized it; the
default absolute tolerance of \scipy's adaptive quadrature truncated
several small integrals (the outer tail of the Jeans velocity
dispersion, the surface density far from the plane); and the \cc\ and
\python\ implementations of the Chandrasekhar dynamical-friction force
interpolated the velocity dispersion with different splines, which
differed by a few parts in a thousand at small radii. The figures of
this paper, which we regenerate from scratch against the current code,
found yet others: the inverse-cumulative sampling of the
spherical DFs biased the sampled velocity dispersion low by about
$0.7\,\%$ at the default grid resolution, and the forces of the two-power
density profiles were off by up to $10^{-3}$ near the inner power-law
index of three. We fixed pre-existing defects of this kind in their
own PRs against the main branch, never quietly in the backend branch,
such that each is documented and their count is known.

\begin{figure*}[t!]
  \includegraphics[width=\textwidth,clip=]{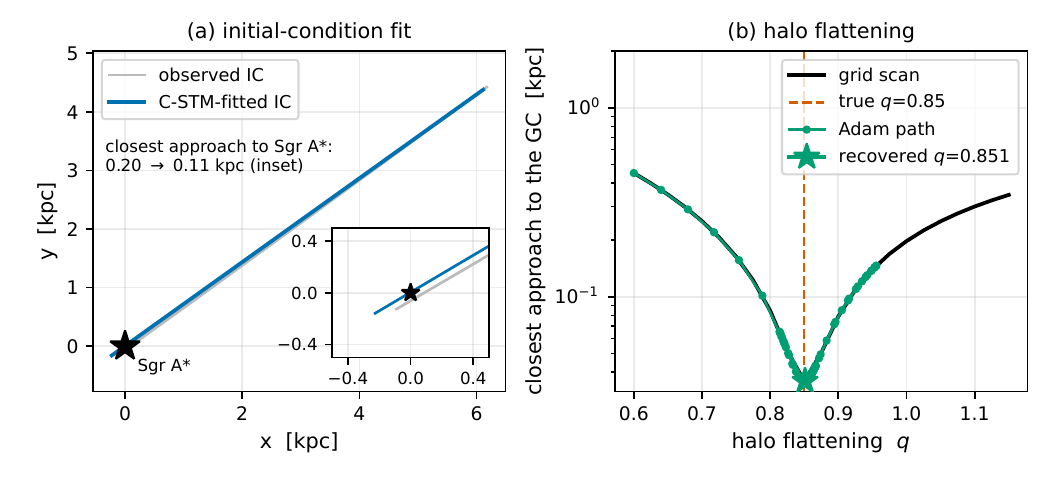}
  \caption{A mock hypervelocity star as a constraint on the Galactic potential.
    Left: the orbit of a star resembling S5-HVS1, integrated backwards
    from its observed phase-space position (grey) and from the position
    within the observational uncertainties that brings the orbit closest
    to Sgr~A$^*$ (blue), found by Gauss--Newton iterations on the
    derivative of the orbit with respect to its initial condition from
    the \cc\ variational integrator. Right: the closest approach to the
    Galactic center of the backward orbit as a function of the
    flattening $q$ of the halo (black, from a grid of integrations) and
    the path of a gradient-descent fit of $q$ using the derivative of the
    closest approach through the in-backend orbit integration (points),
    which recovers the value from which the star was generated (dashed
    line).}\label{fig:hvs}
\end{figure*}

\subsection{Establishing correctness}\label{sec:process-correctness}

JB does not take the code to be correct because an agent reports that
its tests pass; the reasons to believe that it is correct are the
following, roughly in order of weight. The \numpy\ path is byte-identical to the previous
version, measured by checksums of tens of thousands of output values per
PR, such that the entire body of validation that \galpy\ has accumulated
over more than a decade carries over. The backend paths pass the same
tests at the same tolerances, in five modes
(\sectionname~\ref{sec:process-ledger}). We compared every derivative
that the code can compute with a finite difference whose convergence
with step size is demonstrated rather than assumed, or with an
independent reference where one exists: the derivatives of orbits with
respect to their initial conditions, for example, are computed by the
variational equations in \cc\ and by reverse-mode automatic
differentiation through two different ODE solvers, and the three agree
with each other and with finite differences of independently integrated
orbits to a few parts in $10^{10}$ (\figurename~\ref{fig:icjacobian}).
That no derivative was missed is checked by a sweep: for every public
potential, action--angle, and distribution-function class and the
stream-impulse functions, it differentiates each numeric parameter and
one input coordinate, 761 derivatives in 199 configurations, under
compiled \jax\ and eager \torch, with the inputs given as backend
arrays, as \python\ floats, and under a forced backend, and compares
them with converged finite differences of the \numpy\ code. When we
first ran it, about 60 combinations of class and parameter raised an
error or worked only eagerly, and nine families of derivatives were
silently wrong, zero, or disconnected from the parameter; for example,
\torch\ orbit integration with \galpy's \cc\ integrators dropped the
derivatives with respect to the potential's parameters, which made those
of a stream model wrong by up to a factor of 45. We fixed all of them,
and the sweep, 2,791 cases, now runs in the continuous integration; the
one remaining exception is by design: under compilation, the anisotropy
of a constant-anisotropy DF cannot be traced, because it selects the
order of a derivative in the inversion. A second sweep gives each
parameter and input coordinate in turn as a concrete backend array that
is not differentiated, and checks that every output comes back on that
backend with the \numpy\ values (4,249 cases); when we first ran it,
about 1,200 cases returned \numpy\ arrays, raised an error, or returned
an array of the wrong shape, and all of them now pass. Where the
backend code implements a numerical method of its own---a special
function, a quadrature, a root finder---we checked it against \scipy\
and against \texttt{mpmath} at 30 to 50 digits and validated the
reference itself before use, because the analytic formula one reaches
for first is sometimes the one that is wrong. JB, a second AI system,
and adversarial agents reviewed the code; running it on two GPUs, which
the continuous integration cannot do, found the device-placement bugs
that are invisible on a CPU; and the figures of this paper, each of
which is produced by a script that is re-run against the current code
and whose output is checked against a stored hash, found several of the
bugs listed above. None of these checks is sufficient on its own, and
JB would not have accepted the code on the basis of any one of them,
only on their combination together with his own review of the code
itself.

%% file: examples.tex
\section{Examples}\label{sec:examples}

While the sections above illustrate each part of the code on its own,
here we give two examples that combine several of them in a single
analysis. Both are deliberately small:
they use mock data and run in minutes, and their point is to illustrate
the computation rather than the result.

\subsection{A hypervelocity star as a probe of the potential}\label{sec:examples-hvs}

A hypervelocity star ejected from the Galactic center must, when its
orbit is integrated backwards, return to the Galactic center, and the
closest approach of the backward orbit to the center is therefore a
function of the Galactic potential that a measurement of the star's
phase-space position constrains \citep[see, \eg,][]{Gnedin2005,Contigiani2019}.
This is a natural problem for differentiable orbits: the objective is
a function of an integrated orbit, and both the star's initial
condition and the potential's parameters are to be varied.
\figurename~\ref{fig:hvs} displays the two derivatives of
\sectionname~\ref{sec:orbit-paths} at work on a mock star resembling
S5-HVS1 \citep{Koposov2020} in a cored, flattened logarithmic halo. In the left panel, we
hold the potential fixed and adjust the star's phase-space position
within its observational uncertainties to minimize the closest approach,
using the derivative of the orbit with respect to its initial condition
from the \cc\ variational integrator; seven Gauss--Newton iterations
bring the closest approach from $0.2$ to $0.1$\,kpc. In the right
panel, we hold the star fixed and fit the flattening $q$ of the halo by
descending the closest approach with the derivative obtained by
reverse-mode differentiation through the in-backend orbit integration,
which the \cc\ path cannot provide; the recovered $q=0.851$ matches the
true $0.85$ from which the mock star was generated. In code, the first fit uses the derivative of the backward orbit with
respect to the star's observed coordinates,
\begin{code}
def closest_approach_ic(vxvv):
    # vxvv: (ra,dec,d,pmra,pmdec,vlos)
    o= Orbit(vxvv,radec=True,ro=8.,vo=220.)
    o.integrate(-ts,pot,method='dop853_c') # C + STM
    return jnp.min(o.r(-ts))
J= jax.jacrev(closest_approach_ic)(
    jnp.asarray(vxvv_obs))
\end{code}
which a Gauss--Newton step then combines with the covariance of the
observations; the second fit is
\begin{code}
def closest_approach(q):
    pot= LogarithmicHaloPotential(normalize=1.,
                                  core=0.2,q=q)
    o= Orbit(jnp.asarray(vxvv),radec=True,ro=8.,vo=220.)
    o.integrate(-ts,pot,method='diffrax')
    return jnp.min(o.r(-ts))
objective= jax.jit(jax.value_and_grad(closest_approach))
q= adam_descent(objective,q0=1.1) # a few lines of optax
\end{code}
With real
data, one would combine the two fits into one likelihood over the star's
parameters and those of the potential, and the derivative of that
likelihood with respect to all of them would be available for a
gradient-based sampler.

\begin{figure*}[t!]
  \includegraphics[width=\textwidth,clip=]{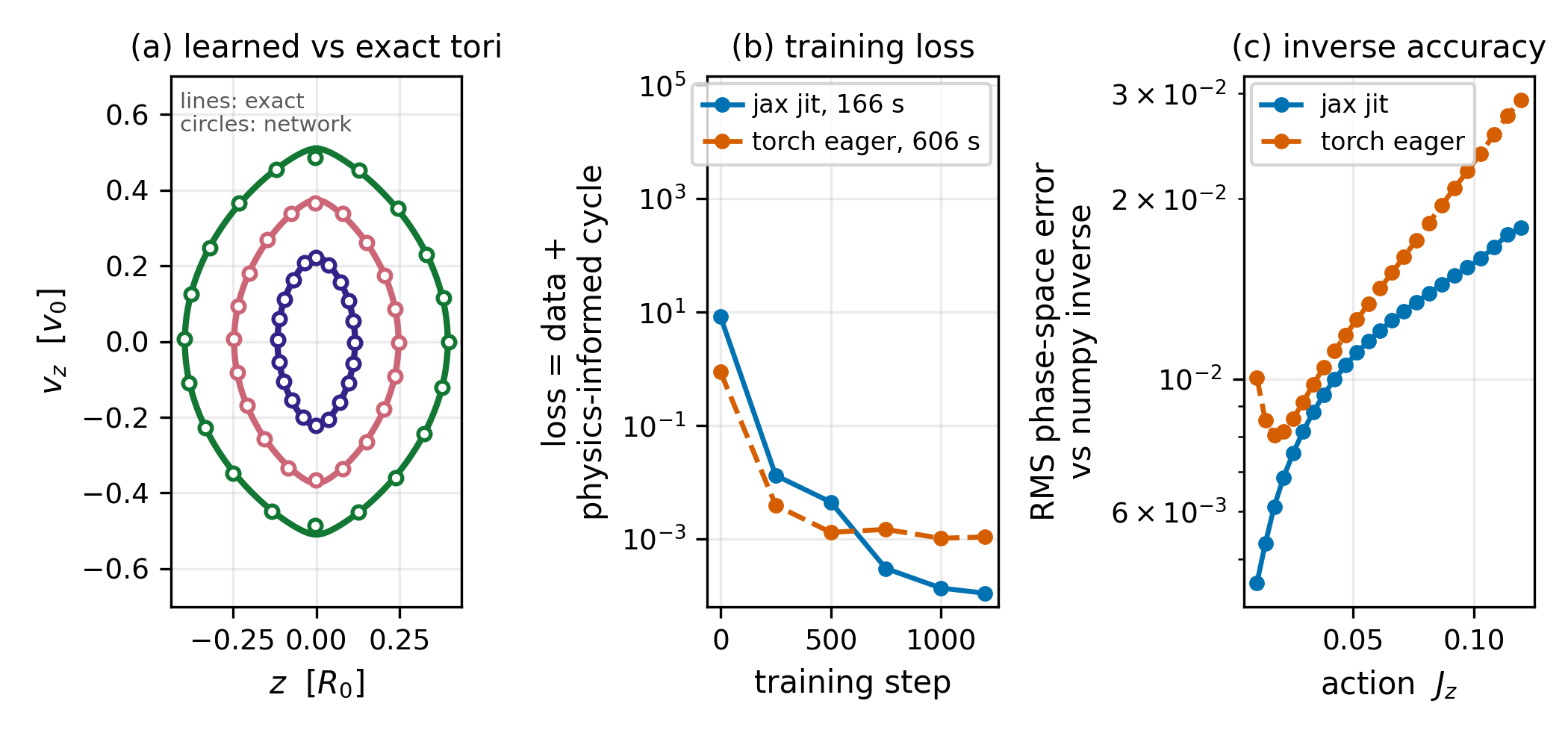}
  \caption{A neural network learns the inverse action--angle
    transformation of the vertical motion in \texttt{MWPotential2014} at
    the solar radius by differentiating through \galpy's forward
    transformation, in \jax\ and in \torch. Left: the exact tori at three
    vertical actions (lines) and the network's output for the same
    actions and a grid of angles (circles). Middle: the training loss,
    the sum of the supervised and the cycle-consistency terms, in
    the two frameworks, with the training time in the legend: \jax\
    (\texttt{flax} and \texttt{optax}) with a compiled training step and
    \torch\ (\texttt{torch.nn}) eagerly, to show both options. Right: the
    root-mean-square phase-space error of
    the network against the exact inverse as a function of the
    action. The network reproduces the exact tori in both
    frameworks.}\label{fig:neuralvertical}
\end{figure*}

\subsection{Learning the inverse action--angle transformation}\label{sec:examples-neural}

The transformation from actions and angles to positions and velocities
has no closed form in a general potential, and \galpy's implementations
of it are currently limited to one-dimensional potentials and to the
torus machinery of \citet{BinneyMcMillan2016}. A differentiable forward transformation makes a different
approach possible: a neural network $G(\vecj,\veca)$ that takes actions
and angles to phase-space positions can be trained by requiring that
\galpy's forward transformation, applied to the network's output,
return the actions and angles that went in. The loss then contains a
cycle-consistency term, $|F(G(\vecj,\veca))-(\vecj,\veca)|^2$, with $F$ the
\texttt{actionAngle} method, whose gradient with respect to
the network's weights flows through $F$. In both demonstrations below,
we add to it a supervised term, the mean-square phase-space error on
training pairs obtained by applying $F$ to phase-space points sampled
in a box around the origin, which keeps the network on bound orbits
early in training (in the second demonstration, the weight of the cycle
term is ramped up from zero over the first 400 steps). Because the
pairs come from the forward transformation, no reference inverse is
needed anywhere in the training. We show this for the vertical motion in
\texttt{MWPotential2014} at the solar radius, where the exact inverse of \sectionname~\ref{sec:aa-other} exists and
serves as a check, but is not used in the training (\figurename~\ref{fig:neuralvertical}), and for the
motion in the meridional plane with the St\"{a}ckel approximation, where
no reference exists (\figurename~\ref{fig:neural}). In the first case a
small network trained in a few minutes, identically under \jax\ and
\torch, reproduces the exact tori to a phase-space error of $0.5$ to
$2\,\%$. In the second, we keep the problem small by fixing the angular
momentum at that of a circular orbit at the solar radius, $L_z = 1$ in
\galpy's natural units, such that the network maps the two actions $J_R$
and $J_z$ and the two meridional angles to $(R,z,v_R,v_z)$; a four-layer
network trained for half an hour on a GPU returns actions on the round
trip with a scatter of $7\times10^{-3}$ and reconstructs meridional
orbits that are not in its training set. A network for the full
three-dimensional transformation would take $L_z$ and the azimuthal
angle as further inputs. The cycle term of the training step is a few lines, whatever the framework,
\begin{code}
def cycle_loss(params,J,theta):
    z,vz= net.apply(params,J,theta) # the network
    Jp,_,thetap= aAV.actionsFreqsAngles(z,vz) # galpy
    return (jnp.mean((Jp-J)**2/J**2)
            +jnp.mean(wrap(thetap-theta)**2))
grads= jax.jit(jax.grad(cycle_loss))(params,J,theta)
\end{code}
where \texttt{net} is a small multi-layer perceptron and \texttt{wrap}
reduces an angle difference to $[-\pi,\pi)$; the supervised term is an
ordinary mean-square error on the network's output. Neither
network is a replacement for a proper inverse, and the point is not the
accuracy of these particular fits but that \galpy's action--angle code
sits inside a training loop, as a layer whose gradient the framework
computes, and that the same code does so in both frameworks.

%% file: conclusion.tex
\section{Conclusion}\label{sec:conclusion}

We have described an extension of \galpy\ that makes it run, compile, and
differentiate under \jax\ and \torch, on CPUs and GPUs, while leaving
its existing \numpy\ and \cc\ implementation byte-for-byte unchanged,
up to the handful of deliberate exceptions of
\sectionname~\ref{sec:design-principles}, and its interface as it was. Potentials, orbit integration, the
action--angle transformations and their inverses, and the distribution
functions, including those of tidal streams, accept the arrays of either
framework and return derivatives with respect to their inputs and
parameters. The existing test suite of \galpy, now 4,462 tests apart
from the 27 of the out-of-scope torus-mapping code, passes under \numpy,
under both frameworks in eager mode, and under both with compilation,
and the dedicated tests of the backend code add 18,807 more. We tested
derivatives against finite differences whose convergence is
demonstrated, against the \cc\ variational equations, and against each
other across the two frameworks, and where \galpy\ implements a
numerical method of its own, we tested it against \scipy\ and against
arbitrary-precision references. The work also found defects in the
\numpy\ and \cc\ code that had been there before, which 40 PRs to the
main branch fixed.

\begin{figure*}[t!]
  \includegraphics[width=\textwidth,clip=]{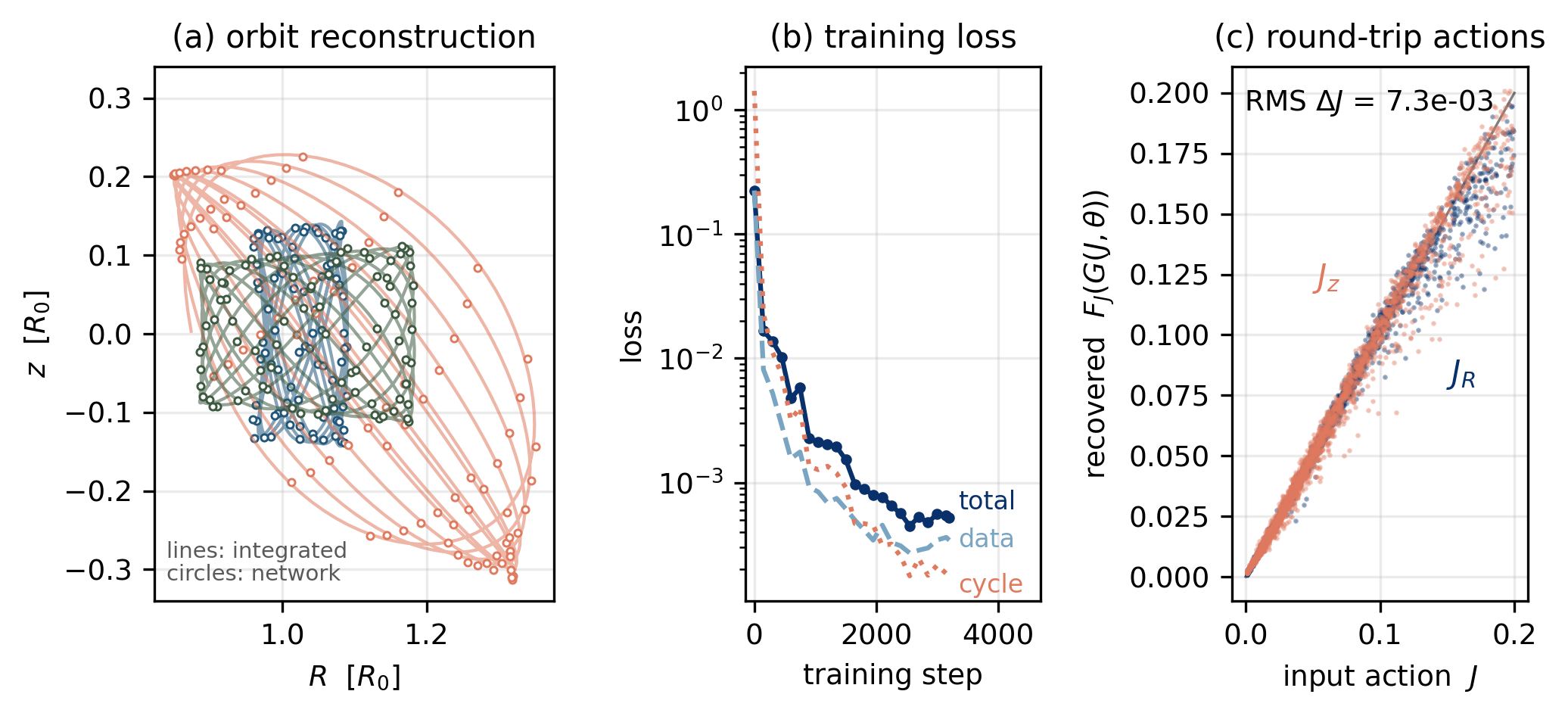}
  \caption{Neural inverse of the St\"{a}ckel action--angle transformation
    in \texttt{MWPotential2014} at fixed $L_z$, learned by differentiating
    through \galpy's forward transformation.
    Left: three orbits integrated in the potential (lines) and the
    positions returned by the network for the orbits' actions and
    angles (circles). Middle: the training loss, with its supervised
    (``data'') and cycle-consistency terms. Right: the actions recovered by applying
    the forward transformation to the network's output against the
    input actions for a test set. The network recovers the input actions
    on the round trip and reconstructs orbits that are not in its
    training set.}\label{fig:neural}
\end{figure*}

What this makes possible is the gradient-based fitting of dynamical
models that \galpy\ already contains, with a likelihood whose derivative
with respect to every parameter---of the potential, of the DF, of the
orbits of the objects being modeled---costs a small multiple of the
likelihood itself, and the use of these models as components of larger
differentiable programs, such as neural networks that represent a
potential or a DF, or a simulation whose output is compared with data.
The speed-ups that a GPU provides over a single CPU core, of one to
five orders of magnitude for large batches, are a second benefit, and one that depends on
keeping the computation on the device by compiling it as a whole.

The GPU speed-ups are not uniform. Evaluating forces and drawing samples
are limited by the memory bandwidth of the GPU and come close to it, but
the in-backend orbit integration reaches only a few percent of the
H100's double-precision peak: the batched array programs that \jax\
and \torch\ generate step all orbits in lockstep, with a loop overhead
per step and branch-heavy special functions, which maps poorly onto a
GPU. GPU-native kernels that compile \galpy's existing \cc\ potentials
and integrators as CUDA device functions, with one thread per orbit,
are a natural complement. A prototype for \texttt{MWPotential2014}
agrees with \galpy's \cc\ integrators to within their truncation error
and, on the H100, integrates $10^6$ disc orbits for 2\,Gyr with
\texttt{dop853} in 2.4\,$\mu$s per orbit with \galpy's arithmetic and in
0.9\,$\mu$s with GPU-tuned special functions, compared to 10\,$\mu$s for
compiled \jax\ on the same GPU. Adaptive high-order integrators are
limited by their per-thread state, and a fixed-step sixth-order
symplectic integrator reaches the same accuracy another 3 to 3.7 times
faster when, as here, the orbits share a dynamical timescale; adaptive
stepping remains necessary when they do not. This speed came from
restructuring the computation rather than from a direct translation,
porting all of \galpy's potentials would be a substantial effort, and
such kernels are not differentiable unless their variational or
adjoint equations are written by hand as well. They would therefore
complement the differentiable backends rather than replace them.

Some things are not done: the torus-mapping code, a wrapper of the
external code of \citet{BinneyMcMillan2016}, is not covered and will not
be, \astropy\ units cannot be used inside compiled programs, the
interpolated velocity sampler of the quasi-isothermal disk DF runs on
the backend but is a rejection sampler and therefore not differentiable,
and the variational
equations and Lyapunov exponents are features of the \cc\
integrators. Derivatives through the \cc\ integrators and the \cc\
action--angle methods are with respect to the initial conditions only
and of first order. A backend is also a moving target: the special-function fallbacks exist
because the frameworks' own versions are incomplete today, and the test
that checks this will say when they are no longer needed. It already
did once: \torch~2.15, released while we revised this paper, made its
modified Bessel functions of the second kind differentiable, the test
failed on the day of the release, and \galpy\ now uses \torch's own
functions from that version on.

The code described here is the start of \galpy's version~2 series,
which for now is distributed as pre-releases, starting with \flag{v2.0.0b1}, that are
installed with \texttt{pip install --pre galpy[jax]} or
\texttt{galpy[torch]}; a plain \texttt{pip install galpy} remains at
version~1.x; the documentation of version~2, which describes
the backends, is at
\\ \centerline{\url{https://docs.galpy.org/en/v2/}~.} All code used to produce
the figures in this paper, together with the working notes and logs of
the development, is available at
\\ \centerline{\url{https://github.com/jobovy/galpy-backends}~.}
\begin{figure*}[t!]
  \includegraphics[width=\textwidth,clip=]{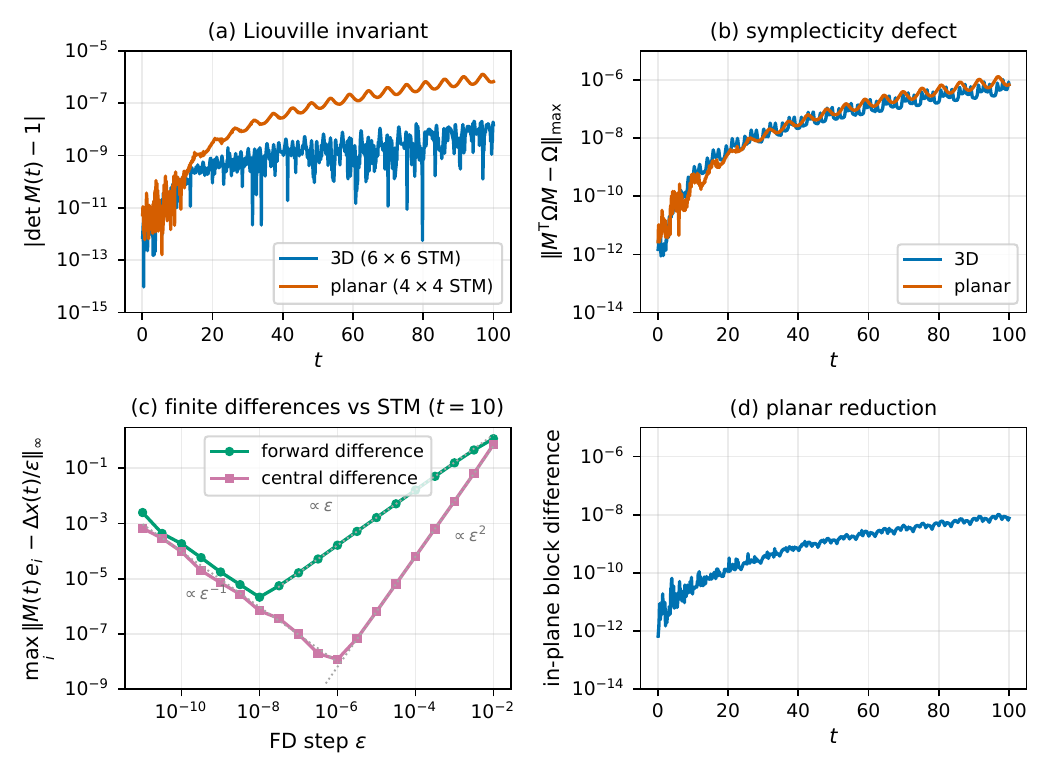}
  \caption{Checks of the \cc\ variational equations for an orbit in
    \texttt{MWPotential2014}, integrated with \texttt{dop853\_c} at
    tolerances of $10^{-12}$. (a) The Liouville defect $|\det\vec{M}-1|$
    of the state-transition matrix $\vec{M}$
    and (b) the symplecticity defect for the three-dimensional and the
    planar state-transition matrix. (c) The difference between the
    columns of $\vec{M}$ at $t=10$ and forward and central finite
    differences of independently integrated orbits as a function of the
    finite-difference step; the variational result agrees with the best
    central difference at its floor. (d) The in-plane block of the
    three-dimensional matrix for a planar initial condition agrees with
    the planar matrix, with no leakage out of the plane.}\label{fig:variational}
\end{figure*}

%% file: acknowledgments.tex
\begin{acknowledgments}
It is a pleasure to thank Anthropic for providing a Claude Max
subscription free of charge for the first part of this work, GitHub
for covering the cost of the continuous integration through its
Education program, and NVIDIA for the TITAN Xp GPU provided through its
Academic Grant Program. We thank the contributors to \galpy's code,
listed here in order of their number of contributions: Aladdin Seaifan,
Ted Mackereth, James Lane, Mathew Bub, Nathaniel Starkman, Henry Leung,
Wilma Trick, Samuel Wong, Jack Hong, Semyeong Oh, Rok Ro\v{s}kar,
Turner Garrow, Jeremy Webb, Jason Sanders, Marius Cautun, Anna Juranova, Morgan
Bennett, James Grimmett, Anya Dovgal, Gus Beane, Andrew Pontzen,
Anthony Mercer, Mark Fardal, John Weatherall, Yingtian Chen, Hood Chatham, Mike
German, Rhitik Bhatt, Junyang Lu, Gabriel Pfaffman, Yansong Qian, Jeffrey
Simpson, Adrian Szpilfidel, Jovan Veljanoski, Carlos Vera-Ciro, and
Jenny Wagner. We also thank the many users of \galpy\ whose
bug reports and questions over more than a decade shaped the code that is
extended here. JB acknowledges financial support from NSERC (funding
reference number \mbox{RGPIN-2026-04837}).
This work was made possible by the open-source codes \numpy\
\citep{Harris2020}, \scipy\ \citep{Virtanen2020}, \matplotlib\
\citep{Hunter2007}, \astropy\ \citep{Astropy2022}, \jax\
\citep{Bradbury2018}, \torch\ \citep{Paszke2019}, \diffrax\ and
\texttt{equinox} \citep{Kidger2021,KidgerGarcia2021}, \texttt{torchdiffeq}
\citep{Chen2018}, \torchode\ \citep{LienenGunnemann2022}, and the GNU Scientific
Library \citep{Galassi2009}. Computations on the H100 GPU were performed
on the Trillium supercomputer at the SciNet HPC Consortium
\citep{Loken2010,Ponce2019}. SciNet is funded by Innovation, Science and
Economic Development Canada; the Digital Research Alliance of Canada; the
Ontario Research Fund: Research Excellence; and the University of Toronto.
This research was enabled in part by support provided by Compute Ontario
(\url{https://www.computeontario.ca}) and the Digital Research Alliance of
Canada (\url{https://alliancecan.ca}).
Version-control, continuous-integration, coverage, and documentation
services are provided by GitHub, Codecov, and Read the Docs and are
gratefully acknowledged.
\end{acknowledgments}

\section*{Statement on the use of AI}

This paper has a single human author, JB, but is written in the first
person plural, because the work that it describes was a collaboration
between JB and Claude, an AI coding agent developed by Anthropic and
used through the \texttt{Claude Code} environment. ``We'' refers to that
collaboration; where the distinction matters, as in
\sectionname~\ref{sec:process}, JB and the agent are named separately.
The agent wrote the code described in this paper under JB's direction
and review between 2026 June 1 and October 30; successive
versions of the Claude models were used over this period, and each
commit records the model that produced it. JB set the goals, the
constraints, and the acceptance criteria, reviewed the pull requests,
directed the design where the agent's choices were not what he wanted,
and merged the code; the agent wrote the code, the tests, the
documentation, and the working notes, and ran the measurements. For
about two weeks in the summer the agent worked without supervision, as
described in \sectionname~\ref{sec:process-autonomous}, and JB reviewed
its work from that period afterwards. JB also used OpenAI's Codex to
review the agent's code at several points. The agent produced the figures
of this paper and the analysis behind them from scripts that are part of
the accompanying repository. The text of this paper was
drafted by the agent from the working notes of the project and from JB's
instructions and was then edited by JB, who is responsible for its
content.

%% file: appendix_dxdv.tex
\section{Variational equations and the state-transition matrix}\label{app:dxdv}

This appendix collects the equations behind \texttt{Orbit.integrate\_dxdv},
\texttt{Orbit.lyapunov}, and the derivative rule of the \cc\ integrators
of \sectionname~\ref{sec:orbit-paths}. None of it is specific to a
backend; it is implemented in \cc\ and in \numpy.

\begin{figure*}[t!]
  \includegraphics[width=\textwidth,clip=]{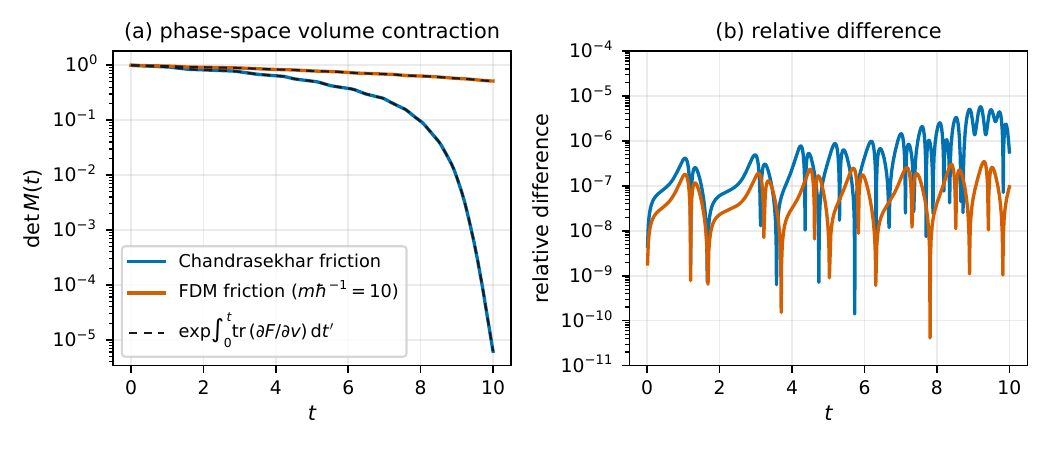}
  \caption{Phase-space volume under dissipative forces. Left:
    $\det\vec{M}(t)$ for an orbit in \texttt{MWPotential2014} with
    Chandrasekhar dynamical friction and with fuzzy-dark-matter
    friction \citep{Szpilfidel2026}, from the \cc\ variational equations (solid) and from the
    integral of the velocity divergence of the force along the orbit
    (dashed). Right: the relative difference between the two. The
    variational equations reproduce the contraction of phase-space
    volume by dissipative forces.}\label{fig:dissipative}
\end{figure*}

Write the phase-space coordinates as $\vecw = (\vecx,\vecv)$ in
rectangular Galactocentric coordinates, in which \galpy's \cc\
integrators work, and the equations of motion as $\dot{\vecw} =
\vec{f}(\vecw,t)$ with $\dot{\vecx} = \vecv$ and $\dot{\vecv} =
\vec{a}(\vecx,\vecv,t)$. A small deviation $\delta\vecw$ from an orbit
evolves according to the linearized equations
\begin{equation}\label{eq:variational}
  \frac{\dd\,\delta\vecw}{\dd t} = \vec{J}(t)\,\delta\vecw\,,\qquad
  \vec{J} = \begin{pmatrix} \vec{0} & \vec{I} \\
    \partial\vec{a}/\partial\vecx & \partial\vec{a}/\partial\vecv\end{pmatrix}\,,
\end{equation}
evaluated along the orbit. For a conservative force,
$\partial\vec{a}/\partial\vecv = \vec{0}$ and $\vec{K} \equiv
\partial\vec{a}/\partial\vecx = -\nabla\nabla\Phi$ is the symmetric
Hessian of the potential. \galpy's potentials provide their second
derivatives in cylindrical coordinates, and $\vec{K}$ is assembled from
them: with $\cos\phi = x/R$ and $\sin\phi = y/R$, the in-plane block is
the one of the planar variational equations that \galpy\ has had since
2011, built from $\partial^2\Phi/\partial R^2$,
$\partial^2\Phi/\partial\phi^2$, and $\partial^2\Phi/\partial R\partial\phi$,
and the vertical couplings are
\begin{align}
  K_{xz} &= -\cos\phi\,\frac{\partial^2\Phi}{\partial R\partial z}
            +\frac{\sin\phi}{R}\,\frac{\partial^2\Phi}{\partial z\partial\phi}\,,\nonumber\\
  K_{yz} &= -\sin\phi\,\frac{\partial^2\Phi}{\partial R\partial z}
            -\frac{\cos\phi}{R}\,\frac{\partial^2\Phi}{\partial z\partial\phi}\,,\\
  K_{zz} &= -\frac{\partial^2\Phi}{\partial z^2}\,.\nonumber
\end{align}
We added the mixed derivatives $\partial^2\Phi/\partial R\partial z$ and
$\partial^2\Phi/\partial z\partial\phi$ to the \cc\ interface of every
potential for this purpose, and the test suite checks the symmetry of
$\vec{K}$. For a velocity-dependent force
$\vec{F}(\vecx,\vecv)$, such as dynamical friction or the Coriolis force
in a rotating frame, the lower-right block $\partial\vec{F}/\partial\vecv$
is non-zero and the lower-left block need not be symmetric; both are
implemented for all three of \galpy's velocity-dependent forces: the
Chandrasekhar dynamical friction, the dynamical friction in fuzzy dark
matter \citep{Szpilfidel2026}, and the fictitious forces of a
non-inertial frame.

The state-transition matrix $\vec{M}(t) = \partial\vecw(t)/\partial\vecw(0)$
is the solution of \eqnname~(\ref{eq:variational}) for six independent
deviation vectors, $\dot{\vec{M}} = \vec{J}\vec{M}$ with
$\vec{M}(0) = \vec{I}$. The Runge--Kutta integrators integrate the orbit
and all six columns as one 42-component state, such that the deviations
see the same steps as the orbit; the symplectic integrators apply to the
deviations the drift ($\delta\vecx \leftarrow \delta\vecx +
h\,\delta\vecv$) and kick ($\delta\vecv \leftarrow \delta\vecv +
h\,\vec{K}\,\delta\vecx$) sub-steps with the same coefficients as the
orbit, which is the tangent map of the integrator and is itself
symplectic. For a conservative force, $\vec{M}$ is symplectic,
$\vec{M}^T\boldsymbol{\Omega}\vec{M} = \boldsymbol{\Omega}$ with
$\boldsymbol{\Omega}$ the standard symplectic form, and in particular
$\det\vec{M} = 1$, which is Liouville's theorem; for a dissipative force,
$\det\vec{M}(t) = \exp\int_0^t \mathrm{tr}\,(\partial\vec{F}/\partial\vecv)\,\dd t'$,
which measures the contraction of phase-space volume. These identities
are necessary but not sufficient checks---a sign error in one block of
$\vec{K}$ can leave the determinant equal to one---and the test that
settles the correctness of $\vec{M}$ is the comparison of its columns
with finite differences of independently integrated orbits, in which the
finite-difference error follows its characteristic V shape with step
size and the variational result sits at the bottom of the V.
\figurename~\ref{fig:variational} shows these checks for a Milky-Way
orbit: the Liouville and symplecticity defects stay below $10^{-6}$ over
a hundred time units, and the state-transition matrix agrees with the
best central difference at the finite-difference floor of $10^{-8}$.
\figurename~\ref{fig:dissipative} shows the dissipative case: for an
orbit that sinks under Chandrasekhar dynamical friction, the phase-space
volume contracts by five orders of magnitude over ten time units, and
$\det\vec{M}$ follows the integral of the velocity divergence of the
force to a few parts in $10^{6}$; the same orbit under the fuzzy-dark-matter
friction of \galpy\ \citep{Szpilfidel2026} contracts far less, because
that force is suppressed at the orbit's velocities.

The derivative rule of \sectionname~\ref{sec:orbit-paths} follows from
the chain rule: for a scalar function $L$ of the orbit sampled at times
$t_k$, $\partial L/\partial\vecw(0) = \sum_k \vec{M}(t_k)^T\,
\partial L/\partial\vecw(t_k)$, which is a single matrix--vector product
per output time once $\vec{M}$ is known. Lyapunov exponents are
computed by integrating a deviation vector with
\eqnname~(\ref{eq:variational}), renormalizing it to unit length every
fixed number of steps, and accumulating the logarithms of the
normalization factors \citep{Benettin1980a,Benettin1980b}; the spectrum follows by
orthonormalizing a set of deviation vectors in the same way.

%% file: ms.bbl
\begin{thebibliography}{}
\expandafter\ifx\csname natexlab\endcsname\relax\def\natexlab#1{#1}\fi
\providecommand{\url}[1]{\href{#1}{#1}}
\providecommand{\dodoi}[1]{doi:~\href{http://doi.org/#1}{\nolinkurl{#1}}}
\providecommand{\doeprint}[1]{\href{http://ascl.net/#1}{\nolinkurl{http://ascl.net/#1}}}
\providecommand{\doarXiv}[1]{\href{https://arxiv.org/abs/#1}{\nolinkurl{https://arxiv.org/abs/#1}}}

\bibitem[{J. An {et~al.}(2021)An, Naik, Evans, \& Burrage}]{An2021}
An, J., Naik, A.~P., Evans, N.~W., \& Burrage, C. 2021,
  \bibinfo{title}{{Charting galactic accelerations: when and how to extract a
  unique potential from the distribution function},} \mnras, 506, 5721,
  \dodoi{10.1093/mnras/stab2049}

\bibitem[{N. Anau~Montel \& F. Schmidt(2026)Anau~Montel \&
  Schmidt}]{AnauMontel2026}
Anau~Montel, N., \& Schmidt, F. 2026, \bibinfo{title}{{A differentiable forward
  model for weakly perturbed stellar streams: substructure forecasts from
  density and kinematics spectra},} arXiv e-prints, arXiv:2606.09629.
\newblock \doarXiv{2606.09629}

\bibitem[{N. Banik {et~al.}(2021)Banik, Bovy, Bertone, Erkal, \&
  deBoer}]{Banik2021}
Banik, N., Bovy, J., Bertone, G., Erkal, D., \& deBoer, T.~J.~L. 2021,
  \bibinfo{title}{{Evidence of a population of dark subhaloes from {G}aia and
  {P}an-{STARRS} observations of the {GD}-1 stream},} \mnras, 502, 2364,
  \dodoi{10.1093/mnras/stab210}

\bibitem[{A.~G. Baydin {et~al.}(2018)Baydin, Pearlmutter, Radul, \&
  Siskind}]{Baydin2018}
Baydin, A.~G., Pearlmutter, B.~A., Radul, A.~A., \& Siskind, J.~M. 2018,
  \bibinfo{title}{{Automatic Differentiation in Machine Learning: a Survey},}
  Journal of Machine Learning Research, 18, 1.
\newblock \doarXiv{1502.05767}

\bibitem[{V. Belokurov {et~al.}(2020)Belokurov, Sanders, Fattahi, Smith,
  Deason, Evans, \& Grand}]{Belokurov2020}
Belokurov, V., Sanders, J.~L., Fattahi, A., {et~al.} 2020, \bibinfo{title}{{The
  biggest splash},} \mnras, 494, 3880, \dodoi{10.1093/mnras/staa876}

\bibitem[{G. Benettin {et~al.}(1980{\natexlab{a}})Benettin, Galgani, Giorgilli,
  \& Strelcyn}]{Benettin1980a}
Benettin, G., Galgani, L., Giorgilli, A., \& Strelcyn, J.-M.
  1980{\natexlab{a}}, \bibinfo{title}{{Lyapunov Characteristic Exponents for
  smooth dynamical systems and for {H}amiltonian systems; a method for
  computing all of them. {P}art 1: Theory},} Meccanica, 15, 9,
  \dodoi{10.1007/BF02128236}

\bibitem[{G. Benettin {et~al.}(1980{\natexlab{b}})Benettin, Galgani, Giorgilli,
  \& Strelcyn}]{Benettin1980b}
Benettin, G., Galgani, L., Giorgilli, A., \& Strelcyn, J.-M.
  1980{\natexlab{b}}, \bibinfo{title}{{Lyapunov Characteristic Exponents for
  smooth dynamical systems and for {H}amiltonian systems; a method for
  computing all of them. {P}art 2: Numerical application},} Meccanica, 15, 21,
  \dodoi{10.1007/BF02128237}

\bibitem[{J. Binney(2010)Binney}]{Binney2010}
Binney, J. 2010, \bibinfo{title}{{Distribution functions for the {M}ilky
  {W}ay},} \mnras, 401, 2318, \dodoi{10.1111/j.1365-2966.2009.15845.x}

\bibitem[{J. Binney(2012)Binney}]{Binney2012}
Binney, J. 2012, \bibinfo{title}{{Actions for axisymmetric potentials},}
  \mnras, 426, 1324, \dodoi{10.1111/j.1365-2966.2012.21757.x}

\bibitem[{J. Binney \& P. McMillan(2011)Binney \&
  McMillan}]{BinneyMcMillan2011}
Binney, J., \& McMillan, P. 2011, \bibinfo{title}{{Models of our {G}alaxy --
  {II}},} \mnras, 413, 1889, \dodoi{10.1111/j.1365-2966.2011.18268.x}

\bibitem[{J. Binney \& P.~J. McMillan(2016)Binney \&
  McMillan}]{BinneyMcMillan2016}
Binney, J., \& McMillan, P.~J. 2016, \bibinfo{title}{{Torus mapper: a code for
  dynamical models of galaxies},} \mnras, 456, 1982,
  \dodoi{10.1093/mnras/stv2734}

\bibitem[{D. Boubert {et~al.}(2018)Boubert, Guillochon, Hawkins, Ginsburg,
  Evans, \& Strader}]{Boubert2018}
Boubert, D., Guillochon, J., Hawkins, K., {et~al.} 2018,
  \bibinfo{title}{{Revisiting hypervelocity stars after {G}aia {DR2}},} \mnras,
  479, 2789, \dodoi{10.1093/mnras/sty1601}

\bibitem[{J. Bovy(2014)Bovy}]{Bovy2014}
Bovy, J. 2014, \bibinfo{title}{{Dynamical Modeling of Tidal Streams},} \apj,
  795, 95, \dodoi{10.1088/0004-637X/795/1/95}

\bibitem[{J. Bovy(2015)Bovy}]{Bovy2015}
Bovy, J. 2015, \bibinfo{title}{{galpy: A {P}ython Library for Galactic
  Dynamics},} \apjs, 216, 29, \dodoi{10.1088/0067-0049/216/2/29}

\bibitem[{J. Bovy {et~al.}(2016)Bovy, Bahmanyar, Fritz, \&
  Kallivayalil}]{Bovy2016}
Bovy, J., Bahmanyar, A., Fritz, T.~K., \& Kallivayalil, N. 2016,
  \bibinfo{title}{{The Shape of the Inner {M}ilky {W}ay Halo from Observations
  of the {P}al 5 and {GD}-1 Stellar Streams},} \apj, 833, 31,
  \dodoi{10.3847/1538-4357/833/1/31}

\bibitem[{J. Bovy \& H.-W. Rix(2013)Bovy \& Rix}]{BovyRix2013}
Bovy, J., \& Rix, H.-W. 2013, \bibinfo{title}{{A Direct Dynamical Measurement
  of the {M}ilky {W}ay's Disk Surface Density Profile, Disk Scale Length, and
  Dark Matter Profile at 4 kpc $\lesssim$ R $\lesssim$ 9 kpc},} \apj, 779, 115,
  \dodoi{10.1088/0004-637X/779/2/115}

\bibitem[{J. Bradbury {et~al.}(2018)Bradbury, Frostig, Hawkins, Johnson, Leary,
  Maclaurin, Necula, Paszke, VanderPlas, Wanderman-Milne, \&
  Zhang}]{Bradbury2018}
Bradbury, J., Frostig, R., Hawkins, P., {et~al.} 2018, {{JAX}: composable
  transformations of {P}ython+{N}um{P}y programs}, 0.11.2
  \url{http://github.com/jax-ml/jax}

\bibitem[{M.~R. Buckley {et~al.}(2023)Buckley, Lim, Putney, \&
  Shih}]{Buckley2023}
Buckley, M.~R., Lim, S.~H., Putney, E., \& Shih, D. 2023,
  \bibinfo{title}{{Measuring {G}alactic dark matter through unsupervised
  machine learning},} \mnras, 521, 5100, \dodoi{10.1093/mnras/stad843}

\bibitem[{J.-E. Campagne {et~al.}(2023)Campagne, Lanusse, Zuntz, Boucaud,
  Casas, Karamanis, Kirkby, Lanzieri, Peel, \& Li}]{Campagne2023}
Campagne, J.-E., Lanusse, F., Zuntz, J., {et~al.} 2023,
  \bibinfo{title}{{{JAX-COSMO}: An End-to-End Differentiable and {GPU}
  Accelerated Cosmology Library},} The Open Journal of Astrophysics, 6, 15,
  \dodoi{10.21105/astro.2302.05163}

\bibitem[{M. Cautun {et~al.}(2020)Cautun, Ben{\'i}tez-Llambay, Deason, Frenk,
  Fattahi, G{\'o}mez, Grand, Oman, Navarro, \& Simpson}]{Cautun2020}
Cautun, M., Ben{\'i}tez-Llambay, A., Deason, A.~J., {et~al.} 2020,
  \bibinfo{title}{{The {M}ilky {W}ay total mass profile as inferred from {G}aia
  {DR2}},} \mnras, 494, 4291, \dodoi{10.1093/mnras/staa1017}

\bibitem[{D. Chemaly {et~al.}(2026)Chemaly, Sola, Belokurov, Koposov, Meyong,
  Zhang, \& Erkal}]{Chemaly2026}
Chemaly, D., Sola, E., Belokurov, V., {et~al.} 2026,
  \bibinfo{title}{{Constraining the population of dark matter halo shapes using
  hierarchical inference with extragalactic stellar streams},} \mnras, 550,
  stag1161, \dodoi{10.1093/mnras/stag1161}

\bibitem[{R.~T.~Q. Chen {et~al.}(2018)Chen, Rubanova, Bettencourt, \&
  Duvenaud}]{Chen2018}
Chen, R.~T.~Q., Rubanova, Y., Bettencourt, J., \& Duvenaud, D. 2018,
  \bibinfo{title}{{Neural Ordinary Differential Equations},} in Advances in
  Neural Information Processing Systems 31.
\newblock \doarXiv{1806.07366}

\bibitem[{Y. Chen {et~al.}(2025)Chen, Valluri, Gnedin, \& Ash}]{Chen2025}
Chen, Y., Valluri, M., Gnedin, O.~Y., \& Ash, N. 2025,
  \bibinfo{title}{{Improved Particle Spray Algorithm for Modeling Globular
  Cluster Streams},} \apjs, 276, 32, \dodoi{10.3847/1538-4365/ad9904}

\bibitem[{O. Contigiani {et~al.}(2019)Contigiani, Rossi, \&
  Marchetti}]{Contigiani2019}
Contigiani, O., Rossi, E.~M., \& Marchetti, T. 2019, \bibinfo{title}{{On
  measuring the {G}alactic dark matter halo with hypervelocity stars},} \mnras,
  487, 4025, \dodoi{10.1093/mnras/stz1547}

\bibitem[{W. {Dehnen}(1999){Dehnen}}]{Dehnen1999}
{Dehnen}, W. 1999, \bibinfo{title}{{Simple Distribution Functions for Stellar
  Disks},} \aj, 118, 1201, \dodoi{10.1086/301010}

\bibitem[{S. Duane {et~al.}(1987)Duane, Kennedy, Pendleton, \&
  Roweth}]{Duane1987}
Duane, S., Kennedy, A.~D., Pendleton, B.~J., \& Roweth, D. 1987,
  \bibinfo{title}{{Hybrid {M}onte {C}arlo},} Physics Letters B, 195, 216,
  \dodoi{10.1016/0370-2693(87)91197-X}

\bibitem[{{Euclid Collaboration} {et~al.}(2026){Euclid Collaboration},
  Starkman, Nibauer, Pearson, Wu, Walmsley, Necib, Bovy, Marleau, Sola, Scott,
  Altieri, {et~al.}}]{Euclid2026}
{Euclid Collaboration}, Starkman, N., Nibauer, J., {et~al.} 2026,
  \bibinfo{title}{{{Euclid} Quick Data Release ({Q}1): The geometry of dark
  matter halos from extragalactic streams},} arXiv e-prints, arXiv:2606.21774.
\newblock \doarXiv{2606.21774}

\bibitem[{N.~W. {Evans} {et~al.}(1997){Evans}, {H{\"a}fner}, \& {de
  Zeeuw}}]{Evans1997}
{Evans}, N.~W., {H{\"a}fner}, R.~M., \& {de Zeeuw}, P.~T. 1997,
  \bibinfo{title}{{Simple three-integral scale-free galaxy models},} \mnras,
  286, 315, \dodoi{10.1093/mnras/286.2.315}

\bibitem[{M.~A. Fardal {et~al.}(2015)Fardal, Huang, \& Weinberg}]{Fardal2015}
Fardal, M.~A., Huang, S., \& Weinberg, M.~D. 2015, \bibinfo{title}{{Generation
  of mock tidal streams},} \mnras, 452, 301, \dodoi{10.1093/mnras/stv1198}

\bibitem[{J.~C. Forbes {et~al.}(2025)Forbes, Bannister, Lintott, Forrest,
  Portegies~Zwart, Dorsey, Albrow, \& Hopkins}]{Forbes2025}
Forbes, J.~C., Bannister, M.~T., Lintott, C., {et~al.} 2025,
  \bibinfo{title}{{He Awa Whiria: The Tidal Streams of Interstellar Objects},}
  \apj, 988, 121, \dodoi{10.3847/1538-4357/adc9ac}

\bibitem[{N. Frankel {et~al.}(2020)Frankel, Sanders, Ting, \&
  Rix}]{Frankel2020}
Frankel, N., Sanders, J., Ting, Y.-S., \& Rix, H.-W. 2020,
  \bibinfo{title}{{Keeping It Cool: Much Orbit Migration, yet Little Heating,
  in the {G}alactic Disk},} \apj, 896, 15, \dodoi{10.3847/1538-4357/ab910c}

\bibitem[{T.~K. Fritz {et~al.}(2018)Fritz, Battaglia, Pawlowski, Kallivayalil,
  van~der Marel, Sohn, Brook, \& Besla}]{Fritz2018}
Fritz, T.~K., Battaglia, G., Pawlowski, M.~S., {et~al.} 2018,
  \bibinfo{title}{{{G}aia {DR2} proper motions of dwarf galaxies within 420
  kpc. Orbits, {M}ilky {W}ay mass, tidal influences, planar alignments, and
  group infall},} \aap, 619, A103, \dodoi{10.1051/0004-6361/201833343}

\bibitem[{{Gaia Collaboration} {et~al.}(2016){Gaia Collaboration}, Prusti,
  de~Bruijne, Brown, Vallenari, Babusiaux, Bailer-Jones, Bastian, Biermann,
  Evans, Eyer, Jansen, {et~al.}}]{Gaia2016}
{Gaia Collaboration}, Prusti, T., de~Bruijne, J.~H.~J., {et~al.} 2016,
  \bibinfo{title}{{The {G}aia mission},} \aap, 595, A1,
  \dodoi{10.1051/0004-6361/201629272}

\bibitem[{{Gaia Collaboration} {et~al.}(2018){Gaia Collaboration}, Brown,
  Vallenari, Prusti, de~Bruijne, Babusiaux, Bailer-Jones, Biermann, Evans,
  Eyer, Jansen, Jordi, {et~al.}}]{Gaia2018}
{Gaia Collaboration}, Brown, A.~G.~A., Vallenari, A., {et~al.} 2018,
  \bibinfo{title}{{{G}aia Data Release 2. Summary of the contents and survey
  properties},} \aap, 616, A1, \dodoi{10.1051/0004-6361/201833051}

\bibitem[{{Gaia Collaboration} {et~al.}(2023){Gaia Collaboration}, Vallenari,
  Brown, Prusti, de~Bruijne, Arenou, Babusiaux, Biermann, Creevey, Ducourant,
  Evans, Eyer, {et~al.}}]{Gaia2023}
{Gaia Collaboration}, Vallenari, A., Brown, A.~G.~A., {et~al.} 2023,
  \bibinfo{title}{{{G}aia Data Release 3. Summary of the content and survey
  properties},} \aap, 674, A1, \dodoi{10.1051/0004-6361/202243940}

\bibitem[{M. {Galassi} {et~al.}(2009){Galassi}, {Davies}, {Theiler}, {Gough},
  {Jungman}, {Alken}, {Booth}, \& {Rossi}}]{Galassi2009}
{Galassi}, M., {Davies}, J., {Theiler}, J., {et~al.} 2009, {GNU Scientific
  Library Reference Manual}, 3rd edn. (Network Theory Ltd.)

\bibitem[{O.~Y. Gnedin {et~al.}(2005)Gnedin, Gould, Miralda-Escud{\'e}, \&
  Zentner}]{Gnedin2005}
Gnedin, O.~Y., Gould, A., Miralda-Escud{\'e}, J., \& Zentner, A.~R. 2005,
  \bibinfo{title}{{Probing the Shape of the {G}alactic Halo with Hypervelocity
  Stars},} \apj, 634, 344, \dodoi{10.1086/496958}

\bibitem[{G.~M. Green {et~al.}(2023)Green, Ting, \& Kamdar}]{Green2023}
Green, G.~M., Ting, Y.-S., \& Kamdar, H. 2023, \bibinfo{title}{{Deep Potential:
  Recovering the Gravitational Potential from a Snapshot of Phase Space},}
  \apj, 942, 26, \dodoi{10.3847/1538-4357/aca3a7}

\bibitem[{C.~R. Harris {et~al.}(2020)Harris, Millman, van~der Walt, Gommers,
  Virtanen, Cournapeau, Wieser, Taylor, Berg, Smith, Kern, Picus,
  {et~al.}}]{Harris2020}
Harris, C.~R., Millman, K.~J., van~der Walt, S.~J., {et~al.} 2020,
  \bibinfo{title}{{Array programming with {NumPy}},} \nat, 585, 357,
  \dodoi{10.1038/s41586-020-2649-2}

\bibitem[{L. Hernquist(1990)Hernquist}]{Hernquist1990}
Hernquist, L. 1990, \bibinfo{title}{{An analytical model for spherical galaxies
  and bulges},} \apj, 356, 359, \dodoi{10.1086/168845}

\bibitem[{M.~D. Hoffman \& A. Gelman(2014)Hoffman \&
  Gelman}]{HoffmanGelman2014}
Hoffman, M.~D., \& Gelman, A. 2014, \bibinfo{title}{{The No-U-Turn Sampler:
  Adaptively Setting Path Lengths in {H}amiltonian {M}onte {C}arlo},} Journal
  of Machine Learning Research, 15, 1593.
\newblock \doarXiv{1111.4246}

\bibitem[{J.~A.~S. Hunt \& J. Bovy(2018)Hunt \& Bovy}]{Hunt2018}
Hunt, J.~A.~S., \& Bovy, J. 2018, \bibinfo{title}{{The 4:1 outer {L}indblad
  resonance of a long-slow bar as an explanation for the {H}ercules stream},}
  \mnras, 477, 3945, \dodoi{10.1093/mnras/sty921}

\bibitem[{J.~D. Hunter(2007)Hunter}]{Hunter2007}
Hunter, J.~D. 2007, \bibinfo{title}{{Matplotlib: A {2D} Graphics Environment},}
  Computing in Science \& Engineering, 9, 90, \dodoi{10.1109/MCSE.2007.55}

\bibitem[{T. Kalda \& G.~M. Green(2025)Kalda \& Green}]{Kalda2025}
Kalda, T., \& Green, G.~M. 2025, \bibinfo{title}{{Deep Potential: Recovering
  the Gravitational Potential and Local Pattern Speed in the Solar Neighborhood
  with {GDR}3 Using Normalizing Flows},} \apj, 992, 84,
  \dodoi{10.3847/1538-4357/adf8ea}

\bibitem[{P. Kidger(2021)Kidger}]{Kidger2021}
Kidger, P. 2021, \bibinfo{title}{{On Neural Differential Equations},} PhD
  thesis, University of Oxford.
\newblock \doarXiv{2202.02435}

\bibitem[{P. Kidger \& C. Garcia(2021)Kidger \& Garcia}]{KidgerGarcia2021}
Kidger, P., \& Garcia, C. 2021, \bibinfo{title}{{Equinox: neural networks in
  {JAX} via callable {PyTrees} and filtered transformations},} arXiv e-prints,
  arXiv:2111.00254.
\newblock \doarXiv{2111.00254}

\bibitem[{H. Kinoshita {et~al.}(1991)Kinoshita, Yoshida, \&
  Nakai}]{Kinoshita1991}
Kinoshita, H., Yoshida, H., \& Nakai, H. 1991, \bibinfo{title}{{Symplectic
  integrators and their application to dynamical astronomy},} Celestial
  Mechanics and Dynamical Astronomy, 50, 59, \dodoi{10.1007/BF00048986}

\bibitem[{S.~E. Koposov {et~al.}(2020)Koposov, Boubert, Li, Erkal, Da~Costa,
  Zucker, Ji, Kuehn, Lewis, Mackey, Simpson, Shipp, {et~al.}}]{Koposov2020}
Koposov, S.~E., Boubert, D., Li, T.~S., {et~al.} 2020,
  \bibinfo{title}{{Discovery of a nearby 1700 km s$^{-1}$ star ejected from the
  {M}ilky {W}ay by {S}gr {A}*},} \mnras, 491, 2465,
  \dodoi{10.1093/mnras/stz3081}

\bibitem[{J.~M.~M. Lane {et~al.}(2022)Lane, Bovy, \& Mackereth}]{Lane2022}
Lane, J.~M.~M., Bovy, J., \& Mackereth, J.~T. 2022, \bibinfo{title}{{The
  kinematic properties of Milky Way stellar halo populations},} \mnras, 510,
  5119, \dodoi{10.1093/mnras/stab3755}

\bibitem[{J.~M.~M. Lane {et~al.}(2023)Lane, Bovy, \& Mackereth}]{Lane2023}
Lane, J.~M.~M., Bovy, J., \& Mackereth, J.~T. 2023, \bibinfo{title}{{The
  stellar mass of the {G}aia-{S}ausage/{E}nceladus accretion remnant},} \mnras,
  526, 1209, \dodoi{10.1093/mnras/stad2834}

\bibitem[{Y. Li {et~al.}(2024)Li, Modi, Jamieson, Zhang, Lu, Feng, Lanusse, \&
  Greengard}]{Li2024}
Li, Y., Modi, C., Jamieson, D., {et~al.} 2024, \bibinfo{title}{{Differentiable
  Cosmological Simulation with the Adjoint Method},} \apjs, 270, 36,
  \dodoi{10.3847/1538-4365/ad0ce7}

\bibitem[{M. Lienen \& S. G{\"u}nnemann(2022)Lienen \&
  G{\"u}nnemann}]{LienenGunnemann2022}
Lienen, M., \& G{\"u}nnemann, S. 2022, \bibinfo{title}{{torchode: A Parallel
  {ODE} Solver for {PyTorch}},} arXiv e-prints, arXiv:2210.12375.
\newblock \doarXiv{2210.12375}

\bibitem[{C. Loken {et~al.}(2010)Loken, Gruner, Groer, Peltier, Bunn, Craig,
  Henriques, Dempsey, Yu, Chen, Dursi, Chong, Northrup, Pinto, Knecht, \& van
  Zon}]{Loken2010}
Loken, C., Gruner, D., Groer, L., {et~al.} 2010, \bibinfo{title}{{SciNet:
  Lessons Learned from Building a Power-efficient Top-20 System and Data
  Centre},} Journal of Physics Conference Series, 256, 012026,
  \dodoi{10.1088/1742-6596/256/1/012026}

\bibitem[{M. Lucey {et~al.}(2023)Lucey, Pearson, Hunt, Hawkins, Ness, Petersen,
  Price-Whelan, \& Weinberg}]{Lucey2023}
Lucey, M., Pearson, S., Hunt, J.~A.~S., {et~al.} 2023,
  \bibinfo{title}{{Dynamically constraining the length of the {M}ilky {W}ay
  bar},} \mnras, 520, 4779, \dodoi{10.1093/mnras/stad406}

\bibitem[{J.~T. Mackereth \& J. Bovy(2018)Mackereth \&
  Bovy}]{MackerethBovy2018}
Mackereth, J.~T., \& Bovy, J. 2018, \bibinfo{title}{{Fast Estimation of Orbital
  Parameters in {M}ilky {W}ay-like Potentials},} \pasp, 130, 114501,
  \dodoi{10.1088/1538-3873/aadcdd}

\bibitem[{J.~T. Mackereth {et~al.}(2019)Mackereth, Bovy, Leung, Schiavon,
  Trick, Chaplin, Cunha, Feuillet, Majewski, Martig, Miglio, Nidever,
  {et~al.}}]{Mackereth2019}
Mackereth, J.~T., Bovy, J., Leung, H.~W., {et~al.} 2019,
  \bibinfo{title}{{Dynamical heating across the {M}ilky {W}ay disc using
  {APOGEE} and {G}aia},} \mnras, 489, 176, \dodoi{10.1093/mnras/stz1521}

\bibitem[{K. Malhan \& R.~A. Ibata(2019)Malhan \& Ibata}]{Malhan2019}
Malhan, K., \& Ibata, R.~A. 2019, \bibinfo{title}{{Constraining the {M}ilky
  {W}ay halo potential with the {GD}-1 stellar stream},} \mnras, 486, 2995,
  \dodoi{10.1093/mnras/stz1035}

\bibitem[{T. Marchetti {et~al.}(2022)Marchetti, Evans, \&
  Rossi}]{Marchetti2022}
Marchetti, T., Evans, F.~A., \& Rossi, E.~M. 2022, \bibinfo{title}{{{G}aia
  {DR3} in 6D: the search for fast hypervelocity stars and constraints on the
  {G}alactic {C}entre environment},} \mnras, 515, 767,
  \dodoi{10.1093/mnras/stac1777}

\bibitem[{C. Modi {et~al.}(2021)Modi, Lanusse, \& Seljak}]{Modi2021}
Modi, C., Lanusse, F., \& Seljak, U. 2021, \bibinfo{title}{{{FlowPM}:
  Distributed {TensorFlow} implementation of the {FastPM} cosmological {N}-body
  solver},} Astronomy and Computing, 37, 100505,
  \dodoi{10.1016/j.ascom.2021.100505}

\bibitem[{L. Necib \& T. Lin(2022)Necib \& Lin}]{Necib2022}
Necib, L., \& Lin, T. 2022, \bibinfo{title}{{Substructure at High Speed. II.
  The Local Escape Velocity and Milky Way Mass with {G}aia eDR3},} \apj, 926,
  189, \dodoi{10.3847/1538-4357/ac4244}

\bibitem[{J. Nibauer \& A. Bonaca(2025)Nibauer \& Bonaca}]{Nibauer2025b}
Nibauer, J., \& Bonaca, A. 2025, \bibinfo{title}{{Galactic Accelerations from
  the {GD}-1 Stream Suggest a Tilted Dark Matter Halo},} \apjl, 985, L22,
  \dodoi{10.3847/2041-8213/add0a9}

\bibitem[{J. Nibauer {et~al.}(2025)Nibauer, Bonaca, Spergel, Price-Whelan,
  Greene, Starkman, \& Johnston}]{Nibauer2025}
Nibauer, J., Bonaca, A., Spergel, D.~N., {et~al.} 2025,
  \bibinfo{title}{{{StreamSculptor}: Hamiltonian Perturbation Theory for
  Stellar Streams in Flexible Potentials with Differentiable Simulations},}
  \apj, 983, 68, \dodoi{10.3847/1538-4357/adb8e8}

\bibitem[{T.~N. O’Doherty {et~al.}(2023)O’Doherty, Bahramian, Miller-Jones,
  Goodwin, Mandel, Willcox, Atri, \& Strader}]{ODoherty2023}
O’Doherty, T.~N., Bahramian, A., Miller-Jones, J.~C.~A., {et~al.} 2023,
  \bibinfo{title}{{An observationally derived kick distribution for neutron
  stars in binary systems},} \mnras, 521, 2504, \dodoi{10.1093/mnras/stad680}

\bibitem[{A.~B. Pace {et~al.}(2022)Pace, Erkal, \& Li}]{Pace2022}
Pace, A.~B., Erkal, D., \& Li, T.~S. 2022, \bibinfo{title}{{Proper Motions,
  Orbits, and Tidal Influences of {M}ilky {W}ay Dwarf Spheroidal Galaxies},}
  \apj, 940, 136, \dodoi{10.3847/1538-4357/ac997b}

\bibitem[{A. Paszke {et~al.}(2019)Paszke, Gross, Massa, Lerer, Bradbury,
  Chanan, Killeen, Lin, Gimelshein, Antiga, Desmaison, K{\"o}pf,
  {et~al.}}]{Paszke2019}
Paszke, A., Gross, S., Massa, F., {et~al.} 2019, \bibinfo{title}{{{PyTorch}: An
  Imperative Style, High-Performance Deep Learning Library},} in Advances in
  Neural Information Processing Systems 32.
\newblock \doarXiv{1912.01703}

\bibitem[{F.~I. Pelupessy {et~al.}(2013)Pelupessy, van Elteren, de~Vries,
  McMillan, Drost, \& Portegies~Zwart}]{Pelupessy2013}
Pelupessy, F.~I., van Elteren, A., de~Vries, N., {et~al.} 2013,
  \bibinfo{title}{{The Astrophysical Multipurpose Software Environment},} \aap,
  557, A84, \dodoi{10.1051/0004-6361/201321252}

\bibitem[{M. Ponce {et~al.}(2019)Ponce, van Zon, Northrup, Gruner, Chen,
  Ertinaz, Fedoseev, Groer, Mao, Mundim, Nolta, Pinto, Saldarriaga, Slavnic,
  Spence, Yu, \& Peltier}]{Ponce2019}
Ponce, M., van Zon, R., Northrup, S., {et~al.} 2019, \bibinfo{title}{{Deploying
  a Top-100 Supercomputer for Large Parallel Workloads: the Niagara
  Supercomputer},} in Proceedings of the Practice and Experience in Advanced
  Research Computing on Rise of the Machines (learning), PEARC '19 (ACM), 1--8,
  \dodoi{10.1145/3332186.3332195}

\bibitem[{A. Pontzen {et~al.}(2013)Pontzen, Ro{\v{s}}kar, Stinson, \&
  Woods}]{Pontzen2013}
Pontzen, A., Ro{\v{s}}kar, R., Stinson, G., \& Woods, R. 2013, {pynbody:
  N-Body/SPH analysis for python},, Astrophysics Source Code Library, record
  ascl:1305.002

\bibitem[{S. Portegies~Zwart \& S. McMillan(2018)Portegies~Zwart \&
  McMillan}]{PortegiesZwart2018}
Portegies~Zwart, S., \& McMillan, S. 2018, {Astrophysical Recipes: The art of
  AMUSE} (IOP Publishing), \dodoi{10.1088/978-0-7503-1320-9}

\bibitem[{S. Portegies~Zwart {et~al.}(2009)Portegies~Zwart, McMillan, Harfst,
  Groen, Fujii, {\'O}~Nuall{\'a}in, Glebbeek, Heggie, Lombardi, Hut, Angelou,
  Banerjee, {et~al.}}]{PortegiesZwart2009}
Portegies~Zwart, S., McMillan, S., Harfst, S., {et~al.} 2009,
  \bibinfo{title}{{A multiphysics and multiscale software environment for
  modeling astrophysical systems},} \na, 14, 369,
  \dodoi{10.1016/j.newast.2008.10.006}

\bibitem[{A.~M. Price-Whelan(2017)Price-Whelan}]{PriceWhelan2017}
Price-Whelan, A.~M. 2017, \bibinfo{title}{{Gala: A {P}ython package for
  galactic dynamics},} The Journal of Open Source Software, 2, 388,
  \dodoi{10.21105/joss.00388}

\bibitem[{A.~M. Price-Whelan {et~al.}(2025)Price-Whelan, Hunt, Horta, Oeur,
  Hogg, Johnston, \& Widrow}]{PriceWhelan2025}
Price-Whelan, A.~M., Hunt, J.~A.~S., Horta, D., {et~al.} 2025,
  \bibinfo{title}{{Data-driven Dynamics with Orbital Torus Imaging: A Flexible
  Model of the Vertical Phase Space of the Galaxy},} \apj, 979, 115,
  \dodoi{10.3847/1538-4357/ad969a}

\bibitem[{Y. Qian {et~al.}(2022)Qian, Arshad, \& Bovy}]{Qian2022}
Qian, Y., Arshad, Y., \& Bovy, J. 2022, \bibinfo{title}{{The structure of
  accreted stellar streams},} \mnras, 511, 2339, \dodoi{10.1093/mnras/stac238}

\bibitem[{S. Reino {et~al.}(2021)Reino, Rossi, Sanderson, Sellentin, Helmi,
  Koppelman, \& Sharma}]{Reino2021}
Reino, S., Rossi, E.~M., Sanderson, R.~E., {et~al.} 2021,
  \bibinfo{title}{{Galactic potential constraints from clustering in action
  space of combined stellar stream data},} \mnras, 502, 4170,
  \dodoi{10.1093/mnras/stab304}

\bibitem[{J.~L. Sanders {et~al.}(2016)Sanders, Bovy, \& Erkal}]{Sanders2016}
Sanders, J.~L., Bovy, J., \& Erkal, D. 2016, \bibinfo{title}{{Dynamics of
  stream-subhalo interactions},} \mnras, 457, 3817,
  \dodoi{10.1093/mnras/stw232}

\bibitem[{F. {Shu}(1969){Shu}}]{Shu1969}
{Shu}, F. 1969, \bibinfo{title}{{Models of Partially Relaxed Stellar Disks},}
  \apj, 158, 505, \dodoi{10.1086/150214}

\bibitem[{N. Starkman {et~al.}(2026{\natexlab{a}})Starkman, Nibauer, Pearson,
  Wu, \& Necib}]{Starkman2026b}
Starkman, N., Nibauer, J., Pearson, S., Wu, S., \& Necib, L.
  2026{\natexlab{a}}, \bibinfo{title}{{When Streams Curve Away: a Test of Dark
  Matter from Extragalactic Stellar Stream Populations},} arXiv e-prints,
  arXiv:2609.40057.
\newblock \doarXiv{2609.40057}

\bibitem[{N. Starkman {et~al.}(2025)Starkman, Price-Whelan, \&
  Nibauer}]{Starkman2025}
Starkman, N., Price-Whelan, A., \& Nibauer, J. 2025, \bibinfo{title}{{unxt: A
  Python package for unit-aware computing with JAX},} Journal of Open Source
  Software, 10, 7771, \dodoi{10.21105/joss.07771}

\bibitem[{N. Starkman {et~al.}(2026{\natexlab{b}})Starkman, Price-Whelan,
  Nibauer, \& Chen}]{Starkman2026}
Starkman, N., Price-Whelan, A., Nibauer, J., \& Chen, Y. 2026{\natexlab{b}},
  {{GalacticDynamics/galax}: v0.0.3}, v0.0.3 Zenodo,
  \dodoi{10.5281/zenodo.21985205}

\bibitem[{G. Stef{\'a}nsson {et~al.}(2023)Stef{\'a}nsson, Mahadevan, Miguel,
  Robertson, Delamer, Kanodia, Ca{\~n}as, Winn, Ninan, Terrien, Holcomb, Ford,
  {et~al.}}]{Stefansson2023}
Stef{\'a}nsson, G., Mahadevan, S., Miguel, Y., {et~al.} 2023,
  \bibinfo{title}{{A {N}eptune-mass exoplanet in close orbit around a very
  low-mass star challenges formation models},} Science, 382, 1031,
  \dodoi{10.1126/science.abo0233}

\bibitem[{C. Swiggum {et~al.}(2024)Swiggum, Alves, Benjamin, Ratzenb{\"o}ck,
  Miret-Roig, Großschedl, Meingast, Goodman, Konietzka, Zucker, Hunt, \&
  Reffert}]{Swiggum2024}
Swiggum, C., Alves, J., Benjamin, R., {et~al.} 2024, \bibinfo{title}{{Most
  nearby young star clusters formed in three massive complexes},} \nat, 631,
  49, \dodoi{10.1038/s41586-024-07496-9}

\bibitem[{A. Szpilfidel {et~al.}(2026)Szpilfidel, Boldrini, Bovy, \&
  Di~Matteo}]{Szpilfidel2026}
Szpilfidel, A., Boldrini, P., Bovy, J., \& Di~Matteo, P. 2026,
  \bibinfo{title}{{Fuzzy dark matter dynamical friction: Stalling of globular
  clusters induced by dynamical heating},} \aap, 710, A189,
  \dodoi{10.1051/0004-6361/202558231}

\bibitem[{P.~J. Teuben(1995)Teuben}]{Teuben1995}
Teuben, P.~J. 1995, \bibinfo{title}{{The Stellar Dynamics Toolbox {NEMO}},} in
  Astronomical Society of the Pacific Conference Series, Vol.~77, Astronomical
  Data Analysis Software and Systems IV, ed. R.~A. Shaw, H.~E. Payne, \&
  J.~J.~E. Hayes, 398

\bibitem[{{The Astropy Collaboration} {et~al.}(2022){The Astropy
  Collaboration}, Price-Whelan, Lim, Earl, Starkman, Bradley, Shupe, Patil,
  Corrales, Brasseur, N{\"o}the, Donath, {et~al.}}]{Astropy2022}
{The Astropy Collaboration}, Price-Whelan, A.~M., Lim, P.~L., {et~al.} 2022,
  \bibinfo{title}{{The Astropy Project: Sustaining and Growing a
  Community-oriented Open-source Project and the Latest Major Release (v5.0) of
  the Core Package},} \apj, 935, 167, \dodoi{10.3847/1538-4357/ac7c74}

\bibitem[{S. Tonnesen(2019)Tonnesen}]{Tonnesen2019}
Tonnesen, S. 2019, \bibinfo{title}{{The Journey Counts: The Importance of
  Including Orbits when Simulating Ram Pressure Stripping},} \apj, 874, 161,
  \dodoi{10.3847/1538-4357/ab0960}

\bibitem[{W.~H. Trick {et~al.}(2016)Trick, Bovy, \& Rix}]{Trick2016}
Trick, W.~H., Bovy, J., \& Rix, H.-W. 2016, \bibinfo{title}{{Action-based
  Dynamical Modeling for the {M}ilky {W}ay Disk},} \apj, 830, 97,
  \dodoi{10.3847/0004-637X/830/2/97}

\bibitem[{E. Vasiliev(2019)Vasiliev}]{Vasiliev2019}
Vasiliev, E. 2019, \bibinfo{title}{{{AGAMA}: action-based galaxy modelling
  architecture},} \mnras, 482, 1525, \dodoi{10.1093/mnras/sty2672}

\bibitem[{P. Virtanen {et~al.}(2020)Virtanen, Gommers, Oliphant, Haberland,
  Reddy, Cournapeau, Burovski, Peterson, Weckesser, Bright, van~der Walt,
  Brett, {et~al.}}]{Virtanen2020}
Virtanen, P., Gommers, R., Oliphant, T.~E., {et~al.} 2020,
  \bibinfo{title}{{{SciPy} 1.0: fundamental algorithms for scientific computing
  in {P}ython},} Nature Methods, 17, 261, \dodoi{10.1038/s41592-019-0686-2}

\bibitem[{G. Viterbo \& T. Buck(2025)Viterbo \& Buck}]{Viterbo2025}
Viterbo, G., \& Buck, T. 2025, \bibinfo{title}{{Differentiable {N}-body code
  for Galactic Dynamics -- {Odisseo}},} arXiv e-prints, arXiv:2511.22468.
\newblock \doarXiv{2511.22468}

\bibitem[{J.~J. {Webb} {et~al.}(2019){Webb}, {Bovy}, {Carlberg}, \&
  {Gieles}}]{Webb2019}
{Webb}, J.~J., {Bovy}, J., {Carlberg}, R.~G., \& {Gieles}, M. 2019,
  \bibinfo{title}{{Modelling the effects of dark matter substructure on
  globular cluster evolution with the tidal approximation},} \mnras, 488, 5748,
  \dodoi{10.1093/mnras/stz2118}

\bibitem[{J.~J. {Webb} {et~al.}(2020){Webb}, {Price-Jones}, {Bovy}, {Portegies
  Zwart}, {Hunt}, {Mackereth}, \& {Leung}}]{Webb2020}
{Webb}, J.~J., {Price-Jones}, N., {Bovy}, J., {et~al.} 2020,
  \bibinfo{title}{{Searching for solar siblings in APOGEE and Gaia DR2 with
  N-body simulations},} \mnras, 494, 2268, \dodoi{10.1093/mnras/staa788}

\bibitem[{S. Wu {et~al.}(2026)Wu, Starkman, Pearson, Nibauer, Miro-Carretero,
  \& Martinez-Delgado}]{Wu2026b}
Wu, S., Starkman, N., Pearson, S., {et~al.} 2026, \bibinfo{title}{{{Potamides}:
  Mapping Dark Matter Halo Shapes from Stellar Stream Tracks in the Local
  Universe},} arXiv e-prints, arXiv:2605.07809.
\newblock \doarXiv{2605.07809}

\bibitem[{M. Xiang \& H.-W. Rix(2022)Xiang \& Rix}]{Xiang2022}
Xiang, M., \& Rix, H.-W. 2022, \bibinfo{title}{{A time-resolved picture of our
  {M}ilky {W}ay's early formation history},} \nat, 603, 599,
  \dodoi{10.1038/s41586-022-04496-5}

\bibitem[{T. Yang {et~al.}(2020)Yang, Boruah, \& Afshordi}]{Yang2020}
Yang, T., Boruah, S.~S., \& Afshordi, N. 2020, \bibinfo{title}{{Gravitational
  potential from small-scale clustering in action space: application to {G}aia
  Data Release 2},} \mnras, 493, 3061, \dodoi{10.1093/mnras/staa441}

\bibitem[{H. Yoshida(1990)Yoshida}]{Yoshida1990}
Yoshida, H. 1990, \bibinfo{title}{{Construction of higher order symplectic
  integrators},} Physics Letters A, 150, 262,
  \dodoi{10.1016/0375-9601(90)90092-3}

\bibitem[{C. Zucker {et~al.}(2022)Zucker, Goodman, Alves, Bialy, Foley,
  Speagle, Groschedl, Finkbeiner, Burkert, Khimey, \& Swiggum}]{Zucker2022}
Zucker, C., Goodman, A.~A., Alves, J., {et~al.} 2022, \bibinfo{title}{{Star
  formation near the {S}un is driven by expansion of the {L}ocal {B}ubble},}
  \nat, 601, 334, \dodoi{10.1038/s41586-021-04286-5}

\end{thebibliography}
